\documentclass[
aps,
prx,
longbibliography,
twocolumn,
showpacs,
superscriptaddress
]{revtex4-2}

\usepackage{graphicx}
\usepackage{dcolumn}
\usepackage{bm}
\usepackage{amsmath,amssymb,mathtools}
\usepackage{color}
\usepackage{appendix}
\usepackage[normalem]{ulem}
\usepackage[version=4]{mhchem}

\definecolor{darkblue}{rgb}{0,0,0.6}
\definecolor{darkred}{rgb}{0.6,0,0}
\definecolor{darkgreen}{rgb}{0,0.6,0}

\usepackage[colorlinks=true,urlcolor=darkblue,citecolor=darkblue,linkcolor=darkred,hyperfootnotes=false]{hyperref}

\usepackage{soul}
\usepackage{multirow}
\usepackage{tikz}
\usepackage{hhline}
\usepackage{algorithm}
\usepackage{algorithmic}

\usepackage[para,online,flushleft]{threeparttable}
\usepackage{booktabs}
\usepackage{multirow}

\usepackage{setspace}

\makeatother

\begin{document}

\title{Nonequilibrium Fluctuation--Response Theory in the Frequency Domain}

\author{Euijoon Kwon}
\affiliation{Quantum Universe Center, Korea Institute for Advanced Study, Seoul 02455, Republic of Korea}

\author{Hyun-Myung Chun}
\affiliation{School of Physics, Korea Institute for Advanced Study, Seoul 02455, Republic of Korea}

\author{Hyunggyu Park}
\affiliation{Quantum Universe Center, Korea Institute for Advanced Study, Seoul 02455, Republic of Korea}

\author{Jae Sung Lee}
\email{jslee@kias.re.kr}
\affiliation{School of Physics, Korea Institute for Advanced Study, Seoul 02455, Republic of Korea}

\date{\today}
 
\begin{abstract}
    We establish a unified fluctuation--response theory in the frequency domain for nonequilibrium steady states governed by overdamped Langevin dynamics and Markov jump processes. 
    The central identity is an exact fluctuation--response relation that reconstructs the power spectrum of general observables from local responses measured at the same frequency. This relation applies to state-dependent observables, current-like observables, and their combinations, and reveals how local dynamics shape global fluctuations.    
    From this identity, we derive a hierarchy of relations in the frequency domain, including response uncertainty relations, kinetic and thermodynamic uncertainty relations, the equilibrium fluctuation--dissipation theorem, and Harada--Sasa relations. This establishes an overarching framework for fluctuation--response theory in the frequency domain.     
    Applications to stochastic networks and driven diffusive systems demonstrate how the theory decomposes fluctuation spectra into local contributions and reveals frequency-dependent tradeoffs between fluctuations, response, and dissipation.
\end{abstract}


\maketitle

\section{Introduction}
\label{sec:introduction}

The relation between a system’s fluctuations and its response to weak perturbations is a central theme in statistical physics. At equilibrium, this relation is expressed by the fluctuation--dissipation theorem (FDT) \cite{Kubo1966fluctuation}. Away from equilibrium, the FDT is generally violated, and several complementary approaches have been developed to characterize fluctuation--response relations, including nonequilibrium FDTs \cite{Agarwal1972fluctuation,Speck2006restoring,Baiesi2009fluctuations,Prost2009generalized,Seifert2010fluctuation,Altaner2016fluctuation} and Harada--Sasa relations that connect FDT violation to dissipation \cite{Harada2005equality,Harada2006energy,Golestanian2025hodrydynamically,Leticia1997}.

In parallel, another line of research has pursued universal constraints on fluctuations. Thermodynamic uncertainty relations (TURs) constrain current fluctuations through entropy production~\cite{Barato2015thermodynamic,Gingrich2016dissipation,Pietzonka2016universal,Pietzonka2017finite,Horowitz2017proof}, while kinetic uncertainty relations (KURs) reveal the role of dynamical activity as a complementary constraint~\cite{DiTerlizzi2019kinetic,Falasco2020unifying,Horowitz2020thermodynamic,Lee2021universal,Vo2022unified,Kwon2024unified}. In contrast to FDT-type relations, which establish equalities linking fluctuations to response, TURs and KURs take the form of inequalities that bound fluctuations. More recently, these ideas have been extended to response bounds, showing that the sensitivity of nonequilibrium systems to kinetic and entropic perturbations is constrained by fluctuations, dissipation, activity, and network topology~\cite{Dechant2020fluctuation,Gao2022thermodynamic,Chun2023trade,FernandesMartins2023topologically,Aslyamov2024nonequilibrium,Gao2024thermodynamic,Ptaszynski2024dissipation,Liu2025dynamical}.

The most recent development is the emergence of fluctuation--response relations (FRRs) far from equilibrium. For Markov jump processes, static FRRs were first derived to express long-time current covariances as quadratic combinations of local responses~\cite{Aslyamov2025nonequilibrium}. This framework was subsequently generalized to state observables and correlations between state and current observables~\cite{Ptaszynski2026nonequilibriumState,Ptaszynski2026nonequilibriumStateCurrent}. Fluctuation--response inequalities (FRIs)~\cite{Kwon2025fluctuation} further extended the theory to more general observables and finite-time dynamics. Together, FRRs and FRIs recover FDT, TURs, and KURs in appropriate limits, revealing a hierarchical structure underlying relations between fluctuations and response. 
This hierarchical framework was recently extended to overdamped Langevin dynamics~\cite{Chun2026fluctuation}.

Formulations of fluctuation--response relations in the frequency domain have also begun to appear, but none provides a fully frequency-resolved equality. Instead, existing approaches either take the form of inequalities~\cite{dechant2026finite}, are restricted to specific settings such as macroscopic Gaussian systems near stable fixed points~\cite{Aslyamov2026macroscopic}, or remain unresolved in frequency, as in finite-time relations for nonautonomous jump processes~\cite{Aslyamov2026dynamical}. From an experimental perspective, reconstructing the power spectrum from the finite-frequency components of response functions is of central importance, as both quantities are directly measurable in the frequency domain. However, such an exact identity connecting spectral power to response at the same finite frequency has not yet been established.

In this work, we derive a frequency-resolved FRR for nonequilibrium steady states in both overdamped Langevin systems and Markov jump processes. We show that the power spectrum of a broad class of observables, including state-dependent and current-like components, can be exactly reconstructed from local linear responses measured at the same frequency. Owing to its decomposition into local response contributions, the resulting FRR makes explicit which spatial locations or transition channels dominate the fluctuation spectrum. From this identity, we further derive a variety of relations and bounds. In particular, we recover the FDT and the Harada--Sasa relation in the frequency domain, derive a frequency-resolved response uncertainty relation (RUR), and obtain a frequency-resolved TUR expressed in terms of the power spectrum. Taken together, these results establish a unified framework for fluctuation--response theory in the frequency domain.

The paper is organized as follows. In Sec.~\ref{sec:main_results}, we introduce the setups for overdamped Langevin dynamics and Markov jump processes, and present the main results. In Sec.~\ref{sec:langevin_frt}, we derive the FRR for overdamped Langevin systems, and in Sec.~\ref{sec:jump_frt} we develop the corresponding edge-resolved formulation for Markov jump processes. In Sec.~\ref{sec:consequences}, we show how RURs, KURs and TURs, the equilibrium FDT, and Harada--Sasa relations follow from this identity. In Sec.~\ref{sec:examples}, we illustrate the theory with examples of Markov jump networks and Langevin systems. Technical details are deferred to the appendices for clarity.

\section{Setups and Main Results}
\label{sec:main_results}

In this section, we first introduce the setups for overdamped Langevin systems and Markov jump processes, including their equations of motion, observables, perturbations, and the corresponding response functions. We then present the main results of the paper: the frequency-resolved FRR and its consequences, including the resulting identities and bounds such as the RURs, TURs, equilibrium FDT, and Harada--Sasa relations in the frequency domain.

\subsection{Setup for overdamped Langevin systems}
\label{subsec:langevin_setup}

We consider an $N$-dimensional overdamped Langevin system described by 
\begin{equation}
    \dot{\bm{x}}(t)
    =
    \textsf M(\bm{x}(t))\bm{F}(\bm{x}(t))
    + \textsf B(\bm{x}(t))\circledast \bm{\xi}(t),
    \label{eq:langevin_antiito}
\end{equation}
where $\textsf M(\bm{x})$ is the mobility matrix, $\bm{F}(\bm{x})$ is the drift force, $\textsf B(\bm{x})$ is the noise-amplitude matrix, and $\bm{\xi}(t)$ is Gaussian white noise satisfying
\begin{equation}
    \langle \bm{\xi}(t)\rangle = \bm{0},
    \qquad
    \langle \xi_i(t)\xi_j(t')\rangle = \delta_{ij}\delta(t-t').
\end{equation}
Throughout, $\langle\cdot\rangle$ denotes an ensemble average over
realizations of the stochastic dynamics.
In Eq.~\eqref{eq:langevin_antiito}, $\circledast$ denotes the anti-It\^o product~\cite{lau2007state}. The following discussion, however, applies to arbitrary stochastic conventions, provided that the drift term is modified by the appropriate conversion factor. For example, under the It\^o convention, the term $\textsf M(\bm{x}) \bm{F}(\bm{x})$ is replaced by $\textsf M(\bm{x})\bm{F}(\bm{x})+ \bigl[\bm{\nabla}_{\bm{x}}^\mathsf{T} \textsf D(\bm{x})\bigr]^{\mathsf T}$, where $\bm{\nabla}_{\bm{x}} \equiv (\partial_{x_1} , \cdots , \partial_{x_N} )^\mathsf{T}$ and $\textsf D(\bm{x})=\frac{1}{2}\textsf B(\bm{x}) \textsf B(\bm{x})^{\mathsf T}$.
The superscript $\mathsf{T}$ denotes transposition of vectors and matrices.
For simplicity, we assume that $\textsf M(\bm{x})$ and $\textsf D(\bm{x})$ are diagonal, i.e., $M_{ij}(\bm{x})=\mu_i (\bm{x})\delta_{ij}$ and $D_{ij}(\bm{x}) =D_i(\bm{x}) \delta_{ij}$. We then introduce the temperature matrix as
\begin{equation}
    \textsf T(\bm{x}) = \mathrm{diag} \bigl(T_1(\bm{x}),\dots,T_N(\bm{x})\bigr),
\end{equation}
and assume the Einstein relation $\textsf D(\bm{x}) = \textsf M(\bm{x}) \textsf T(\bm{x})$. 
We further assume that the dynamics admits a unique nonequilibrium steady state with probability density $\pi(\bm{x})$, with the corresponding steady-state probability current denoted by $\bm{j}_{\mathrm{ss}}(\bm{x})$. The extension to non-diagonal matrices is straightforward.

We consider $N_{\rm O}$ observables, $\bm{\theta} = (\theta_1, \cdots, \theta_{N_{\rm O}})^{\textsf T}$, defined as
\begin{equation}
    \bm{\theta}(t) =
    \bm{G} (\bm{x}(t)) +
    \textsf L(\bm{x}(t))\circ \dot{\bm{x}}(t),
    \label{eq:langevin_observable_vector}
\end{equation}
where $\bm{G}(\bm{x})$ is an $N_{\rm O}$-dimensional vector and $\textsf L(\bm{x})$ is an $N_{\rm O}\times N$ matrix. The product $\circ$ denotes the Stratonovich product. This representation includes purely state-dependent observables $\bm{G}(\bm{x})$, purely current-like observables $\textsf L(\bm{x})\circ \dot{\bm{x}}$, as well as arbitrary linear combinations of the two.

We consider a small perturbation to the function $\phi_k$, the $k$th component of $\bm{\phi}$, where $\bm{\phi}\in\{\bm{F},\ln \textsf{M},\textsf{T}\}$. Since $\textsf{M}$ and $\textsf{T}$ are assumed to be diagonal, each of these quantities has $N$ independent components. With a slight abuse of notation, we identify $\ln\textsf{M}$ and $\textsf{T}$ with the vectors of their diagonal components. Accordingly, we write $\bm{\phi} = (\phi_1,\ldots,\phi_N)^{\mathsf T}$. A local impulsive perturbation of $\phi_k$ at $\bm x = \bm z$ and $t=s$ is given by
\begin{equation}
    \phi_k(\bm{x})
    \mapsto
    \phi_k(\bm{x})
    +
    \epsilon\,\delta(\bm{x}-\bm{z})\delta(t-s),
    \label{eq:langevin_local_impulsive_perturbation}
\end{equation}
where $\epsilon$ is infinitesimal. Note that mobility perturbations induce no response at equilibrium, as they modify only the characteristic timescale of the dynamics. However, such perturbations provide a means to probe nonequilibrium features and are not reducible to conventional force perturbations~\cite{Kwon2025fluctuation,Chun2026fluctuation,Gao2022thermodynamic,Gao2024thermodynamic}.
The response of the mean observable $\langle \theta_i (t)\rangle$ to the perturbation~\eqref{eq:langevin_local_impulsive_perturbation} is defined by the functional derivative
\begin{equation}
    R_{\phi_k (\bm z)}^{\theta_i} (t,s)
    \equiv \frac{\delta \langle \theta_i (t) \rangle}{\delta \phi_k (\bm z, s)}.
    \label{eq:langevin_local_response}
\end{equation}
By causality, $R_{\phi_k (\bm z)}^{\theta_i} (t,s) = 0$ for $t<s$. These local perturbations and responses serve as the elementary building blocks of the FRR. 
For the local impulsive perturbation~\eqref{eq:langevin_local_impulsive_perturbation} applied at time $s=0$, the spectral response of an observable $\theta_i$ is defined as 
\begin{equation}
    R_{\phi_k(\bm z)}^{\theta_i} (\omega) \equiv \int_0^\infty R_{\phi_k (\bm z)}^{\theta_i} (t,0) e^{i \omega t} dt.
    \label{eq:spectral_response_Langevin}
\end{equation}

Instead of the local impulsive perturbation, we may also consider a global perturbation applied to all components indexed by $k$:
\begin{equation}
    \phi_k (\bm{x})\mapsto \phi_k (\bm{x})+\epsilon \psi_k (\bm{x},t).
    \label{eq:global_pert_Lan}
\end{equation}
Since $\psi_k (\bm{x},t)$ can be decomposed into the local impulsive perturbations as $\psi_k (\bm{x},t) = \int d{\bm z} \int ds \psi_k (\bm{z},s) \delta (\bm{x}-\bm{z}) \delta(t-s)$, the response to the global perturbation~\eqref{eq:global_pert_Lan} is given by
\begin{equation}
    R^{\theta_i}_{\bm\phi, \bm\psi}(t)
    \equiv \sum_{k=1}^N \int d\bm{x} \int_{-\infty}^t ds~    \psi_k(\bm{x},s)\,R^{\theta_i}_{\phi_k (\bm x)}(t,s).   \label{eq:global_response_definition_Langevin}
\end{equation}

\subsection{Setup for Markov jump systems}
\label{subsec:jump_setup}

We consider a continuous-time Markov jump process on a finite set of states $n\in\{1,2,\dots,S\}$. Let $p_n(t)$ denote the probability of occupying state $n$ at time $t$. The dynamics is governed by the master equation
\begin{equation}
    \partial_t p_n(t)
    =
    \sum_{m(\neq n)}
    \bigl[
        W_{nm}p_m(t)-W_{mn}p_n(t)
    \bigr],
    \label{eq:jump_master_equation_component}
\end{equation}
where $W_{nm}$ denotes the transition rate from state $m$ to state $n$. 
Introducing the rate matrix $\textsf{W}$ with off-diagonal elements $[\textsf{W}]_{nm}=W_{nm}$ for $n \neq m$, and diagonal elements
$[\mathsf{W}]_{nn} = -\sum_{m(\neq n)}W_{mn}$, Eq.~\eqref{eq:jump_master_equation_component} can be written compactly as
\begin{equation}
    \partial_t \bm{p}(t)=\textsf{W}\bm{p}(t),
    \label{eq:jump_master_equation_matrix}
\end{equation}
where $ \bm{p}(t) = \bigl(p_1(t),\dots,p_S(t)\bigr)^{\mathsf T}$.

To separate the kinetic and entropic aspects of the dynamics, we parametrize the transition rates as
\begin{equation}
    W_{nm}
    =
    \exp\!\left(
        B_{nm}+\frac{F_{nm}}{2}
    \right),
    \label{eq:jump_rate_parametrization}
\end{equation}
where $B_{nm}=B_{mn}$ and $F_{nm}=-F_{mn}$ for all pairs $n\neq m$. Here, $B_{nm}$ represents the symmetric part of the logarithmic rates and controls the kinetic timescale of the transition, while $F_{nm}$ represents their antisymmetric part and encodes the thermodynamic bias associated with the transition. This decomposition has been commonly used in recent nonequilibrium fluctuation--response theory for Markov jump processes~\cite{Aslyamov2025nonequilibrium,Kwon2025fluctuation,Ptaszynski2026nonequilibriumState,Ptaszynski2026nonequilibriumStateCurrent}.

We assume that the process is irreducible and admits a unique stationary distribution $\pi_n$, satisfying
\begin{equation}
    \sum_m W_{nm}\pi_m=0.
    \label{eq:jump_stationary_distribution}
\end{equation}
The nonequilibrium character of the stationary state is reflected in the steady-state current along the edge $m\leftrightarrow n$,
\begin{equation}
    j_{nm}
    =
    W_{nm}\pi_m-W_{mn}\pi_n,
    \label{eq:jump_steady_current}
\end{equation}
which characterizes the net probability flow. By contrast, the corresponding traffic~\cite{maes2008canonical},
\begin{equation}
    a_{nm}
    =
    W_{nm}\pi_m+W_{mn}\pi_n,
    \label{eq:jump_activity}
\end{equation}
quantifies the total transition activity along the edge, irrespective of direction.

To define observables, we first introduce the state indicator
\begin{equation}
    \eta_n(t)\equiv \delta_{X(t),n},
    \label{eq:jump_state_indicator}
\end{equation}
where $X(t)$ denotes the state occupied by the system at time $t$. The variable $\eta_n(t)$ takes the value $1$ when the system is in state $n$ and $0$ otherwise. To describe transitions, let $N_{nm}(t)$ denote the accumulated number of jumps from state $m$ to state $n$ up to time $t$. The corresponding empirical current on the edge $m\leftrightarrow n$ is defined as
\begin{equation}
    \jmath_{nm}(t)\equiv \dot N_{nm}(t)-\dot N_{mn}(t),
    \label{eq:jump_empirical_current}
\end{equation}
where the overdot denotes the time derivative. This quantity is antisymmetric under the exchange of $n$ and $m$, and measures the instantaneous net jump current along the edge. Note that $\langle \eta_n\rangle_{\rm ss} = \pi_n$ and $\langle \jmath_{nm} \rangle_{\rm ss} = j_{nm}$, where $\langle\cdot\rangle_{\rm ss}$ denotes the steady-state ensemble average.

Similar to the Langevin case, we consider a set of $N_{\rm O}$ observables $\bm{\theta} = (\theta_1, \cdots, \theta_{N_{\rm O}})^{\textsf T}$, defined as
\begin{equation}
    \bm{\theta}(t) =
    \textsf G \bm{\eta}(t) +
    \textsf L \bm{\jmath}(t),
    \label{eq:jump_observable_vector}
\end{equation}
where $\textsf G$ is an $N_{\rm O} \times S$ matrix and $\textsf L$ is an $N_{\rm O}\times N_{\rm E}$ matrix, with $N_{\rm E}$ denoting the total number of edges.
Here,  $\bm{\eta}(t)=\bigl(\eta_1(t),\dots,\eta_S(t)\bigr)^{\mathsf T}$ is the state-indicator vector, and $\bm{\jmath}(t)$ is the $N_E$-dimensional vector of empirical edge currents, whose components are indexed by unoriented edges with a fixed convention, such as $n>m$. In Eq.~\eqref{eq:jump_observable_vector}, $\textsf G \bm \eta (t)$ and $\textsf L \bm \jmath (t)$ represent the state-dependent and current-like components of the observable, respectively.

For Markov jump processes, we consider local impulsive perturbations to the edge parameter $B_{nm}$ or $F_{nm}$. For a perturbation applied to a specific edge $n \leftrightarrow m$, we take
\begin{equation}
    \phi_{nm}\mapsto \phi_{nm}+\epsilon\,\delta(t-s),
    \label{eq:jump_local_phi_perturbation}
\end{equation}
where $\phi_{nm}$ denotes either $B_{nm}$ or $F_{nm}$. 
These local impulsive perturbations are the discrete counterparts of the spatially localized perturbations introduced in Eq.~\eqref{eq:langevin_local_impulsive_perturbation}. Similar to the Langevin case, the response of $\langle \theta_i \rangle$ to the perturbation~\eqref{eq:jump_local_phi_perturbation} is defined by the functional derivative
\begin{equation}
    R_{\phi_{nm}}^{\theta_i} (t,s)
    \equiv \frac{\delta \langle \theta_i (t) \rangle}{\delta \phi_{nm}  (s)}.
    \label{eq:Markov_local_response}
\end{equation}
For the local impulsive perturbation~\eqref{eq:jump_local_phi_perturbation} applied at time $s=0$, the spectral response of an observable $\theta_i$ is defined as 
\begin{equation}
    R_{\phi_{nm}}^{\theta_i} (\omega) \equiv \int_0^\infty R_{\phi_{nm}}^{\theta_i} (t,0) e^{i \omega t} dt. 
    \label{eq:spectral_response_Markov}
\end{equation}

For a global perturbation where all edges are simultaneously perturbed as
\begin{equation}
    \phi_{nm}\mapsto \phi_{nm}+\epsilon\,\psi_{nm}(t),
    \label{eq:jump_global_phi_perturbation}
\end{equation}
the response to this global perturbation is given by
\begin{equation}
    R^{\theta_i}_{\bm\phi, \bm\psi}(t)
    \equiv \sum_{n>m} \int_{-\infty}^t ds ~   \psi_{nm}(s) R^{\theta_i}_{\phi_{nm}}(t,s).   \label{eq:global_response_definition_jump}
\end{equation}

\subsection{Main results}
\label{subsec:unified_ffrr}

To characterize fluctuations, we define the spectral covariance between two observables $\theta_1$ and $\theta_2$ as
\begin{equation}
    C_{\theta_1,\theta_2}(\omega)
    \equiv
    \int_{-\infty}^{\infty}
    \Bigl[
        \langle \theta_1(t)\theta_2(0)\rangle_{\rm ss}
        -
        \langle \theta_1\rangle_{\mathrm{ss}}\langle \theta_2\rangle_{\mathrm{ss}}
    \Bigr]
    e^{i\omega t}\,dt.
    \label{eq:covariance_definition_main}
\end{equation}
In particular, for a vector observable $\bm{\theta}(t)$, $\textsf C_{\bm{\theta},\bm{\theta}^{\mathsf T}}(\omega)$ denotes the power spectrum matrix, whose $(i,j)$ element is given by $[\textsf C_{\bm{\theta},\bm{\theta}^{\mathsf T}}(\omega)]_{ij} = C_{\theta_i,\theta_j}(\omega)$.

We now present the first main result of this work: a frequency-resolved FRR for both overdamped Langevin dynamics and Markov jump processes. First, for overdamped Langevin systems, the frequency-resolved FRR takes the form
\begin{equation}
    \textsf C_{\bm{\theta},\bm{\theta}^{\mathsf T}}(\omega)
    =
    \sum_{k,l=1}^{N}
    \int d\bm{z} 
    \bm R^{\bm{\theta}}_{\phi_k(\bm{z})}(\omega) 
    \bigl[\textsf A^{\bm\phi}(\bm{z})^{-1}\bigr]_{kl}
    \bigl[
        \bm R^{\bm{\theta}}_{\phi_l(\bm{z})}(\omega)
    \bigr]^\dagger,
    \label{eq:ffrr_langevin_main}
\end{equation}
where $\textsf A^{\bm\phi}(\bm z)$ is a positive--definite $N\times N$ matrix associated with the perturbation type $\bm\phi$. Its explicit form is given in Eq.~\eqref{eq:langevin_Aphi_definition}. The $N_{\rm O}$-dimensional response vector $ \bm R^{\bm{\theta}}_{\phi_k(\bm z)}(\omega) = \bigl( R^{\theta_1}_{\phi_k(\bm z)}(\omega), \ldots, R^{\theta_{N_{\rm O}}}_{\phi_k(\bm z)}(\omega) \bigr)^{\mathsf T} $ collects the frequency-domain responses of all observable components, and the superscript $\dagger$ denotes the Hermitian conjugate.

Next, for Markov jump systems, the corresponding relation takes the form
\begin{equation}
    \textsf C_{\bm{\theta},\bm{\theta}^{\mathsf T}}(\omega)
    =
    \sum_{n>m}
    \frac{1}{A^{\phi}_{nm}}\,
    \bm R^{\bm{\theta}}_{\phi_{nm}}(\omega)  \bigl[
        \bm R^{\bm{\theta}}_{\phi_{nm}}(\omega)
    \bigr]^\dagger,
    \label{eq:ffrr_jump_main}
\end{equation}
where $A^{\phi}_{nm}$ is a positive scalar weight associated with the perturbation type $\phi \in \{B, F \}$. Its explicit form is given in Eq.~\eqref{eq:jump_Aphi_definition}. The $N_{\rm O}$-dimensional response vector
$\bm R^{\bm{\theta}}_{\phi_{nm}}(\omega) = \bigl(R^{\theta_1}_{\phi_{nm}}(\omega), \ldots, R^{\theta_{N_{\rm O}}}_{\phi_{nm}}(\omega) \bigr)^{\mathsf T}$ collects the responses of all observable components.

Equations~\eqref{eq:ffrr_langevin_main} and \eqref{eq:ffrr_jump_main} constitute the central results of this work: finite-frequency fluctuation--response identities for nonequilibrium steady states. Despite the differences between the two classes of systems, the resulting frequency-resolved FRRs share the same mathematical structure: the power spectrum of an observable can be exactly reconstructed as a quadratic form of local linear responses measured at the same frequency. This decomposition is spatial for Langevin systems and edge-resolved for Markov jump processes, thereby identifying which spatial regions or transition channels contribute most strongly to fluctuations at a given frequency. Because these relations apply to a broad class of observables and perturbations, the frequency-resolved FRR provides a unified framework for relating spontaneous fluctuations to controlled perturbations far from equilibrium.

\begin{figure*}
    \centering
    \includegraphics[scale=0.55]{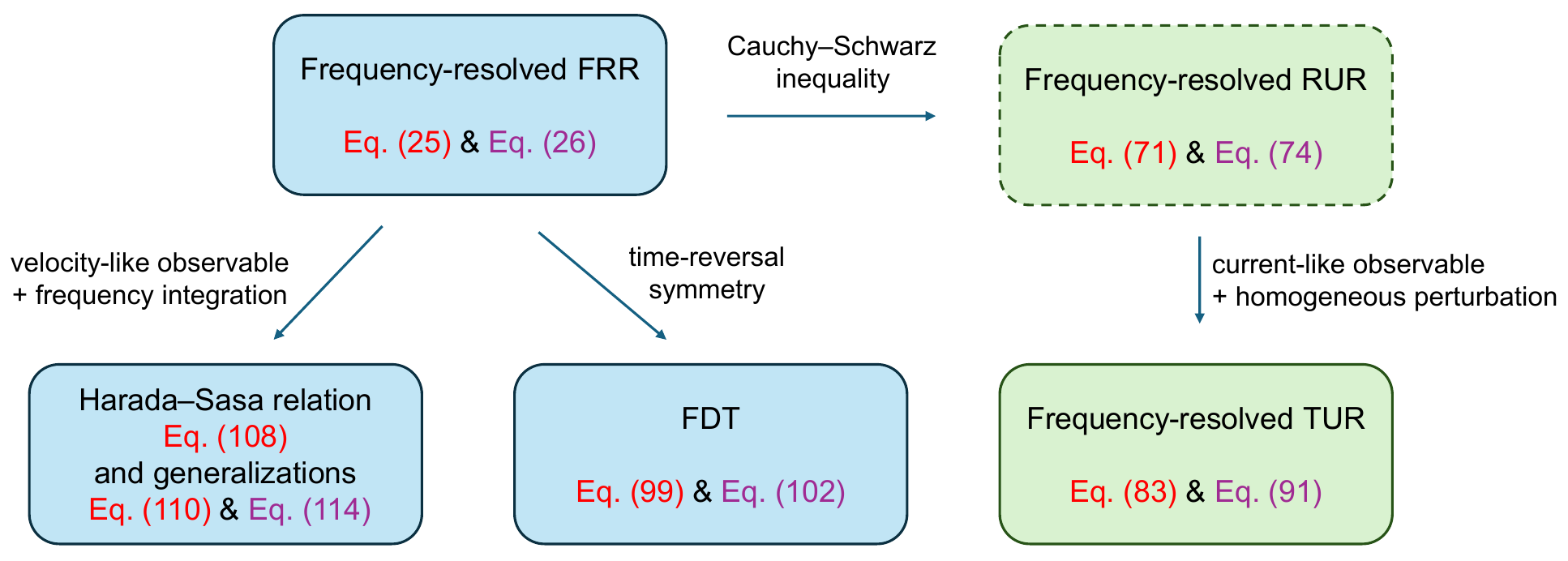}
    \caption{Hierarchy of various fluctuation--response relations in the frequency domain. Various fluctuation--response relations can be derived from the frequency-resolved FRR. Blue boxes denote exact identities, whereas green boxes represent inequalities. Each arrow indicates how a lower-level relation follows from a higher-level one. Equation numbers are shown in red for overdamped Langevin systems and in purple for Markov jump processes. The dotted box denotes a relation that reduces to previously reported results in the appropriate limits (Refs.~\cite{dechant2026finite,Aslyamov2026macroscopic}).}
    \label{fig:hierarchy}
\end{figure*}

The significance of the frequency-resolved FRR lies not only in the identity itself, but also in the hierarchy of results that follow from it. Once the exact identity is established, the RUR and related inequalities in the frequency domain arise from applications of the Cauchy--Schwarz inequality; the response KUR and TUR follow from the RUR through specific perturbations. Especially, the frequency-resolved TUR is derived from the homogeneous mobility perturbation for current-like observables.
In the equilibrium limit, the FRR reduces to the FDT. This reduction is nontrivial, as the frequency-resolved FRR is structurally different from the equilibrium FDT: at equilibrium, fluctuations and responses are related by a direct proportionality~\cite{Kubo1966fluctuation}, whereas away from equilibrium the power spectrum is expressed as a weighted quadratic form of local responses. In addition, the nonequilibrium remainder yields Harada--Sasa dissipation relations~\cite{Harada2005equality,Harada2006energy,Golestanian2025hodrydynamically}. Together, these results form a hierarchy rooted in the frequency-resolved FRR. This hierarchy constitutes the second main result of this work and is illustrated in Fig.~\ref{fig:hierarchy}.

It is important to understand the significance of the present result in the context of recent developments in frequency-domain fluctuation--response theory. Recent results have taken several forms, including finite-frequency fluctuation--response inequalities~\cite{dechant2026finite} and macroscopic Gaussian theories valid near stable fixed points~\cite{Aslyamov2026macroscopic}. These approaches provide valuable insights, but either constrain response functions relative to the power spectrum without reconstructing it, or rely on Gaussian approximations that are valid only for weak fluctuations around stable deterministic states. By contrast, the present work establishes an exact finite-frequency fluctuation--response identity that reconstructs the power spectrum from response functions without relying on a Gaussian approximation.

A complementary direction was recently developed for finite-time nonautonomous jump processes~\cite{Aslyamov2026dynamical}, where the covariance is expressed in terms of Fourier components of response functions. While this theory extends fluctuation--response relations beyond stationary states, it does not provide a frequency-resolved reconstruction of the power spectrum.

\section{Fluctuation--response theory in overdamped Langevin systems}
\label{sec:langevin_frt}

In this section, we derive the frequency-resolved FRR for overdamped Langevin systems. We begin by introducing the empirical density and current fields. The central object of the derivation is a frequency-domain excess propagator, in terms of which both local response functions and two-point covariance functions can be expressed. This common structure leads first to a FRR for the empirical density and current fields and, by linearity, to the FRR for general observables. Technical details of the derivation are deferred to Appendix~\ref{app:langevin}.

\subsection{Empirical fields and excess propagator}
\label{subsec:langevin_observables}

The local fields underlying the observable representation~\eqref{eq:langevin_observable_vector} are the empirical density and empirical current,
\begin{equation}
    \rho(\bm{x},t)
    \equiv
    \delta(\bm{x}-\bm{x}(t)),
    ~~~
    \bm{\jmath}(\bm{x},t)
    \equiv
    \delta(\bm{x}-\bm{x}(t))\circ \dot{\bm{x}}(t).
    \label{eq:langevin_empirical_fields}
\end{equation}
In terms of these fields, the observable vector can be written as
\begin{equation}
    \bm{\theta}(t)
    =
    \int d\bm{x}\,
    \Bigl[
        \bm G(\bm{x})\rho(\bm{x},t)
        +
        \textsf L(\bm{x})\bm{\jmath}(\bm{x},t)
    \Bigr].
    \label{eq:langevin_observable_from_empirical_fields}
\end{equation}
Consequently, the fluctuations and responses of $\bm{\theta}(t)$ are completely determined by those of $\rho(\bm{x},t)$ and $\bm{\jmath}(\bm{x},t)$ through the weight functions $\bm G$ and $\textsf L$.
Using the notation introduced in Eq.~\eqref{eq:covariance_definition_main}, we denote the spectral covariances of the empirical fields by $C_{\rho(\bm{x}),\rho(\bm{y})}(\omega)$, $\textsf C_{\rho(\bm{x}),\bm{\jmath}(\bm{y})^{\mathsf T}}(\omega)$, and $\textsf C_{\bm{\jmath}(\bm{x}),\bm{\jmath}(\bm{y})^{\mathsf T}}(\omega)$. The power spectrum matrix of the observable vector $\bm \theta$ is denoted by $\textsf C_{\bm{\theta},\bm{\theta}^{\mathsf T}}(\omega)$. Similarly, the corresponding local response functions are denoted by $R^{\rho(\bm{x})}_{\phi_k(\bm{z})}(\omega)$, $\bm R^{\bm{\jmath}(\bm{x})}_{\phi_k(\bm{z})}(\omega)$, and $\bm R^{\bm{\theta}}_{\phi_k(\bm{z})}(\omega)$.

For the derivation of the frequency-resolved FRR, we introduce the frequency-domain excess propagator~\cite{aslyamov2026integrated},
\begin{equation}
    H(\bm{x},\bm{z};\omega)
    \equiv
    \int_0^\infty
    \bigl[
        P(\bm{x},t|\bm{z},0)-\pi(\bm{x})
    \bigr]
    e^{i\omega t}\,dt,
    \label{eq:langevin_excess_propagator}
\end{equation}
where $P(\bm{x},t|\bm{y},s)$ denotes the transition probability density of the unperturbed dynamics and $\pi(\bm{x})$ is the steady-state distribution. More specifically, $P(\bm{x},t|\bm{y},s)$ satisfies the Fokker--Planck equation
\begin{equation}
    \partial_t P(\bm{x},t|\bm{y},s)
    = \hat{\mathcal{L}}_{\bm{x}}P(\bm{x},t|\bm{y},s),
    \label{eq:langevin_fp_equation}
\end{equation}
where $\hat{\mathcal{L}}_{\bm{x}}$ is the Fokker--Planck generator acting on the variable $\bm{x}$. We write
\begin{equation}
    \hat{\mathcal{L}}_{\bm{x}}
    =
    -\nabla_{\bm{x}}^{\mathsf T}\hat{\bm J}_{\bm{x}},
    \qquad
    \hat{\bm{J}}_{\bm{x}}
    =
    \mathsf{M}(\bm{x})\bigl[\bm{F}(\bm{x})-\mathsf{T}(\bm{x})\nabla_{\bm{x}}\bigr].
    \label{eq:langevin_fp_generator}
\end{equation}
The steady-state density $\pi(\bm x)$ satisfies
\begin{equation}
    \hat{\mathcal{L}}_{\bm{x}}\pi(\bm{x})=0,
\end{equation}
and the corresponding steady-state current is
\begin{equation}
    \bm{j}_{\mathrm{ss}}(\bm{x})
    =
    \hat{\bm{J}}_{\bm{x}}\pi(\bm{x}).
    \label{eq:langevin_ss_current_operator}
\end{equation}
We note that, in the steady state, $\langle \rho(\bm x,t) \rangle_{\rm ss} = \pi(\bm x)$ and $\langle \bm \jmath(\bm x,t) \rangle_{\rm ss} = \bm j_{\rm ss}(\bm x)$.

The excess propagator~\eqref{eq:langevin_excess_propagator} quantifies the excess probability density at $\bm{x}$ relative to the steady state, generated by a localized initial condition at $\bm{z}$ and resolved at frequency $\omega$. It satisfies the following identities:
\begin{align}
    &\int d\bm{z}\,H(\bm{x},\bm{z};\omega)\pi(\bm{z})=0,
    \label{eq:langevin_H_identity_1} \\   
    &\bigl(i\omega+\hat{\mathcal{L}}_{\bm{x}}\bigr)H(\bm{x},\bm{z};\omega)
    =
    -\delta(\bm{x}-\bm{z})+\pi(\bm{x}),
    \label{eq:langevin_H_identity_2} \\
    &\bigl(i\omega+\hat{\mathcal{L}}_{\bm{z}}^{\dagger}\bigr)H(\bm{x},\bm{z};\omega)
    =
    -\delta(\bm{x}-\bm{z})+\pi(\bm{x}).
    \label{eq:langevin_H_identity_3}
\end{align}
These identities show that $H$ is the reduced resolvent of the Fokker--Planck operator, with the stationary mode projected out. Their derivations are provided in Appendix~\ref{appsubsec:langevin_H}.

\subsection{Response functions of empirical fields}
\label{subsec:langevin_response}

We consider a local impulsive perturbation of $\phi_k(\bm{z})$ applied at time $s=0$, as defined in Eq.~\eqref{eq:langevin_local_impulsive_perturbation}. At the level of the Fokker--Planck generator, this perturbation takes the form
\begin{equation}
    \hat{\mathcal{L}}_{\bm{x}}
    \mapsto
    \hat{\mathcal{L}}_{\bm{x}}
    -
    \epsilon \delta(t) \,
    \nabla_{\bm{x}}^{\mathsf T}
    \delta(\bm{x}-\bm{z})\,
    \hat{\bm{K}}_{\phi_k,\bm{x}}\,,
    \label{eq:langevin_generator_perturbation}
\end{equation}
where the operator-valued vector $\hat{\bm{K}}_{\phi_k,\bm{x}}$ depends on the choice of the perturbed quantity $\phi_k$ (see Eqs.~\eqref{eq:K_F}--\eqref{eq:K_T}). We further define the associated local prefactor vector
\begin{equation}
    {\bm N}_{\phi_k}(\bm{z})
    \equiv \hat{\bm{K}}_{\phi_k,\bm{z}}\pi(\bm{z}) \,,
    \label{eq:langevin_Nphi_definition}
\end{equation}
whose explicit forms depend on the choice of $\phi_k$ and are given in Eqs.~\eqref{eq:N_F}--\eqref{eq:N_T}. 

The spectral response function of the empirical density reads
\begin{equation}
    R^{\rho(\bm{x})}_{\phi_k(\bm{z})}(\omega)
    = \bigl[\bm N_{\phi_k}(\bm{z}) \bigr]^{\mathsf T}
    \nabla_{\bm{z}}H(\bm{x},\bm{z};\omega).
    \label{eq:langevin_density_response}
\end{equation}
Similarly, the spectral response function of the empirical current is given by
\begin{equation}
    \bm R^{\bm{\jmath}(\bm{x})}_{\phi_k(\bm{z})}(\omega)
    =
    \textsf{P}(\bm{x},\bm{z};\omega)\,\bm N_{\phi_k}(\bm{z}),
    \label{eq:langevin_current_response}
\end{equation}
where
\begin{equation}
    \textsf{P}(\bm{x},\bm{z};\omega)
    =
    \textsf{I}\,\delta(\bm{x}-\bm{z})
    + \hat{\bm{J}}_{\bm{x}}\nabla_{\bm{z}} ^\mathsf{T} H(\bm{x},\bm{z};\omega).
    \label{eq:langevin_P_definition}
\end{equation}
Here, $\textsf I$ denotes the $N\times N$ identity matrix.
Equations~\eqref{eq:langevin_density_response} and \eqref{eq:langevin_current_response} show that both density and current responses are determined by the same excess propagator, with the dependence on the perturbation type entering solely through the local prefactor vector $\bm N_{\phi_k}(\bm z)$. The derivations of \eqref{eq:langevin_density_response} and \eqref{eq:langevin_current_response} are provided in Appendix~\ref{appsubsec:langevin_response}.

\subsection{FRRs for empirical fields}
\label{subsec:langevin_local_ffrr}

We now show that the spectral covariance functions of the empirical density and current can be expressed in terms of the same excess propagator~\eqref{eq:langevin_excess_propagator}. This shared structure provides the key link between fluctuations and responses of the empirical fields in the frequency domain, ultimately leading to the frequency-resolved FRR.

First, the key step in deriving the density--density covariance is the following identity:
\begin{align} \label{eq:langevin_rho_rho_cov_quadratic}
    &C_{\rho(\bm{x}),\rho(\bm{y})}(\omega)
    \\ \nonumber &=
    2\int d\bm{z}\,
    \pi(\bm{z})\,
    \nabla_{\bm{z}} ^\mathsf{T}H(\bm{x},\bm{z};\omega)\,
    \textsf D(\bm{z})\,
    \nabla_{\bm{z}}H(\bm{y},\bm{z};-\omega).
\end{align}
The derivation of Eq.~\eqref{eq:langevin_rho_rho_cov_quadratic} is provided in Appendix~\ref{appsubsec:langevin_quadratic_identity}.
Using Eq.~\eqref{eq:langevin_P_definition}, the density--current covariance can be written in a parallel form as
\begin{align} \label{eq:langevin_rho_j_cov_quadratic}
    &\textsf C_{\rho(\bm{x}),\bm{\jmath}(\bm{y})^{\mathsf T}}(\omega)
    \\ \nonumber &=
    2\int d\bm{z}\,
    \pi(\bm{z})\,
    \nabla_{\bm{z}}^\mathsf{T}H(\bm{x},\bm{z};\omega)\,
    \textsf D(\bm{z})\,
    \bigl[ 
    \textsf{P}(\bm{y},\bm{z};\omega)
    \bigr]^\dagger,
\end{align}
and the current--current covariance is given by
\begin{equation}
    \textsf C_{\bm{\jmath}(\bm{x}),\bm{\jmath}(\bm{y})^{\mathsf T}}(\omega)
    =
    2\int d\bm{z}\,
    \pi(\bm{z})\,
    \textsf{P}(\bm{x},\bm{z};\omega)\,
    \textsf D(\bm{z})\,
    \bigl[\textsf{P}(\bm{y},\bm{z};\omega
    )\bigr]^\dagger.
    \label{eq:langevin_j_j_cov_quadratic}
\end{equation}
Equations~\eqref{eq:langevin_rho_rho_cov_quadratic}--\eqref{eq:langevin_j_j_cov_quadratic} show that the local covariance functions are constructed from the same quantities that appear in the response formulas, Eqs.~\eqref{eq:langevin_density_response} and \eqref{eq:langevin_current_response}.
The derivations of Eqs.~\eqref{eq:langevin_rho_j_cov_quadratic} and \eqref{eq:langevin_j_j_cov_quadratic} are provided in Appendix~\ref{appsubsec:langevin_covariances}.

To make this connection explicit, we define the $N\times N$ prefactor matrix
\begin{equation}\label{eq:prefactor_vec_langevin}
    \textsf N^{\bm \phi}(\bm{z})
    \equiv
    \bigl(
        \bm N_{\phi_1}(\bm{z}),
        \dots,
        \bm N_{\phi_N}(\bm{z})
    \bigr),
\end{equation}
and the associated weight matrix
\begin{equation}
    \textsf A^{\bm \phi}(\bm{z})
    \equiv
    \frac{1}{2\pi(\bm{z})}
    \bigl[\textsf N^{\bm \phi}(\bm{z})\bigr]^{\mathsf T}
    \textsf D(\bm{z})^{-1}
    \textsf N^{\bm \phi}(\bm{z}).
    \label{eq:langevin_Aphi_definition}
\end{equation}
The explicit forms of $\textsf A^{\bm \phi}(\bm{z})$ for $\bm \phi \in \{ 
\bm{F}, \ln \textsf M, \textsf T \}$ are provided in Appendix~\ref{appsubsec:langevin_Aphi}.
Provided that $\textsf N^{\bm\phi}(\bm z)$ is nonsingular,
 Eq.~\eqref{eq:langevin_Aphi_definition} gives
\[
\textsf N^{\bm\phi}
(\textsf A^{\bm\phi})^{-1}
(\textsf N^{\bm\phi})^{\mathsf T}
=
2\pi\,\textsf D,
\]
which, together with Eqs.~\eqref{eq:langevin_density_response} and \eqref{eq:langevin_rho_rho_cov_quadratic},
yields 
\begin{align} \label{eq:langevin_local_ffrr_rho_rho}
    & C_{\rho(\bm{x}),\rho(\bm{y})}(\omega)
    \\ \nonumber &=
    \sum_{k,l=1}^{N}
    \int d\bm{z}\,
    R^{\rho(\bm{x})}_{\phi_k(\bm{z})}(\omega)\,
    \bigl[\textsf A^{\bm \phi}(\bm{z})^{-1}\bigr]_{kl}\,
    R^{\rho(\bm{y})}_{\phi_l(\bm{z})}(-\omega).
\end{align}
At singular points of $\textsf N^{\bm\phi}(\bm{z})$, the FRR is understood in the limiting sense: although the component of $\textsf{A}^{\bm{\phi}}(\bm{z})^{-1}$ diverges, the corresponding responses vanish proportionally, leaving their products finite.

Likewise, combining the response representations
\eqref{eq:langevin_density_response} and
\eqref{eq:langevin_current_response}
with the covariance identities
\eqref{eq:langevin_rho_j_cov_quadratic} and
\eqref{eq:langevin_j_j_cov_quadratic}
yields the fluctuation--response relations
\begin{align} 
    & \textsf C_{\rho(\bm{x}),\bm{\jmath}(\bm{y})^{\mathsf T}}(\omega) \label{eq:langevin_local_ffrr_rho_j}
    \\ \nonumber &=
    \sum_{k,l=1}^{N}
    \int d\bm{z}\,
    R^{\rho(\bm{x})}_{\phi_k(\bm{z})}(\omega)\,
    \bigl[\textsf A^{\bm \phi}(\bm{z})^{-1}\bigr]_{kl}\,
    \bigl[\bm R^{\bm{\jmath}(\bm{y})}_{\phi_l(\bm{z})}(\omega)
    \bigr]^\dagger,  \\
    & \textsf C_{\bm{\jmath}(\bm{x}),\bm{\jmath}(\bm{y})^{\mathsf T}}(\omega)
    \label{eq:langevin_local_ffrr_j_j}
    \\ \nonumber &=
    \sum_{k,l=1}^{N}
    \int d\bm{z}\,
    \bm R^{\bm{\jmath}(\bm{x})}_{\phi_k(\bm{z})}(\omega)\,
    \bigl[\textsf A^{\bm \phi}(\bm{z})^{-1}\bigr]_{kl}\,
    \bigl[\bm R^{\bm{\jmath}(\bm{y})}_{\phi_l(\bm{z})}(\omega)
    \bigr]^\dagger.  
\end{align}
Equations~\eqref{eq:langevin_local_ffrr_rho_rho}--\eqref{eq:langevin_local_ffrr_j_j} constitute the frequency-resolved FRR for empirical fields for overdamped Langevin systems. They show that the spectral covariance functions of empirical density and current are given by integrals of products of spectral responses over the perturbation point $\bm{z}$, weighted by the inverse of the matrix $\textsf A^{\bm \phi}(\bm{z})$.

\subsection{Frequency-resolved FRR for general observables}
\label{subsec:langevin_general_ffrr}

Given the local fluctuation--response relations~\eqref{eq:langevin_local_ffrr_rho_rho}--\eqref{eq:langevin_local_ffrr_j_j} for empirical fields, the extension to general observables is straightforward. Since the observable vector $\bm{\theta}$ defined in Eq.~\eqref{eq:langevin_observable_from_empirical_fields} is linear in the empirical density and current, its local response is given by
\begin{equation}
    \bm R^{\bm{\theta}}_{\phi_k(\bm{z})}(\omega)
    =
    \int d\bm{x}\,
    \Bigl[
        \bm G(\bm{x})\,R^{\rho(\bm{x})}_{\phi_k(\bm{z})}(\omega)
        +
        \textsf L(\bm{x})\,\bm R^{\bm{\jmath}(\bm{x})}_{\phi_k(\bm{z})}(\omega)
    \Bigr].
    \label{eq:langevin_theta_response}
\end{equation}
Similarly, substituting Eq.~\eqref{eq:langevin_observable_from_empirical_fields} into Eq.~\eqref{eq:covariance_definition_main} yields the power spectrum matrix of $\bm{\theta}(t)$:
\begin{equation}
    \begin{aligned}
        \textsf C_{\bm{\theta},\bm{\theta}^{\mathsf T}}(\omega)
        =
        \int d\bm{x}\,d\bm{y}\,
        \Big[
        &
        \bm G(\bm{x})\,C_{\rho(\bm{x}),\rho(\bm{y})}(\omega)\,\bm G(\bm{y})^{\mathsf T}
        \\
        &+
        \bm G(\bm{x})\,\textsf C_{\rho(\bm{x}),\bm{\jmath}(\bm{y})^{\mathsf T}}(\omega)\,\textsf L(\bm{y})^{\mathsf T}
        \\
        &+
        \textsf L(\bm{x})\,\textsf C_{\bm{\jmath}(\bm{x}),\rho(\bm{y})}(\omega)\,\bm G(\bm{y})^{\mathsf T}
        \\
        &+
        \textsf L(\bm{x})\,\textsf C_{\bm{\jmath}(\bm{x}),\bm{\jmath}(\bm{y})^{\mathsf T}}(\omega)\,\textsf L(\bm{y})^{\mathsf T}
        \Big].
    \end{aligned}
    \label{eq:langevin_theta_covariance}
\end{equation}
Substitution of Eqs.~\eqref{eq:langevin_local_ffrr_rho_rho}--\eqref{eq:langevin_local_ffrr_j_j} into Eq.~\eqref{eq:langevin_theta_covariance}, together with Eq.~\eqref{eq:langevin_theta_response}, combines the integrations over $\bm{x}$ and $\bm{y}$ into the local responses of the observable $\bm{\theta}$. This yields
\begin{equation}
    \textsf C_{\bm{\theta},\bm{\theta}^{\mathsf T}}(\omega)
    =
    \sum_{k,l=1}^{N}
    \int d\bm{z}\,
    \bm R^{\bm{\theta}}_{\phi_k(\bm{z})}(\omega)\,
    \bigl[\textsf A^{\bm \phi}(\bm{z})^{-1}\bigr]_{kl}\,
    \bigl[\bm R^{\bm{\theta}}_{\phi_l(\bm{z})}(\omega)
    \bigr]^\dagger.
    \label{eq:langevin_general_ffrr}
\end{equation}
Equation~\eqref{eq:langevin_general_ffrr} is the frequency-resolved FRR for general observables in overdamped Langevin systems. Its physical interpretation is straightforward: the power spectrum is reconstructed from local responses to perturbations applied at each point in configuration space.
The relation applies equally to purely state-dependent observables, purely current-like observables, and their linear combinations. Moreover, its structure is independent of the choice of perturbation; whether the perturbation acts on the force, mobility, or temperature, the FRR retains the same form, with the weight matrix $\textsf A^{\bm \phi}(\bm{z})$ changing accordingly. This completes the derivation of the frequency-resolved FRR for overdamped Langevin systems.

\section{Fluctuation--response theory in Markov jump systems}
\label{sec:jump_frt}

In this section, we construct the frequency-resolved FRR for Markov jump systems. The derivation closely parallels that of Sec.~\ref{sec:langevin_frt}, but the local structure is now organized over the edges of the transition network rather than over points in configuration space. The central object is the discrete excess propagator, which determines both local responses and two-point covariances. This common structure first yields local FRRs for the empirical state indicators and empirical currents, and then, by linearity, extends to the FRR for general observables. Technical derivations are deferred to Appendix~\ref{app:jump}.

\subsection{Excess propagator}
\label{subsec:jump_observables}

We first introduce the excess propagator~\cite{aslyamov2026integrated} for Markov jump processes:
\begin{equation}
    H_{nk}(\omega)
    =
    \int_0^\infty
    \Bigl(
        [e^{\mathsf{W}t}]_{nk}-\pi_n
    \Bigr)e^{i\omega t}\,dt,
    \label{eq:jump_excess_propagator}
\end{equation}
where $\bigl[e^{\mathsf{W}t}\bigr]_{nk} \equiv P(n,t|k,0)$ is the conditional probability of finding the system in state $n$ at time $t$, given that it started in state $k$ at time $0$. $H_{nk}(\omega)$ is the discrete counterpart of the excess propagator~\eqref{eq:langevin_excess_propagator}: it measures the excess occupation of state $n$, relative to the stationary state, resulting from a localized initial condition at state $k$, resolved at frequency $\omega$.

The excess propagator satisfies the following identities:
\begin{align}
    &\sum_m H_{nm}(\omega)\pi_m=0,
    \label{eq:jump_H_identity_1} \\
    &\sum_m
    \bigl(
        i\omega\delta_{nm}+ [\mathsf{W}]_{nm}
    \bigr)H_{mk}(\omega)
    =
    -\delta_{nk}+\pi_n,
    \label{eq:jump_H_identity_2} \\
    &\sum_m
    H_{nm}(\omega)
    \bigl(
        i\omega\delta_{mk}+ [\mathsf{W}]_{mk}
    \bigr)
    =
    -\delta_{nk}+\pi_n.
    \label{eq:jump_H_identity_3}
\end{align}
These identities are the discrete counterparts of Eqs.~\eqref{eq:langevin_H_identity_1}--\eqref{eq:langevin_H_identity_3}. They show that $H_{nk}(\omega)$ is the matrix representation of the reduced resolvent of the Markov generator. The derivations of Eqs.~\eqref{eq:jump_H_identity_1}--\eqref{eq:jump_H_identity_3} are provided in Appendix~\ref{appsubsec:jump_H}.

\subsection{Response functions of empirical state indicators and currents} \label{subsec:jump_local_Rf_empirical}

We first consider an entropic perturbation of the edge $k\leftrightarrow l$, given by $F_{kl}\mapsto F_{kl}+\epsilon\,\delta(t)$. The corresponding response of the state indicator $\eta_n$~\eqref{eq:jump_state_indicator} is 
\begin{equation}
    R^{\eta_n}_{F_{kl}}(\omega)
    =
    \frac{a_{kl}}{2}
    \Bigl[
        H_{nk}(\omega)-H_{nl}(\omega)
    \Bigr].
    \label{eq:jump_eta_response_F}
\end{equation}
Similarly, for a kinetic perturbation, $B_{kl}\mapsto B_{kl}+\epsilon\,\delta(t)$, the response is
\begin{equation}
    R^{\eta_n}_{B_{kl}}(\omega)
    =
    j_{kl}
    \Bigl[
        H_{nk}(\omega)-H_{nl}(\omega)
    \Bigr].
    \label{eq:jump_eta_response_B}
\end{equation}
The derivations of Eqs.~\eqref{eq:jump_eta_response_F} and \eqref{eq:jump_eta_response_B} are provided in Appendix~\ref{appsubsec:jump_response}. Thus, the two response functions are proportional~\cite{Aslyamov2025nonequilibrium,Chun2026fluctuation,Aslyamov2026dynamical}:
\begin{equation}
    R^{\eta_n}_{B_{kl}}(\omega)
    =
    \frac{2 j_{kl}}{a_{kl}}R^{\eta_n}_{F_{kl}}(\omega).
    \label{eq:jump_eta_response_proportionality}
\end{equation}

We next consider the response of the empirical current $\jmath_{nm}$~\eqref{eq:jump_empirical_current}. For an entropic perturbation of the edge $k\leftrightarrow l$, given by $F_{kl}\mapsto F_{kl}+\epsilon\,\delta(t)$, the corresponding response is 
\begin{align} \label{eq:jump_current_response_explicit}
    &\frac{2}{a_{kl}}R^{\jmath_{nm}}_{F_{kl}}(\omega)
    =
    \delta_{nk}\delta_{ml}
    -
    \delta_{nl}\delta_{mk}
    \\ \nonumber &+
    W_{nm}\Bigl[
        H_{mk}(\omega)-H_{ml}(\omega)
    \Bigr]
    -
    W_{mn}\Bigl[
        H_{nk}(\omega)-H_{nl}(\omega)
    \Bigr]. 
\end{align}
The first term, $\delta_{nk}\delta_{ml} - \delta_{nl}\delta_{mk}$, represents the direct local contribution from the impulsive perturbation acting on the perturbed edge, while the remaining terms describe how the perturbation propagates through the network and contributes to the current on the observed edge. 
As in the case of the state indicator, the responses to kinetic and entropic perturbations are proportional:
\begin{equation}
    R^{\jmath_{nm}}_{B_{kl}}(\omega)
    =
    \frac{2 j_{kl}}{a_{kl}}R^{\jmath_{nm}}_{F_{kl}}(\omega).
    \label{eq:jump_current_response_proportionality}
\end{equation}
The derivations of Eqs.~\eqref{eq:jump_current_response_explicit} and \eqref{eq:jump_current_response_proportionality} are provided in Appendix~\ref{appsubsec:jump_response}.
Equations~\eqref{eq:jump_eta_response_F}--\eqref{eq:jump_current_response_proportionality} show that both the state and current responses are determined by the same excess propagator $H_{nk}(\omega)$. For the state response, the relevant quantity is the difference $H_{nk}(\omega)-H_{nl}(\omega)$. For the current response, such differences appear in a rate-weighted combination.

\subsection{FRRs for empirical state indicators and currents}
\label{subsec:jump_local_ffrr}

As in Sec.~\ref{subsec:langevin_local_ffrr}, we express the covariance functions of the elementary empirical observables in terms of the response functions derived in Sec.~\ref{subsec:jump_local_Rf_empirical}. This provides the local building blocks from which the frequency-resolved FRR for general observables will be obtained.

The key step in deriving the state--state covariance is the following quadratic identity:
\begin{align} \label{eq:jump_eta_eta_cov_quadratic}
    &C_{\eta_n,\eta_m}(\omega)
    \\ \nonumber &=
    \sum_{k>l}
    a_{kl}
    \Bigl[
        H_{nk}(\omega)-H_{nl}(\omega)
    \Bigr]
    \Bigl[
        H_{mk}(-\omega)-H_{ml}(-\omega)
    \Bigr].
\end{align}
This identity is the discrete counterpart of Eq.~\eqref{eq:langevin_rho_rho_cov_quadratic}. Its derivation is deferred to Appendix~\ref{appsubsec:jump_quadratic_identity}.
To express Eq.~\eqref{eq:jump_eta_eta_cov_quadratic} in terms of response functions, we define the edge weight factors
\begin{equation}
    A^{F}_{kl} \equiv \frac{a_{kl}}{4},
    \qquad
    A^{B}_{kl} \equiv \frac{j_{kl}^2}{a_{kl}}.
    \label{eq:jump_Aphi_definition}
\end{equation}
Using Eqs.~\eqref{eq:jump_eta_response_F} and \eqref{eq:jump_eta_response_B}, Eq.~\eqref{eq:jump_eta_eta_cov_quadratic} can be rewritten as
\begin{equation}
    C_{\eta_n,\eta_m}(\omega)
    =
    \sum_{k>l}
    \frac{1}{A^\phi_{kl}}\,
    R^{\eta_n}_{\phi_{kl}}(\omega)\,
    R^{\eta_m}_{\phi_{kl}}(-\omega),
    \label{eq:jump_local_ffrr_eta_eta}
\end{equation}
where $\phi\in\{F,B\}$, with $A^\phi_{kl}$ defined accordingly.
When $A^B_{kl}=0$, the corresponding terms in the FRRs are understood in the limiting sense: although $(A^B_{kl})^{-1}$ diverges, the kinetic responses vanish proportionally, leaving their products finite.

The same reasoning applies to the state--current covariance, yielding
\begin{equation}
    C_{\eta_n,\jmath_{n'm'}}(\omega)
    =
    \sum_{k>l}
    \frac{1}{A^\phi_{kl}}\,
    R^{\eta_n}_{\phi_{kl}}(\omega)\,
    R^{\jmath_{n'm'}}_{\phi_{kl}}(-\omega).
    \label{eq:jump_local_ffrr_eta_j}
\end{equation}
Likewise, the current--current covariance satisfies
\begin{equation}
    C_{\jmath_{nm},\jmath_{n'm'}}(\omega)
    =
    \sum_{k>l}
    \frac{1}{A^\phi_{kl}}\,
    R^{\jmath_{nm}}_{\phi_{kl}}(\omega)\,
    R^{\jmath_{n'm'}}_{\phi_{kl}}(-\omega).
    \label{eq:jump_local_ffrr_j_j}
\end{equation}
Equations~\eqref{eq:jump_local_ffrr_eta_eta}--\eqref{eq:jump_local_ffrr_j_j} constitute the frequency-resolved FRR for the elementary empirical observables in Markov jump systems. The derivations of Eqs.~\eqref{eq:jump_local_ffrr_eta_j} and \eqref{eq:jump_local_ffrr_j_j} are deferred to Appendix~\ref{appsubsec:jump_covariances}.
These relations express each spectral covariance as a sum of contributions from individual network edges, where each contribution is determined by the response to a local perturbation applied to a single edge at the same frequency. Consequently, the power spectrum admits a natural edgewise decomposition, directly identifying which transition channels contribute most strongly to a given spectral feature.

\subsection{Frequency-resolved FRR for general observables}
\label{subsec:jump_general_ffrr}

With the local FRRs~\eqref{eq:jump_local_ffrr_eta_eta}--\eqref{eq:jump_local_ffrr_j_j} established, the extension to general observables follows immediately by linearity, as in Sec.~\ref{subsec:langevin_general_ffrr}. 
By linearity, the response of the observable vector to a local perturbation of the edge parameter $\phi_{kl}$ is
\begin{equation}
    \bm R^{\bm{\theta}}_{\phi_{kl}}(\omega)
    =
    \textsf G\,\bm R^{\bm{\eta}}_{\phi_{kl}}(\omega)
    +
    \textsf L\,\bm R^{\bm{\jmath}}_{\phi_{kl}}(\omega),
    \label{eq:jump_theta_response}
\end{equation}
where $\bm R^{\bm{\eta}}_{\phi_{kl}}(\omega) = \bigl( R^{\eta_1}_{\phi_{kl}}(\omega), \dots, R^{\eta_S}_{\phi_{kl}}(\omega) \bigr)^{\mathsf T}$ and  $\bm R^{\bm{\jmath}}_{\phi_{kl}}(\omega)$ is defined analogously.

The covariance of $\bm{\theta}(t)$ follows from Eq.~\eqref{eq:jump_observable_vector}:
\begin{equation}
    \begin{aligned}
        \textsf C_{\bm{\theta},\bm{\theta}^{\mathsf T}}(\omega)
        =
        {}&
        \textsf G\, \textsf C_{\bm{\eta},\bm{\eta}^{\mathsf T}}(\omega)\,\textsf G^{\mathsf T}
        +
        \textsf G\,\textsf C_{\bm{\eta},\bm{\jmath}^{\mathsf T}}(\omega)\,\textsf L^{\mathsf T}
        \\
        &+
        \textsf L\,\textsf C_{\bm{\jmath},\bm{\eta}^{\mathsf T}}(\omega)\,\textsf G^{\mathsf T}
        +
        \textsf L\,\textsf C_{\bm{\jmath},\bm{\jmath}^{\mathsf T}}(\omega)\,\textsf L^{\mathsf T}.
    \end{aligned}
    \label{eq:jump_theta_covariance}
\end{equation}
Substituting Eqs.~\eqref{eq:jump_local_ffrr_eta_eta}--\eqref{eq:jump_local_ffrr_j_j} into Eq.~\eqref{eq:jump_theta_covariance} and using Eq.~\eqref{eq:jump_theta_response}, we obtain
\begin{equation}
    \textsf C_{\bm{\theta},\bm{\theta}^{\mathsf T}}(\omega)
    =
    \sum_{n>m}
    \frac{1}{A^\phi_{nm}}\,
    \bm R^{\bm{\theta}}_{\phi_{nm}}(\omega)\,
    \bigl[
        \bm R^{\bm{\theta}}_{\phi_{nm}}(\omega)
    \bigr]^\dagger.
    \label{eq:jump_general_ffrr}
\end{equation}
Equation~\eqref{eq:jump_general_ffrr} is the frequency-resolved FRR for general observables in Markov jump systems. It shows that the fluctuation spectrum can be reconstructed from local responses to perturbations applied to individual edges. As in the Langevin case, the relation applies to purely state-dependent observables, purely current-like observables, and their combinations. The form of the FRR remains unchanged regardless of whether the perturbation is entropic or kinetic.

\section{Consequences of the frequency-resolved fluctuation--response relation}
\label{sec:consequences}

In this section, we derive the main consequences of the frequency-resolved FRR. Rather than presenting these results as independent relations, we show that they emerge from a unified hierarchy. The RUR follows directly from the quadratic structure of the frequency-resolved FRR; the response KUR and TUR follow from the RUR by choosing specific perturbations; 
for current-like observables, the FDT and Harada--Sasa relations arise from the same FRR identity under equilibrium and nonequilibrium conditions, respectively.
To keep the section focused on the physical content, detailed algebraic manipulations are deferred to Appendix~\ref{app:consequences}.

\subsection{Frequency-resolved response uncertainty relation}
\label{subsec:frur}

A direct consequence of the frequency-resolved FRR, which expresses fluctuations in terms of responses to local perturbations, is a bound involving the response to an arbitrary global perturbation. This bound follows from the Cauchy--Schwarz inequality applied to the quadratic response representation of the power spectrum.

We first consider the overdamped Langevin case. For a global perturbation of $\bm \phi\in \{\bm{F},\ln \textsf M,\textsf T \}$ of the form $\phi_k(\bm{x})\mapsto \phi_k(\bm{x})+\epsilon \, \psi_k(\bm{x},t)$, the corresponding response of the observable vector $\bm{\theta}(t)$ is, using Eq.~\eqref{eq:global_response_definition_Langevin}, given by
\begin{align}
    \bm R^{\bm{\theta}}_{\bm \phi,\bm{\psi}}(\omega)
    &\equiv \int_{-\infty}^\infty dt\, \bm R^{\bm\theta}_{\bm\phi, \bm\psi}(t) e^{i \omega t} \nonumber \\ &=
    \sum_{k=1}^{N}
    \int d\bm{x}\,
    \psi_k(\bm{x},\omega)\,
    \bm R^{\bm{\theta}}_{\phi_k(\bm{x})}(\omega) .
    \label{eq:langevin_global_response}
\end{align}
where $\psi_k(\bm{x},\omega) \equiv \int_{-\infty}^{\infty} \psi_k(\bm{x},t)e^{i\omega t}\,dt$.
Applying the Cauchy--Schwarz inequality to the quadratic representation of the FRR in Eq.~\eqref{eq:langevin_general_ffrr} yields (see Appendix~\ref{appsubsec:frur_derivation} for details)
\begin{align} \label{eq:langevin_frur}
    &\bigl[
        \bm R^{\bm{\theta}}_{\bm \phi,\bm{\psi}}(\omega)
    \bigr]^{\dagger}
    \textsf C_{\bm{\theta},\bm{\theta}^{\mathsf T}}(\omega)^{-1}
    \bm R^{\bm{\theta}}_{\bm \phi,\bm{\psi}}(\omega)
    \\ \nonumber 
    &\le
    \sum_{k,l=1}^{N}
    \int d\bm{x}\,
    \psi_k(\bm{x},-\omega)
    [\textsf A^{\bm{\phi}}(\bm{x})]_{kl}
    \psi_l(\bm{x},\omega) \;.
\end{align}
Throughout, all spectral covariance matrices appearing with an inverse are assumed to be invertible at the frequencies under consideration. For a single observable $\theta(t)$, this inequality reduces to
\begin{equation}
    \begin{aligned}
    &\bigl|
        R^{\theta}_{\bm \phi,\bm{\psi}}(\omega)
    \bigr|^2
    \\ &\le
    C_{\theta,\theta}(\omega)
    \sum_{k,l=1}^{N}
    \int d\bm{x}\,
    \psi_k(\bm{x},-\omega)
    [\textsf A^{\bm{\phi}}(\bm{x})]_{kl}
    \psi_l(\bm{x},\omega).
    \end{aligned}
    \label{eq:langevin_frur_scalar}
\end{equation}

The same structure applies to Markov jump systems. 
For a global perturbation of either the symmetric or antisymmetric edge parameter, $\phi_{nm}\mapsto \phi_{nm}+\epsilon\, \psi_{nm}(t)$ ($\phi\in \{ B, F \}$), the corresponding response of the observable vector $\bm{\theta}(t)$ is, using Eq.~\eqref{eq:global_response_definition_jump}, given by
\begin{equation}
    \bm R^{\bm{\theta}}_{\bm \phi,\bm{\psi}}(\omega)
    \equiv \int_{-\infty}^\infty dt\, \bm R^{\bm\theta}_{\bm\phi, \bm\psi}(t) e^{i \omega t} =
    \sum_{n>m}
    \psi_{nm}(\omega)
    \bm R^{\bm{\theta}}_{\phi_{nm}}(\omega).
    \label{eq:jump_global_response}
\end{equation}
Applying the Cauchy--Schwarz inequality to the quadratic representation of the FRR in Eq.~\eqref{eq:jump_general_ffrr} yields (see Appendix~\ref{appsubsec:frur_derivation} for details)
\begin{align} \label{eq:jump_frur}
    \bigl[
        \bm R^{\bm{\theta}}_{\bm \phi,\bm{\psi}}(\omega)
    \bigr]^{\dagger}
    \textsf C_{\bm{\theta},\bm{\theta}^{\mathsf T}}(\omega)^{-1}
    \bm R^{\bm{\theta}}_{\bm \phi,\bm{\psi}}(\omega)
    \le \sum_{n>m}
    A^\phi_{nm}\,
    |\psi_{nm}(\omega)|^2
    \;.
\end{align}
For a single observable $\theta(t)$, this reduces to
\begin{equation}
    \bigl|
        R^{\theta}_{\bm \phi,\bm{\psi}}(\omega)
    \bigr|^2
    \le
    C_{\theta,\theta}(\omega)
    \sum_{n>m}
    A^\phi_{nm}\,
    |\psi_{nm}(\omega)|^2.
    \label{eq:jump_frur_scalar}
\end{equation}

Equations~\eqref{eq:langevin_frur} and \eqref{eq:jump_frur} represent the frequency-resolved RURs. In both continuous and discrete dynamics, the response at frequency $\omega$ is constrained by two ingredients: the fluctuation spectrum of the observable at that frequency and a quadratic cost associated with the perturbation. The only difference lies in how locality is organized: over configuration space for Langevin dynamics and over edges for jump processes.

\subsection{Frequency-resolved response kinetic and thermodynamic uncertainty relations}
\label{subsec:ftur}

The bounds derived in Sec.~\ref{subsec:frur} become more informative once the perturbation type is specified. Depending on this choice, the quadratic perturbation cost associated with \(A^\phi\) acquires a direct physical interpretation, leading to kinetic and thermodynamic forms of the response uncertainty relation.

We begin with overdamped Langevin systems. For a force perturbation, $\bm{F}(\bm{x})\mapsto \bm{F}(\bm{x})+\epsilon \, \bm{\psi}(\bm{x},t)$, using Eq.~\eqref{eq:N_F}, the weight matrix in Eq.~\eqref{eq:langevin_Aphi_definition} becomes
\begin{equation}
    [\textsf A^{\bm{F}}(\bm{x})]_{kl}
    =
    \delta_{kl}\,
    \frac{\pi(\bm{x})\mu_k(\bm{x})}{2T_k(\bm{x})}.
    \label{eq:langevin_A_force}
\end{equation}
Substituting Eq.~\eqref{eq:langevin_A_force} into Eq.~\eqref{eq:langevin_frur} gives
\begin{align} \label{eq:langevin_rkur}
    &\bigl[
        \bm R^{\bm{\theta}}_{\bm{F},\bm{\psi}}(\omega)
    \bigr]^{\dagger}
    \textsf C_{\bm{\theta},\bm{\theta}^{\mathsf T}}(\omega)^{-1}
    \bm R^{\bm{\theta}}_{\bm{F},\bm{\psi}}(\omega)
    \\ \nonumber 
    & \le
    \sum_{k=1}^{N}
    \int d\bm{x}\,
    \frac{\pi(\bm{x})\mu_k(\bm{x})}{2T_k(\bm{x})}
    |\psi_k(\bm{x},\omega)|^2.
\end{align}
Bounding the perturbation amplitude by $|\psi_k(\bm{x},\omega)| \le \psi_{\max}$, the inequality in Eq.~\eqref{eq:langevin_rkur} further reduces to the compact, albeit looser, bound
\begin{align} \label{eq:langevin_rkur_simple}
    &\bigl[
        \bm R^{\bm{\theta}}_{\bm{F},\bm{\psi}}(\omega)
    \bigr]^{\dagger}
    \textsf C_{\bm{\theta},\bm{\theta}^{\mathsf T}}(\omega)^{-1}
    \bm R^{\bm{\theta}}_{\bm{F},\bm{\psi}}(\omega)
    \\ \nonumber 
    &\le
    \psi_{\max}^2
    \sum_{k=1}^{N}
    \left\langle
        \frac{\mu_k(\bm{x})}{2T_k(\bm{x})}
    \right\rangle_{\mathrm{ss}}.
\end{align}
This bound provides a frequency-resolved extension of the response KUR, which was originally formulated for Markov jump processes in the time domain~\cite{Liu2025dynamical,Kwon2025fluctuation} and later extended to overdamped Langevin dynamics~\cite{Chun2026fluctuation}. 
The finite-frequency fluctuation--response inequality derived in Ref.~\cite{dechant2026finite} is recovered as a special case for separable perturbations of the form $\bm{\psi}(\bm{x},t)=\sum_q \bm{g}_q(\bm{x})\phi_q(t)$, with spatial modes $\bm{g}_q(\bm{x})$ and temporal amplitudes $\phi_q(t)$.

For linear Langevin systems with spatially uniform $\textsf M$ and $\textsf T$, and a force $\bm{F}(\bm{x}) = -\textsf A \bm{x}$, the power spectrum is known to be $\textsf C_{\bm{x},\bm{x}^{\mathsf{T}}}(\omega) =  2(\textsf M \textsf A - i\omega \textsf I)^{-1} \textsf D[(\textsf M \textsf A + i\omega \textsf I)^{\mathsf{T}}]^{-1}$~\cite{keizer2012statistical,Aslyamov2026macroscopic}.
The response function to a homogeneous force perturbation $\bm{F}(\bm{x}) \mapsto \bm{F}(\bm{x}) + \epsilon \bm{\psi}$ is readily obtained as $\bm R_{\bm{F},\bm{\psi}}^{\bm{x}}(\omega) = (\textsf{M} \textsf{A} - i\omega \textsf{I})^{-1} \textsf{M} \bm{\psi}$. Thus, the response KUR is saturated in this case:
$[\bm{R}_{\bm{F}, \bm{\psi}}^{\bm{x}}(\omega)]^\dagger \textsf{C}_{\bm{x},\bm{x}^{\mathsf{T}}}(\omega)^{-1} \bm{R}_{\bm{F}, \bm{\psi}}^{\bm{x}}(\omega) = \bm{\psi}^{\mathsf{T}} \textsf{M}(2\textsf{T})^{-1} \bm{\psi}$.
This saturation has recently been verified by direct algebraic calculation in Refs.~\cite{dechant2026finite,Aslyamov2026macroscopic}. 
Within the present framework, however, the equality condition of the Cauchy--Schwarz inequality [Eqs.~\eqref{eq:CS_ineq_langevin} and \eqref{eq:CS_ineq_jump}] reveals that this saturation is a structural consequence of the frequency-resolved FRR. This perspective not only explains why the response KUR is saturated for linear Langevin systems, but also provides a potential route to identifying saturation conditions for other response uncertainty relations and more general settings.

A second important choice is the mobility perturbation, $\ln \textsf{M}(\bm{x})\mapsto \ln \textsf{M}(\bm{x})+\epsilon \, \bm{\psi}(\bm{x},t)$.
In this case, the weight matrix is determined by the steady-state current as shown in Eq.~\eqref{eq:app_langevin_A_lnM}, and Eq.~\eqref{eq:langevin_frur} becomes
\begin{align} \label{eq:langevin_rtur}
    & \bigl[
        \bm{R}^{\bm{\theta}}_{\ln \textsf{M},\bm{\psi}}(\omega)
    \bigr]^{\dagger}
    \textsf{C}_{\bm{\theta},\bm{\theta}^{\mathsf T}}(\omega)^{-1}
    \bm{R}^{\bm{\theta}}_{\ln \textsf{M},\bm{\psi}}(\omega)
    \\ \nonumber
    & \le
    \sum_{k=1}^{N}
    \int d\bm{x}\,
    \frac{[j_k^{\mathrm{ss}}(\bm{x})]^2}{2\pi(\bm{x})D_k(\bm{x})}
    |\psi_k(\bm{x},\omega)|^2.
\end{align}
Bounding the perturbation amplitude as $|\psi_k(\bm{x},\omega)| \le \psi_{\max}$, this implies
\begin{align}\label{eq:langevin_rtur_simple}
    & \bigl[
        \bm{R}^{\bm{\theta}}_{\ln \textsf{M},\bm{\psi}}(\omega)
    \bigr]^{\dagger}
    \textsf{C}_{\bm{\theta},\bm{\theta}^{\mathsf T}}(\omega)^{-1}
    \bm{R}^{\bm{\theta}}_{\ln \textsf{M},\bm{\psi}}(\omega)
    \le
    \psi_{\max}^2\,\frac{\sigma}{2},
\end{align}
where the entropy production rate $\sigma$ is defined as
\begin{equation}
    \sigma
    \equiv
    \sum_{k=1}^{N}
    \int d\bm{x}\,
    \frac{[j_k^{\mathrm{ss}}(\bm{x})]^2}{\pi(\bm{x})D_k(\bm{x})}.
    \label{eq:langevin_entropy_production}
\end{equation}
This bound provides a frequency-resolved extension of the response TUR for overdamped Langevin dynamics~\cite{Kwon2025fluctuation,Chun2026fluctuation}. Together, the response KUR and TUR bound the response relative to
fluctuations in terms of complementary kinetic and thermodynamic properties of nonequilibrium systems.

A particularly important specialization of Eq.~\eqref{eq:langevin_rtur_simple} is obtained for current-like observables under a homogeneous mobility perturbation, $\psi_k(\bm{x},\omega)= 1$ (for all $k$, $\bm{x}$, and $\omega$). In this case, the global response reduces to the steady-state mean of the observable (see Appendix~\ref{appsubsec:ftur_derivation} for more details):
\begin{equation}
    \bm R^{\bm{\theta}}_{\ln \mathsf{M}, \bm{1}}(\omega)
    =
    \langle \bm{\theta}\rangle_{\mathrm{ss}} \quad
    {\rm for} ~ \mathsf{G} = \mathsf{0}.
    \label{eq:langevin_homogeneous_response}
\end{equation}
Equation~\eqref{eq:langevin_rtur_simple} then becomes
\begin{equation}
    \langle \bm{\theta}\rangle_{\mathrm{ss}}^{\mathsf T}
    \textsf C_{\bm{\theta},\bm{\theta}^{\mathsf T}}(\omega)^{-1}
    \langle \bm{\theta}\rangle_{\mathrm{ss}}
    \le
    \frac{\sigma}{2} \quad
    {\rm for} ~ \mathsf{G}= \mathsf{0}.
    \label{eq:langevin_ftur}
\end{equation}
Equation~\eqref{eq:langevin_ftur} provides a frequency-resolved counterpart of the multidimensional TUR of Ref.~\cite{Dechant2019multidimensional}. Whereas the conventional TUR constrains the covariance of time-integrated observables, our result imposes a separate constraint on each frequency component of the power spectrum.

We now turn to Markov jump systems. For an entropic perturbation, $F_{nm}\mapsto F_{nm}+\epsilon\,\psi_{nm}(t)$, substituting Eq.~\eqref{eq:jump_Aphi_definition} into Eq.~\eqref{eq:jump_frur} yields
\begin{align} \label{eq:jump_activity_bound}
    & \bigl[
        \bm{R}^{\bm{\theta}}_{\bm{F},\bm{\psi}}(\omega)
    \bigr]^{\dagger}
    \textsf{C}_{\bm{\theta},\bm{\theta}^{\mathsf T}}(\omega)^{-1}
    \bm{R}^{\bm{\theta}}_{\bm{F},\bm{\psi}}(\omega)
    \le
    \sum_{n>m}
    \frac{a_{nm}}{4}\,
    |\psi_{nm}(\omega)|^2.
\end{align}
Bounding the perturbation amplitude as $|\psi_{nm}(\omega)|\le \psi_{\max}$, Eq.~\eqref{eq:jump_activity_bound} reduces to
\begin{align}
    & \bigl[
        \bm{R}^{\bm{\theta}}_{\bm{F},\bm{\psi}}(\omega)
    \bigr]^{\dagger}
    \textsf{C}_{\bm{\theta},\bm{\theta}^{\mathsf T}}(\omega)^{-1}
    \bm{R}^{\bm{\theta}}_{\bm{F},\bm{\psi}}(\omega)
    \le
    \psi_{\max}^2\,\frac{\mathcal{A}}{4}.
\end{align}
Here, the dynamical activity $\mathcal{A}$, which characterizes the time-symmetric aspect of the dynamics~\cite{maes2020frenesy}, is defined as 
\begin{equation}
    \mathcal{A} \equiv \sum_{n>m} a_{nm} \;.
\end{equation}
This bound provides a frequency-resolved extension of the response KUR for Markov jump systems~\cite{Liu2025dynamical,Kwon2025fluctuation}.

For a kinetic perturbation, $B_{nm}\mapsto B_{nm}+\epsilon\,\psi_{nm}(t)$,
substituting Eq.~\eqref{eq:jump_Aphi_definition} into Eq.~\eqref{eq:jump_frur} yields
\begin{align} \label{eq:jump_pseudoep_bound}
    & \bigl[
        \bm{R}^{\bm{\theta}}_{\bm{B},\bm{\psi}}(\omega)
    \bigr]^{\dagger}
    \textsf{C}_{\bm{\theta},\bm{\theta}^{\mathsf T}}(\omega)^{-1}
    \bm{R}^{\bm{\theta}}_{\bm{B},\bm{\psi}}(\omega)
    \le
    \sum_{n>m}
    \frac{j_{nm}^2}{a_{nm}}\,
    |\psi_{nm}(\omega)|^2.
\end{align}
Bounding the perturbation amplitude as $|\psi_{nm}(\omega)|\le \psi_{\max}$ yields
\begin{equation}
    \bigl[
        \bm{R}^{\bm{\theta}}_{\bm{B},\bm{\psi}}(\omega)
    \bigr]^{\dagger}
    \textsf{C}_{\bm{\theta},\bm{\theta}^{\mathsf T}}(\omega)^{-1}
    \bm{R}^{\bm{\theta}}_{\bm{B},\bm{\psi}}(\omega)
    \le
    \psi_{\max}^2\,\frac{\sigma^{\mathrm{ps}}}{2},
    \label{eq:jump_pseudoep_bound_simple}
\end{equation}
where $\sigma^{\mathrm{ps}}$ is the pseudo-entropy production rate defined as
\begin{equation}
    \sigma^{\mathrm{ps}}
    \equiv
    2\sum_{n>m}
    \frac{j_{nm}^2}{a_{nm}}.
    \label{eq:jump_pseudo_entropy_production}
\end{equation}
This bound provides a frequency-resolved extension of the response TUR for Markov jump systems~\cite{Kwon2025fluctuation}.

As in the Langevin case, a direct bound on the mean value of a current-like observable is obtained under a homogeneous kinetic perturbation, $\psi_{nm}(\omega)=1$ (for all $n>m$ and $\omega$). 
For current-like observables, the corresponding global response reduces to the steady-state mean (see Appendix~\ref{appsubsec:ftur_derivation} for details),
\begin{equation}
    \bm{R}^{\bm{\theta}}_{\bm{B},\bm{1}}(\omega)
    =
    \langle \bm{\theta}\rangle_{\mathrm{ss}}
    \quad
    {\rm for} ~ \mathsf{G} = \mathsf{0} \;,
    \label{eq:jump_homogeneous_response}
\end{equation}
and Eq.~\eqref{eq:jump_pseudoep_bound_simple} becomes
\begin{equation}
    \langle \bm{\theta}\rangle_{\mathrm{ss}}^{\mathsf T}
    \textsf{C}_{\bm{\theta},\bm{\theta}^{\mathsf T}}(\omega)^{-1}
    \langle \bm{\theta}\rangle_{\mathrm{ss}}
    \le
    \frac{\sigma^{\mathrm{ps}}}{2} \leq \frac{\sigma}{2} \quad
    {\rm for} ~ \mathsf{G} = \mathsf{0},
    \label{eq:jump_ftur}
\end{equation}
where $\sigma^{\rm{ps}} \le \sigma$ follows from the log-mean inequality~\cite{Vo2022unified}.
This bound provides a frequency-resolved extension of the multidimensional TUR for Markov jump systems~\cite{Dechant2019multidimensional}.

It is useful to clarify the relation between the frequency-resolved TUR and the finite-time TUR. For the time-averaged current
\(\bar{\bm\theta}_\tau=\tau^{-1}\int_0^\tau dt\,\bm\theta(t)\) with $\mathsf{G}= \mathsf{0}$, the
multidimensional finite-time TUR reads
\begin{equation}
    \langle\bm\theta\rangle_{\rm ss}^{\mathsf T}
    \Xi_{\bm\theta,\bm\theta^{\mathsf T}}(\tau)^{-1}
    \langle\bm\theta\rangle_{\rm ss}
    \le \frac{\sigma}{2} \quad
    {\rm for} ~ \mathsf{G} = \mathsf{0} \;,
\end{equation}
where
\(\Xi_{\bm\theta,\bm\theta^{\mathsf T}}(\tau)
\equiv \tau\,{\rm Cov}(\bar{\bm\theta}_\tau,\bar{\bm\theta}_\tau^{\mathsf T})\) is the scaled covariance matrix of $\bar{\bm{\theta}}_\tau$~\cite{Dechant2019multidimensional}.
Using the Fourier representation and the identity $\int_0^\tau dt \int_0^\tau ds ~ e^{-i\omega(t-s)} = 4\sin^2(\omega\tau/2)/\omega^2$, we obtain
\begin{equation} \label{eq:finite_time_cov_rep}
    \Xi_{\bm\theta,\bm\theta^{\mathsf T}}(\tau)
    =
    \int_{-\infty}^{\infty}
    \frac{d\omega}{2\pi}\,
    \mathsf{C}_{\bm\theta,\bm\theta^{\mathsf T}}(\omega)
    \frac{4\sin^2(\omega\tau/2)}{\tau\omega^2}.
\end{equation}
The kernel $4\sin^2(\omega\tau/2)/(\tau \omega^2)$ in this integral is non-negative and normalized to unity.
Since
the matrix inverse is operator convex on positive definite matrices~\cite{boyd2004convex}, we have
\begin{equation}
    \Xi_{\bm\theta,\bm\theta^{\mathsf T}}(\tau)^{-1}
    \preceq
    \int_{-\infty}^{\infty}
    \frac{d\omega}{2\pi}\,
    \mathsf{C}_{\bm\theta,\bm\theta^{\mathsf T}}(\omega)^{-1}
    \frac{4\sin^2(\omega\tau/2)}{\tau\omega^2}
\end{equation} 
where $\mathsf A\preceq\mathsf B$ means that
$\mathsf B-\mathsf A$ is positive semidefinite.
Using this relation and the normalization of the kernel, for any observation time \(\tau\), we obtain
\begin{equation}
    \langle\bm\theta\rangle_{\rm ss}^{\mathsf T}
    \Xi_{\bm\theta,\bm\theta^{\mathsf T}}(\tau)^{-1}
    \langle\bm\theta\rangle_{\rm ss}
    \le
    \sup_\omega
    \langle\bm\theta\rangle_{\rm ss}^{\mathsf T}
    \textsf{C}_{\bm\theta,\bm\theta^{\mathsf T}}(\omega)^{-1}
    \langle\bm\theta\rangle_{\rm ss}
    \le
    \frac{\sigma}{2} \;,
    \label{eq:time_TUR_frequency_TUR}
\end{equation}
for $\mathsf{G} = \mathsf{0}$.
This hierarchy shows that the frequency-resolved TUR provides an entropy-production estimator that is at least as tight as any finite-time TUR. Physically, finite-time current fluctuations combine different frequency components through the temporal observation window as seen in Eq.~\eqref{eq:finite_time_cov_rep}, whereas the frequency-resolved TUR isolates individual spectral components and allows one to identify the frequency that yields the tightest bound. We illustrate this advantage in Sec.~\ref{sec:examples}.

The various inequalities derived in this section demonstrate that the general response bound in Sec.~\ref{subsec:frur} generates a family of concrete uncertainty relations whose right-hand sides are expressed in terms of mobility-related quantities, dynamical activity, entropy production, and pseudo-entropy production, depending on the type of dynamics and perturbation. These results show that familiar uncertainty relations of nonequilibrium thermodynamics can be understood as different manifestations of a common frequency-resolved fluctuation--response framework obtained through appropriate choices of perturbations and observables~\cite{Ptaszynski2024dissipation,Aslyamov2025nonequilibrium,Kwon2025fluctuation,Chun2026fluctuation}.

\subsection{Equilibrium limit: fluctuation--dissipation theorem}
\label{subsec:fdt}

The frequency-resolved FRR reduces to the equilibrium FDT when the steady-state currents vanish and the local responses obey the reciprocity relation. We begin with the overdamped Langevin case.
For force perturbations, combining Eqs.~\eqref{eq:langevin_current_response}, \eqref{eq:langevin_P_definition}, and \eqref{eq:N_F} yields
\begin{align}
    \textsf{R}^{\bm{\jmath}(\bm{x})}_{\bm{F}(\bm{z})}(\omega)
    &\equiv 
    \bigl(\bm{R}^{\bm{\jmath}(\bm{x})}_{F_1(\bm{z})}(\omega), \cdots, \bm{R}^{\bm{\jmath}(\bm{x})}_{F_N(\bm{z})}(\omega) \bigr) \nonumber \\
    &=
    \textsf{P}(\bm{x},\bm{z};\omega)\,
    \textsf{M}(\bm{z})\pi(\bm{z}).
    \label{eq:langevin_current_response_force}
\end{align}
Appendix~\ref{appsubsec:alternative_cov_response} shows that the current--current covariance can be rewritten as
\begin{equation}
    \begin{aligned}
       & \textsf{C}_{\bm{\jmath}(\bm{x}),\bm{\jmath}(\bm{z})^{\mathsf T}}(\omega)
        =
        \textsf{R}^{\bm{\jmath}(\bm{x})}_{\bm{F}(\bm{z})}(\omega)\,\textsf{T}(\bm{z})
        +
        \textsf{T}(\bm{x})
        \bigl[
            \textsf{R}^{\bm{\jmath}(\bm{z})}_{\bm{F}(\bm{x})}(\omega)
        \bigr]^{\dagger}
        \\
        &+
        \bigl[
            \hat{\bm{J}}_{\bm{x}}H(\bm{x},\bm{z};\omega)
        \bigr]
        \bm{j}_{\mathrm{ss}}(\bm{z})^{\mathsf T}
        +
        \bm{j}_{\mathrm{ss}}(\bm{x})
        \bigl[
            \hat{\bm{J}}_{\bm{z}}H(\bm{z},\bm{x};\omega)
        \bigr]^{\dagger}.
    \end{aligned}
    \label{eq:langevin_covariance_alternative}
\end{equation} 
At equilibrium, the last two terms vanish because $\bm{j}_{\mathrm{ss}}(\bm{x})=0$. In addition, time-reversal symmetry implies the reciprocity relation (see Appendix~\ref{appsubsec:fdt_reciprocity} for details)
\begin{equation}
    \bigl[
        \textsf{R}^{\bm{\jmath}(\bm{x})}_{\bm{F}(\bm{z})}(\omega)
    \bigr]_{\mathrm{eq}}
    =
    \bigl[
        \textsf{R}^{\bm{\jmath}(\bm{z})}_{\bm{F}(\bm{x})}(\omega)
    \bigr]_{\mathrm{eq}}^{\mathsf{T}}.
    \label{eq:langevin_equilibrium_reciprocity}
\end{equation} 
Therefore, for a spatially homogeneous temperature $\textsf T(\bm{x})=T \textsf I$, Eq.~\eqref{eq:langevin_covariance_alternative} reduces at equilibrium to the local FDT
\begin{equation}
    \bigl[
        \textsf{C}_{\bm{\jmath}(\bm{x}),\bm{\jmath}(\bm{z})^{\mathsf T}}(\omega)
    \bigr]_{\mathrm{eq}}
    =
    2T\,
    \mathrm{Re}\!\left[
        \textsf{R}^{\bm{\jmath}(\bm{x})}_{\bm{F}(\bm{z})}(\omega)
    \right]_{\mathrm{eq}}.
    \label{eq:langevin_local_fdt}
\end{equation}
This is the local frequency-resolved FDT for empirical currents. Since any current-like observable is obtained as a linear functional of the empirical current, the corresponding observable-level FDT follows by multiplying Eq.~\eqref{eq:langevin_local_fdt} with the weight functions \(\textsf L(\bm{x})\) and \(\textsf L(\bm{z})\) and integrating over \(\bm{x}\) and \(\bm{z}\). 

An analogous argument applies to Markov jump systems. Appendix~\ref{appsubsec:alternative_cov_response} shows that the local current--current covariance can be rewritten as 
\begin{equation}
    C_{\jmath_{nm},\jmath_{n'm'}}(\omega)
    =
    R^{\jmath_{nm}}_{F_{n'm'}}(\omega)
    +
    R^{\jmath_{n'm'}}_{F_{nm}}(-\omega)
    +
    \Delta_{nm,n'm'}(\omega),
    \label{eq:jump_covariance_alternative}
\end{equation} 
where $\Delta_{nm,n'm'}(\omega)$ is a term proportional to the steady-state currents. Its explicit form is given in Eqs.~\eqref{eq:def_Z} and \eqref{eq:def_Delta}. At equilibrium, $j_{nm}=0$ for all edges, and therefore $\Delta_{nm,n'm'}(\omega)$ vanishes. In addition, in equilibrium, time-reversal symmetry implies the reciprocity relation (see Appendix~\ref{appsubsec:fdt_reciprocity} for details) 
\begin{equation}
    \bigl[
        R^{\jmath_{nm}}_{F_{n'm'}}(\omega)
    \bigr]_{\mathrm{eq}}
    =
    \bigl[
        R^{\jmath_{n'm'}}_{F_{nm}}(\omega)
    \bigr]_{\mathrm{eq}}.
    \label{eq:jump_equilibrium_reciprocity}
\end{equation} 
Substituting Eq.~\eqref{eq:jump_equilibrium_reciprocity} into Eq.~\eqref{eq:jump_covariance_alternative} gives
\begin{equation}
    \bigl[
        C_{\jmath_{nm},\jmath_{n'm'}}(\omega)
    \bigr]_{\mathrm{eq}}
    =
    2 {\rm Re}\,
    \bigl[
       R^{\jmath_{nm}}_{F_{n'm'}}(\omega)
    \bigr]_{\mathrm{eq}}.
    \label{eq:jump_local_fdt}
\end{equation}
This is the equilibrium FDT for local edge currents in Markov jump systems. Since any current-like observable is a linear combination of edge currents, the observable-level equilibrium FDT follows by multiplying both sides of Eq.~\eqref{eq:jump_local_fdt} by the coefficients $L_{nm}$ and $L_{n'm'}$ and summing over all pairs of edges.

\subsection{FDT violation and Harada--Sasa relations
}
\label{subsec:harada_sasa}

The terms proportional to the steady-state currents in Eqs.~\eqref{eq:langevin_covariance_alternative} and \eqref{eq:jump_covariance_alternative} vanish at equilibrium, reducing the frequency-resolved FRR to the equilibrium FDT. In nonequilibrium steady states, however, these terms remain finite and quantify the violation of the FDT, providing a direct signature of irreversible dynamics. In this section, we show that this FDT violation is directly related to heat dissipation, leading to Harada--Sasa relations~\cite{Harada2005equality,Harada2006energy}.

We first consider the overdamped Langevin case. Under spatially homogeneous mobility and temperature,
\begin{equation}
    \textsf{M}(\bm{x})=\mu \textsf{I},
    \qquad
    \textsf{T}(\bm{x})=T\textsf{I},
\end{equation}
integrating Eq.~\eqref{eq:langevin_covariance_alternative} over $\bm{x}$ and $\bm{z}$ yields the covariance--response identity for the velocity observable $\dot{\bm{x}}(t)$:
\begin{align} \label{eq:langevin_fdt_violation_matrix}
    &\textsf{C}_{\dot{\bm{x}},\dot{\bm{x}}^{\mathsf T}}(\omega)
    -
    T \textsf{R}^{\dot{\bm{x}}}_{\bm{F},\bm 1}(\omega)
    -
    T 
    \bigl[
        \textsf{R}^{\dot{\bm{x}}}_{\bm{F},\bm 1}(\omega)
    \bigr]^{\dagger}
    \\ \nonumber 
    & =
    \textsf{Z}(\omega)
    +
    \textsf{Z}(\omega)^{\dagger},
\end{align}
where $\textsf{Z}(\omega)$ denotes the term proportional to the steady-state current, defined as 
\begin{align}
    \textsf{Z}(\omega)  \equiv \int d\bm{x} \int d\bm{z} \bigl[
            \hat{\bm{J}}_{\bm{x}}H(\bm{x},\bm{z};\omega)
        \bigr]
        \bm{j}_{\mathrm{ss}}(\bm{z})^{\mathsf T}.
\end{align}
Here, $\textsf{R}^{\dot{\bm{x}}}_{\bm{F},\bm 1}(\omega)$ denotes the response matrix to a spatially uniform global force perturbation, defined by
\begin{equation}
    \textsf{R}^{\dot{\bm{x}}}_{\bm{F},\bm 1}(\omega) 
    \equiv
    \bigl(\bm{R}^{\dot{\bm{x}}}_{F_1,1}(\omega), \cdots, \bm{R}^{\dot{\bm{x}}}_{F_N,1}(\omega) \bigr)
\end{equation}
with $\bm{R}^{\dot{\bm{x}}}_{F_k,a}(\omega) = 
\int d\bm{x}\,a \bm R^{\dot{\bm{x}}}_{F_k(\bm{x})}(\omega)$. The left-hand side of Eq.~\eqref{eq:langevin_fdt_violation_matrix} quantifies the frequency-resolved violation of the equilibrium FDT for the velocity fluctuations.

Integrating Eq.~\eqref{eq:langevin_fdt_violation_matrix} over frequency $\omega$ and taking the trace over the coordinate indices, together with the identity 
\begin{equation}
    \int_{-\infty}^{\infty}
    H(\bm{x},\bm{z};\omega)\,
    \frac{d\omega}{2\pi}
    =
    \frac{1}{2}
    \bigl[
        \delta(\bm{x}-\bm{z})-\pi(\bm{x})
    \bigr],
    \label{eq:langevin_H_frequency_integral}
\end{equation}
we obtain
\begin{align} \label{eq:langevin_harada_sasa}
    &\sum_{k=1}^{N}
    \int_{-\infty}^{\infty}
    \Bigl[
        C_{\dot{x}_k,\dot{x}_k}(\omega)
        -
        2T\,\mathrm{Re}\,
        R^{\dot{x}_k}_{F_k,1}(\omega)
    \Bigr]
    \frac{d\omega}{2\pi}
    \\ \nonumber 
    &=
    \mu\,
    \langle \dot Q\rangle_{\mathrm{ss}}
    -
    \bigl|
        \langle \dot{\bm{x}}\rangle_{\mathrm{ss}}
    \bigr|^2, 
\end{align}
where the steady-state heat dissipation rate is defined as
\begin{equation}
    \langle \dot Q\rangle_{\mathrm{ss}}
    =
    \int d\bm{x}\,
    \bm{F}(\bm{x})^{\mathsf T}\bm{j}_{\mathrm{ss}}(\bm{x}),
    \label{eq:langevin_heat_dissipation}
\end{equation}
and $\langle \dot{\bm{x}}\rangle_{\mathrm{ss}} = \int d\bm{x}\, \bm{j}_{\mathrm{ss}}(\bm{x})$ is used.
Equation~\eqref{eq:langevin_harada_sasa} is the Harada--Sasa relation, showing that the frequency-integrated FDT violation quantifies heat dissipation, with a correction associated with the steady-state mean drift~\cite{Harada2005equality,Harada2006energy}.

A more general relation can be obtained without assuming spatially homogeneous mobility by introducing the current-like observable $\bm{w}(t)=\textsf{M}(\bm{x}(t))^{-1}\circ \dot{\bm{x}}(t)$. For a divergence-free mobility matrix, i.e., $\nabla \cdot \textsf{M} = \bm{0}$, Eq.~\eqref{eq:langevin_covariance_alternative} yields the generalized Harada--Sasa relation \begin{align} \label{eq:langevin_harada_sasa_general}
    & \sum_{k=1}^{N}
    \int_{-\infty}^{\infty}
    \mathrm{Re}\Bigl[
        C_{\dot{x}_k,w_k}(\omega)
        -
        2T
        R^{w_k}_{F_k,1}(\omega)
    \Bigr]
    \frac{d\omega}{2\pi}
    \\ \nonumber & =
    \langle \dot Q\rangle_{\mathrm{ss}}
    -
    \langle \bm{w}\rangle_{\mathrm{ss}}^{\mathsf T}
    \langle \dot{\bm{x}}\rangle_{\mathrm{ss}}.
\end{align} 
The detailed derivation is given in Appendix~\ref{appsubsec:general_harada_sasa}. This relation was previously studied in Ref.~\cite{Golestanian2025hodrydynamically}, where the spatial dependence of the mobility arises from hydrodynamic interactions.

Although an exact analog of the Harada--Sasa relation is absent for Markov jump systems, a closely related identity can be derived from Eq.~\eqref{eq:jump_covariance_alternative}, where the violation of the equilibrium FDT is encoded in the term $\Delta_{nm,n'm'}(\omega)$.
As an example, we consider a directed cycle $\mathcal{C}$, consisting of $n_{\mathcal{C}}$ nodes, embedded in a larger network. The corresponding cycle-summed empirical current is defined by
\begin{equation}
    \jmath^{\mathcal C}(t)
    =
    \sum_{i=1}^{n_{\mathcal C}}
    \jmath_{i+1,i}(t),
    \label{eq:cycle_current_definition}
\end{equation}
where the node index is understood modulo \(n_{\mathcal C}\). Summing Eq.~\eqref{eq:jump_covariance_alternative} over all pairs of edges in the cycle yields the covariance--response identity for the cycle-summed current,
\begin{equation}
    C_{\jmath^{\mathcal C},\jmath^{\mathcal C}}(\omega)
    =
    R^{\jmath^{\mathcal C}}_{\bm{F}^{\mathcal C},\bm{1}^{\mathcal C}}(\omega)
    +
    R^{\jmath^{\mathcal C}}_{\bm{F}^{\mathcal{C}},\bm{1}^{\mathcal C}}(-\omega)
    +
    \Delta_{\mathcal C}(\omega),
    \label{eq:jump_cycle_covariance_identity}
\end{equation}
where $R^{\jmath^{\mathcal C}}_{\bm{F}^{\mathcal C},\bm{1}^{\mathcal C}}(\omega) \equiv \sum_{i=1}^{n_{\mathcal{C}}} R^{\jmath^{\mathcal C}}_{F_{i+1,i}}(\omega) $ denotes the response to a homogeneous entropic perturbation applied to all ordered pairs of oriented edges of the cycle $\mathcal{C}$, and $\Delta_{\mathcal C}(\omega) \equiv \sum_{i=1}^{n_{\mathcal{C}}} \sum_{i^\prime=1}^{n_{\mathcal{C}}} \Delta_{(i+1,i),(i^\prime+1,i^\prime)} (\omega)$  is the corresponding cycle-summed violation term.

Using the identity
\begin{equation}
    \int_{-\infty}^{\infty}
    H_{nm}(\omega)\,\frac{d\omega}{2\pi}
    =
    \frac{1}{2}
    (\delta_{nm}-\pi_n),
    \label{eq:jump_H_frequency_integral}
\end{equation}
the frequency-integrated form of Eq.~\eqref{eq:jump_cycle_covariance_identity} becomes
\begin{align} \label{eq:jump_harada_sasa}
    & \int_{-\infty}^{\infty}
    \Bigl[
        C_{\jmath^{\mathcal C},\jmath^{\mathcal C}}(\omega)
        -
        2\,\mathrm{Re}\,
        R^{\jmath^{\mathcal C}}_{\bm{F}^{\mathcal C},\bm{1}^{\mathcal C}}(\omega)
    \Bigr]
    \frac{d\omega}{2\pi}
    \\ \nonumber 
    & =
    \frac{1}{2}
    \sum_{i=1}^{n_{\mathcal C}}
    \Bigl[
        W_{i+1,i}
        -
        W_{i-1,i}
    \Bigr]
    \Bigl[
        j_{i+1,i}
        +
        j_{i,i-1}
    \Bigr]
    -
    \bigl(
        J^{\mathcal C}
    \bigr)^2,
\end{align}
where the stationary cycle-summed current is defined as
\begin{equation}
    J^{\mathcal C}
    \equiv
    \sum_{i=1}^{n_{\mathcal C}}
    j_{i+1,i}.
    \label{eq:cycle_mean_current}
\end{equation}
Equation~\eqref{eq:jump_harada_sasa} can therefore be regarded as a Harada--Sasa-type relation for Markov jump systems. Unlike the Langevin case, however, the integrated violation of the equilibrium FDT is not directly related to heat dissipation. Instead, it is expressed in terms of the irreversible steady-state currents and transition rates along the cycle.

\begin{figure*}
    \centering
    \includegraphics[width=\textwidth]{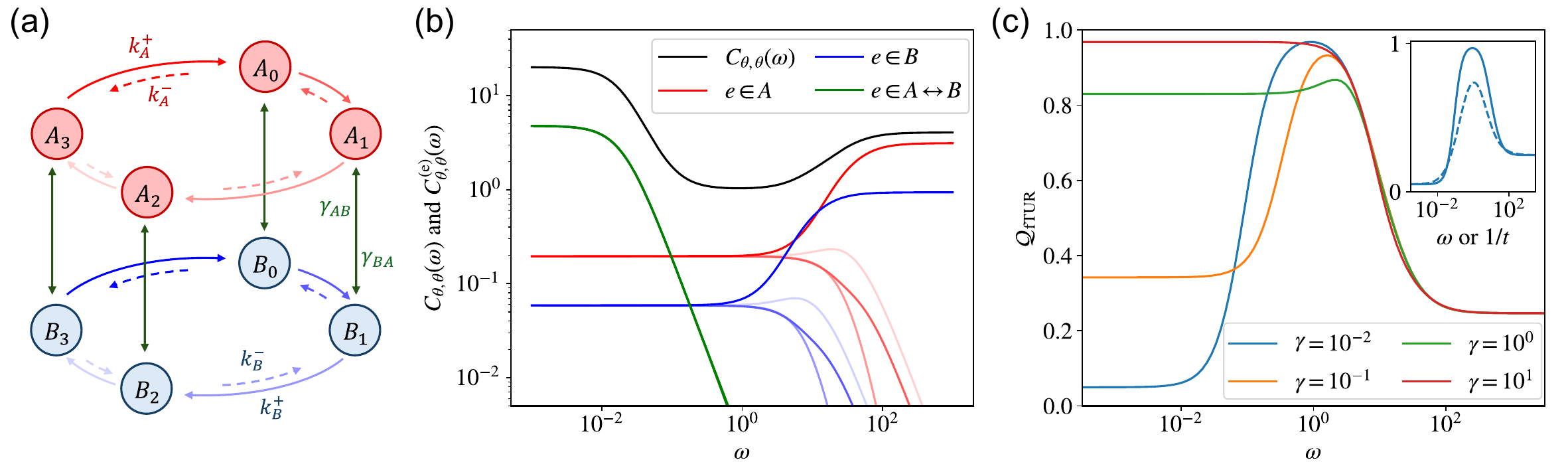}
    \caption{Dynamic-disorder model with two conformational states and its frequency-resolved FRR and TUR.
    (a) Schematic illustration of the dynamic-disorder model with two conformational states. The system consists of two conformations, \(A\) and \(B\), each containing \(N=4\) substates arranged in a periodic cycle. 
    The color intensity of the cycle edges indicates their distance from the observed boundary edge: the darkest edge corresponds to \(3\leftrightarrow0\), followed by \(0\leftrightarrow1\), \(1\leftrightarrow2\), and \(2\leftrightarrow3\).
    (b) Edgewise contributions to the current power spectrum. The black curve shows the total spectrum, \(C_{\theta,\theta}(\omega)\). The colored curves show the edgewise contributions, \(C_{\theta,\theta}^{(e)}(\omega)=|R^\theta_{F_e}(\omega)|^2/A_e^F\) associated with local entropic perturbations applied to individual edges. Red curves correspond to edges in the \(A\)-cycle, blue curves to edges in the \(B\)-cycle, and green curves to the switching edges \(A_i\leftrightarrow B_i\). Within each color, the color intensity follows the convention used in panel (a). The contributions from the four switching edges overlap.
   (c) Frequency-resolved TUR quality factor,
    \(Q_{\rm fTUR}(\omega)=
    2\langle\theta\rangle_{\rm ss}^2/
    [\sigma C_{\theta,\theta}(\omega)]\)
    for several switching rates \(\gamma_{AB}=\gamma_{BA}\equiv\gamma\).
    Inset: comparison of the frequency-resolved TUR quality factor (solid line) with the finite-time TUR quality factor
    (dashed line) for \(\gamma=10^{-2}\).
    }
    \label{fig:two_conf}
\end{figure*}

\section{Examples}
\label{sec:examples}

We illustrate the frequency-resolved FRR and TUR using two representative models. The first is a Markov jump model of a molecular motor with two conformational states, and the second is an overdamped Langevin particle in a tilted periodic potential. 

\subsection{Dynamic-disorder model with two conformational states}
\label{subsec:dynamic_disorder_motor}

The Markov jump model is schematically illustrated in Fig.~\ref{fig:two_conf}(a). The system has two conformational states, denoted by \(A\) and \(B\). Each conformation consists of \(N\) substates arranged on a periodic cycle. Within conformation \(\alpha=A,B\), transitions \(i\leftrightarrow i+1\) occur with rates \(k_\alpha^+\) and \(k_\alpha^-\), where the state index is understood modulo \(N\). The two conformations are connected by switching transitions \(A_i\leftrightarrow B_i\) with rates \(\gamma_{AB}\) and \(\gamma_{BA}\). In the numerical examples below, we set \(\gamma_{AB}=\gamma_{BA}\equiv\gamma\).

This model is motivated by dynamic disorder observed in single-molecule enzyme and motor kinetics. Dynamic disorder refers to temporal modulation of the apparent transition rates by slowly evolving internal or conformational degrees of freedom~\cite{Zwanzig1990rate,Lu1998singlemolecule,English2006everfluctuating}. In our model, switching between conformations A and B changes the transition rates between substeps, i.e., $k_A^\pm \leftrightarrow k_B^\pm$. Although the dynamics in the full state space is Markovian, eliminating the hidden conformational state can produce correlated turnovers and non-exponential waiting-time statistics~\cite{Flomenbom2005stretched,Hwang2016decoding}.

Here, we consider the current-like observable \(\theta(t)\), defined as the net instantaneous current through the boundary transitions. Specifically, forward transitions \(A_{N-1}\to A_0\) and \(B_{N-1}\to B_0\) contribute positively, whereas the reverse transitions \(A_0\to A_{N-1}\) and \(B_0\to B_{N-1}\) contribute negatively. Thus, \(\theta(t)\) does not distinguish which conformation generates the current; it only measures the net cycle-completion rate. This choice reflects typical single-molecule experiments, where the internal conformational dynamics are hidden and only completed turnover or stepping events are directly observed.

We first examine the frequency-resolved FRR using this model. The power spectrum of the current can be decomposed into contributions from individual edges. For definiteness, we consider entropic perturbations \(F_e\) applied to each edge \(e\). The frequency-resolved FRR then reads \begin{equation}
    C_{\theta,\theta}(\omega)
    =
    \sum_e C_{\theta,\theta}^{(e)}(\omega), ~~{\rm where~~} C_{\theta,\theta}^{(e)}(\omega)
    \equiv
    \frac{|R^\theta_{F_e}(\omega)|^2}{A_e^F}.
    \label{eq:edge_sum_motor}
\end{equation}
The quantity \(C_{\theta,\theta}^{(e)}(\omega)\) quantifies the contribution of edge \(e\) to the power spectrum at frequency \(\omega\). Equivalently, it represents the sensitivity of the current to a local entropic perturbation on edge \(e\), normalized by the weight \(A_e^F\). We classify the edges into three categories: edges within the \(A\) conformation, edges within the \(B\) conformation, and switching edges connecting the two conformations.

Figure~\ref{fig:two_conf}(b) shows this decomposition. The parameters used are \(N=4\), \(k_A^+=15\), \(k_A^-=10\), \(k_B^+=4.5\), \(k_B^-=3\), and \(\gamma=10^{-2}\). 
The switching edges dominate the low-frequency regime, \(\omega \lesssim \gamma\). This behavior is expected because perturbing a switching edge changes the relative occupation probabilities of the two conformations, and the effect of this change becomes apparent only on timescales long enough to resolve conformational switching. As the frequency increases beyond the switching rate, this contribution is progressively suppressed. By contrast, at higher frequencies the dominant contributions come from edges within the conformational cycles, particularly those adjacent to the observed boundary transition. Perturbations on these edges directly affect the short-time dynamics of the current and therefore dominate the high-frequency spectrum. The edge-resolved FRR thus identifies which microscopic transitions govern different frequency ranges of the current fluctuations.

We next examine the frequency-resolved TUR for the same current. For a homogeneous kinetic perturbation, the frequency-resolved TUR derived in Eq.~\eqref{eq:jump_ftur} reads
\begin{equation}
    Q_{\rm fTUR}(\omega)
    \equiv
    \frac{2\langle\theta\rangle_{\rm ss}^2}
    {\sigma C_{\theta,\theta}(\omega)}
    \le 1.
    \label{eq:Q_fTUR_motor}
\end{equation}
The quality factor $Q_{\rm fTUR}(\omega)$ is therefore determined entirely by the frequency dependence of the current power spectrum, since both \(\langle\theta\rangle_{\rm ss}\) and \(\sigma\) are frequency independent. Figure~\ref{fig:two_conf}(c) shows \(Q_{\rm fTUR}(\omega)\) for several values of the switching rate \(\gamma\). The parameters are the same as those used in Fig.~\ref{fig:two_conf}(b), except that the switching rate $\gamma$ is varied. 
When $\gamma$ is small, the quality factor is low in the low-frequency limit. In this regime, long-time current fluctuations are strongly influenced by the slow switching between the two conformations. The bound becomes significantly tighter at intermediate frequencies, where the contribution from conformational switching has already decayed, while the high-frequency contribution from the boundary transitions has not yet become dominant. As \(\gamma\) increases, the two conformations are mixed, and the low-frequency quality factor approaches unity. At high frequencies, the curves become nearly independent of \(\gamma\), consistent with the fact that the short-time current fluctuations are governed primarily by the local boundary transitions rather than by the relatively slow conformational switching.

It is also instructive to compare the tightness of the finite-time and frequency-resolved TURs, as anticipated by Eq.~\eqref{eq:time_TUR_frequency_TUR}. The inset of Fig.~\ref{fig:two_conf}(c) compares the two TUR bounds for \(\gamma=10^{-2}\). As expected, the finite-time bound is always looser than the optimized frequency-resolved bound. 
This illustrates the general hierarchy discussed in Sec.~\ref{subsec:ftur}: finite-time current fluctuations average over all frequency components, whereas the frequency-resolved TUR can identify the spectral window where the power spectrum is smallest and the bound is therefore the tightest. The frequency-resolved TUR therefore provides a more effective way to obtain a tighter lower bound on the entropy production.

\begin{figure*}[t]
    \centering
    \includegraphics[width=\textwidth]{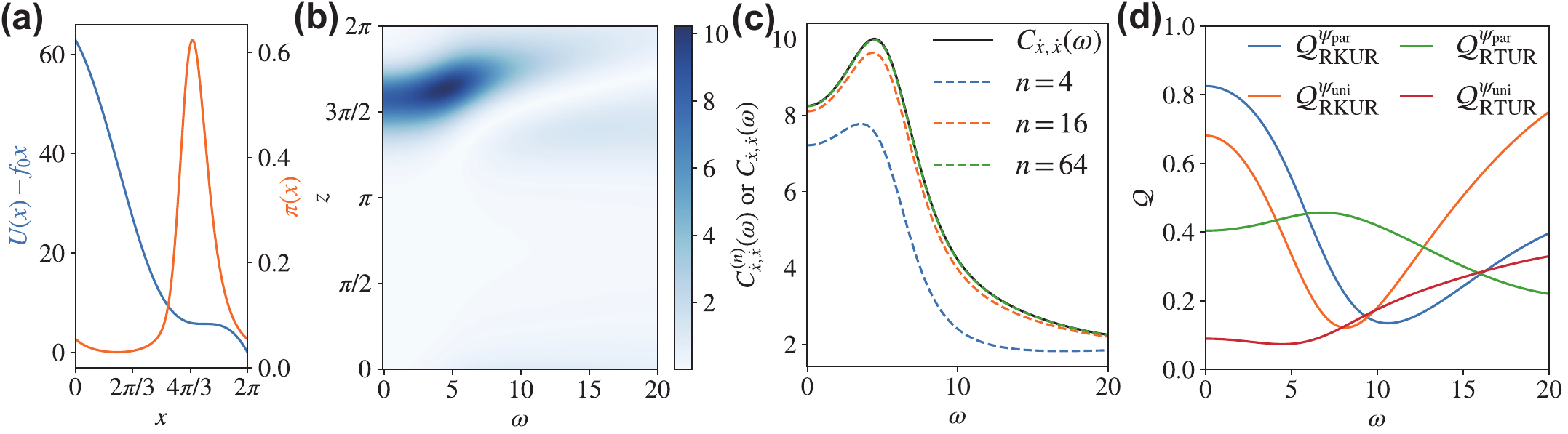}
    \caption{Brownian particle in a tilted periodic potential and its frequency-resolved FRR, response KUR, and response TUR.
    (a) Tilted periodic potential \(U(x)-f_0x\) and the corresponding steady-state probability density \(\pi(x)\) within one spatial period.
    (b) Local contribution \(c_F(z,\omega)\) to the velocity power spectrum, defined in Eq.~\eqref{eq:cF_definition_example} from the response to a local force perturbation.
    (c) Reconstruction of the velocity power spectrum from the responses to the coarse-grained force perturbations. The black curve shows the exact spectrum \(C_{\dot x,\dot x}(\omega)\), whereas the dashed curves show the reconstructed spectra \(C_{\dot x,\dot x}^{(n)}(\omega)\), defined in Eq.~\eqref{eq:coarse_grained_frr}, obtained by dividing the ring into \(n=4\), \(16\), and \(64\) equal intervals.
    (d) Response KUR and TUR quality factors for the uniform perturbation profile, \(\psi_{\rm uni}(x)=1\), and the partial perturbation profile, \(\psi_{\rm par}(x)=1\) for \(4\pi/3<x<2\pi\) and \(0\) otherwise.    
    }
    \label{fig:tilted}
\end{figure*}

\subsection{Spatially resolved response bounds in a tilted periodic potential}
\label{subsec:tilted_potential}

We next consider an overdamped Langevin particle moving on a ring of length \(L \) in a tilted periodic potential, 
\begin{equation}
    \dot x(t)
    =
    \mu\left[f_0-U'(x(t))\right]
    +
    \sqrt{2\mu T}\,\xi(t),
    \label{eq:tilted_potential_example}
\end{equation}
where \(x\) is understood modulo \(L\), and $U(x)=U_0\cos\left(\frac{2\pi x}{L}\right)$ is the periodic potential. The observable is chosen to be the particle velocity, \(\theta(t)=\dot x(t)\). This model is essentially the same as the minimal model used to describe the motion of polystyrene beads attached to the F$_1$-ATPase molecular motor~\cite{Chun2026fluctuation,Hayashi2015giant}.
Unless otherwise specified, the parameter values \(U_0=10\), \(f_0=10\), \(L=2\pi\), \(\mu=1\), and \(T=1\) are used. Figure~\ref{fig:tilted}(a) shows the tilted potential together with the corresponding steady-state density.

For a local force perturbation, Eqs.~\eqref{eq:langevin_general_ffrr} and \eqref{eq:app_langevin_A_F} yield the following FRR:
\begin{equation}
    C_{\dot x,\dot x}(\omega)
    =
    \int_0^L dz\, c_F(z,\omega),
    \label{eq:cF_definition_example}
\end{equation}
where $c_F(z,\omega)
    =
    \frac{2T}{\mu\pi(z)}
    \left|
    R^{\dot x}_{F(z)}(\omega)
    \right|^2 $.
Here, \(R^{\dot x}_{F(z)}(\omega)\) denotes the response of the mean velocity to a local force perturbation applied at position \(z\), and \(\pi(z)\) is the steady-state density. Thus, \(c_F(z,\omega)\) quantifies the local contribution of position \(z\) to the velocity power spectrum at frequency \(\omega\).

Figure~\ref{fig:tilted}(b) shows \(c_F(z,\omega)\). At low and intermediate frequencies ($\omega \lesssim 10$), the contribution is concentrated within a limited spatial region of the ring, which corresponds to a region of relatively high steady-state density. At higher frequencies, the spatial contrast becomes weaker and the contribution is more broadly distributed. Thus, \(c_F(z,\omega)\) provides information beyond the total power spectrum by revealing which spatial regions contribute most strongly to fluctuations at a given frequency.

Direct experimental implementation of Eq.~\eqref{eq:cF_definition_example}
is impractical because it requires responses to pointwise force perturbations. To relate this decomposition to experimentally accessible perturbations, we introduce a coarse-grained version of the FRR. We divide the ring into \(n\) equal intervals \(I_i=[iL/n,(i+1)L/n]\), for $i=0,\dots,n-1$.
Applying the Cauchy--Schwarz inequality to each interval yields
\begin{equation}
    \frac{2T}{\mu}\frac{\bigl| \int_{I_i} dz~ R_{F(z)}^{\dot{x}}(\omega) \bigr|^2}{\int_{I_i} dz~\pi(z)} \le \int_{I_i} dz~ c_F(z,\omega). 
\end{equation}
Summing over all intervals gives $C_{\dot x,\dot x}^{(n)}(\omega) \le C_{\dot x,\dot x}(\omega)$, where
\begin{equation}
    C_{\dot x,\dot x}^{(n)}(\omega)
    \equiv
    \sum_{i=0}^{n-1}
    \frac{2T}{\mu}
    \frac{
    \left|
    \int_{I_i} dz\,
    R^{\dot x}_{F(z)}(\omega)
    \right|^2
    }{
    \int_{I_i} dz\,\pi(z)
    }.
    \label{eq:coarse_grained_frr}
\end{equation}
The equality $C_{\dot x,\dot x}^{(n)}(\omega) = C_{\dot x,\dot x}(\omega)$ is recovered in the limit $n \rightarrow \infty $. Here, $\int_{I_i} dz\, R^{\dot x}_{F(z)}(\omega)$ represents the response to a spatially uniform force perturbation applied over the interval $I_i$. Therefore, \(C_{\dot x,\dot x}^{(n)}(\omega)\) can be obtained from the responses to a finite number of uniform perturbations, rather than requiring infinitely many pointwise perturbations. Figure~\ref{fig:tilted}(c) demonstrates the accuracy of this coarse-grained reconstruction. Even for \(n=4\), the main spectral features are well reproduced. As \(n\) increases, the reconstructed spectrum rapidly converges to the exact one. These results show that the FRR can be implemented with finite spatial resolution,
yielding a systematic hierarchy of lower bounds
that converges to the exact spectrum under successive partition refinement.

We next examine the frequency-resolved response KUR and TUR. For a force perturbation \(F(x)\to F(x)+\epsilon\psi(x)\), the response KUR~\eqref{eq:langevin_rkur} can be expressed as
\begin{equation}
    Q_{\rm RKUR}^{\psi}(\omega)
    \equiv
    \frac{
    |R^{\dot x}_{F,\psi}(\omega)|^2
    }{
    C_{\dot x,\dot x}(\omega)
    \int_0^L dx\,A^F(x)\psi(x)^2
    } \le 1,
    \label{eq:Q_RKUR_example}
\end{equation}
where $A^F(x)=\frac{\mu\pi(x)}{2T}$. Similarly, for a mobility perturbation \(\ln\mu(x)\to\ln\mu(x)+\epsilon\psi(x)\), the response TUR~\eqref{eq:langevin_rtur} can be expressed as
\begin{equation}
    Q_{\rm RTUR}^{\psi}(\omega)
    \equiv
    \frac{
    |R^{\dot x}_{\ln\mu,\psi}(\omega)|^2
    }{
    C_{\dot x,\dot x}(\omega)
    \int_0^L dx\,A^{\ln\mu}(x)\psi(x)^2
    } \le 1,
    \label{eq:Q_RTUR_example}
\end{equation}
where $A^{\ln\mu}(x)=\frac{[j^{\rm ss}(x)]^2}{2\pi(x)D}$.

Here, we consider two perturbation profiles $\psi$. The first is a spatially uniform perturbation, \(\psi_{\rm uni}(x)=1\), and the second is a partial perturbation,
\[
    \psi_{\rm par}(x)
    =
    \begin{cases}
        1, & 4\pi/3 < x < 2\pi,\\
        0, & \text{otherwise}.
    \end{cases}
\]
The partial perturbation is applied over the region where \(c_F(z,\omega)\) is large at low and intermediate frequencies.

Figure~\ref{fig:tilted}(d) shows the $Q$ factors for the response KUR and TUR under the two perturbation profiles. The partial perturbation yields tighter bounds in the low- and intermediate-frequency regime, where the local contribution \(c_F(z,\omega)\) is spatially concentrated. At higher frequencies, where the local contribution becomes more broadly distributed, the uniform perturbation becomes more effective and can provide the tighter bound. This indicates that $c_F(z,\omega)$ provides a useful guide for designing perturbation profiles that yield tighter response KUR and TUR bounds.
In particular, under a spatially uniform mobility perturbation, Eq.~\eqref{eq:Q_RTUR_example} reduces to the frequency-resolved TUR for the velocity current by virtue of Eq.~\eqref{eq:langevin_homogeneous_response}.
The red curve in Fig.~\ref{fig:tilted}(d) can therefore be interpreted as the $Q$ factor of the frequency-resolved TUR for this example.

\section{Conclusion}

In this work, we have developed a fluctuation--response theory in the frequency domain for nonequilibrium steady states. The central result is a frequency-resolved FRR that expresses the power spectrum of general observables in terms of local linear responses at the same frequency. We established this relation for both overdamped Langevin and Markov jump systems. It applies to state-dependent observables, current-like observables, and their combinations, providing a unified framework for describing fluctuations of occupations and currents in continuous- and discrete-state stochastic dynamics. Unlike previous nonequilibrium FRRs, which were formulated primarily for static or long-time covariances~\cite{Aslyamov2025nonequilibrium,Chun2026fluctuation}, the present theory applies directly at finite frequency.

We further showed that the frequency-resolved FRR provides a unified foundation for several fundamental fluctuation--response relations. Applying the Cauchy--Schwarz inequality yields frequency-resolved RURs, from which the response KUR and TUR follow for appropriate perturbation choices. 
For current-like observables, suitable homogeneous
perturbations further yield the frequency-resolved TUR, while the FRR reduces to the FDT at equilibrium and gives rise to Harada--Sasa relations away from equilibrium.
These results reveal that uncertainty relations, the FDT, and Harada--Sasa relations are all different manifestations of the same underlying frequency-resolved fluctuation--response structure.

The examples demonstrate that the frequency-resolved FRR is not merely a formal identity but also a practical tool for analyzing nonequilibrium fluctuations. In the Markov jump example, the edge-resolved decomposition reveals how different transition channels contribute to fluctuations at different frequencies and identifies the frequency range where the TUR gives the tightest bound. In the Langevin example, the spatial decomposition identifies the regions of state space that contribute most strongly to the velocity spectrum and shows that the spectrum can be accurately reconstructed from responses
to a finite number of coarse-grained perturbations. The same spatial information also guides the choice of perturbation profiles for the response KUR and TUR, revealing when partial or uniform perturbations provide tighter response bounds.
Together, these examples illustrate how the theory identifies the edges and spatial regions that contribute strongly to fluctuations and responses, as well as the frequency ranges that yield the tightest uncertainty bounds in nonequilibrium systems.

Several directions remain open. First, the present formulation suggests an experimental route for reconstructing fluctuation spectra from local or controlled finite-frequency responses, especially in systems where direct spectral measurements and perturbative measurements are both accessible. Second, extending the exact frequency-resolved relation to underdamped Langevin dynamics would broaden its applicability to inertial stochastic systems and molecular-scale transport. Third, it would be interesting to investigate whether analogous frequency-resolved fluctuation--response structures can be formulated for open quantum systems, where response, noise, and dissipation are constrained by both stochastic and quantum effects.

\section*{Acknowledgments}
This research was supported by NRF Grants No.~2017R1D1A1B06035497 (H.P.) and individual KIAS Grants Nos.~PG064902 (J.S.L.), PG089402 (H.-M.C.), QP013602 (H.P.), and QP10301 (E.K.) at the Korea Institute for Advanced Study.

\section*{Data Availability}
The data that support the findings of this article are available upon request.

\begin{appendix}

\appendix

\section{Technical details for overdamped Langevin systems}
\label{app:langevin}

\subsection{Perturbed Fokker--Planck generator and local response formulas}
\label{appsubsec:langevin_response}

In this subsection, we derive the local response formulas presented in Sec.~\ref{sec:langevin_frt}. We consider the overdamped Langevin dynamics
\begin{equation}
    \dot{\bm{x}}(t)
    =
    \textsf M (\bm{x}(t))\bm{F}(\bm{x}(t))
    +
    \textsf B (\bm{x}(t))\circledast \bm{\xi}(t),
\end{equation}
where $\circledast$ denotes the anti-It\^o product. The corresponding Fokker--Planck generator $\hat{\mathcal L}_{\bm{x}}$ is 
\begin{equation}
    \hat{\mathcal L}_{\bm{x}}
    =
    -\nabla_{\bm{x}}^{\mathsf T}\hat{\bm J}_{\bm{x}},
    \qquad
    \hat{\bm J}_{\bm{x}}
    =
    \textsf M(\bm{x})\bigl[\bm{F}(\bm{x})-\textsf T(\bm{x})\nabla_{\bm{x}}\bigr],
\end{equation}
where the relation $\frac{1}{2} \textsf B(\bm{x}) \textsf B(\bm{x})^{\mathsf{T}} = \textsf D(\bm{x}) = \textsf M(\bm{x}) \textsf T(\bm{x})$ is used.
The steady-state distribution $\pi(\bm{x})$ satisfies
\begin{equation}
    \hat{\mathcal L}_{\bm{x}}\pi(\bm{x})=0,
\end{equation}
and the corresponding steady-state probability current is
\begin{equation}
    \bm{j}_{\mathrm{ss}}(\bm{x})
    =
    \hat{\bm J}_{\bm{x}}\pi(\bm{x}).
\end{equation}

We consider a local perturbation of the $k$th component of $\bm \phi\in\{\bm{F},\ln \textsf M,\textsf T\}$ at position $\bm{z}$ and time $t=0$:
\begin{equation}
    \phi_k(\bm{x})
    \mapsto
    \phi_k(\bm{x})+\epsilon\,\delta(\bm{x}-\bm{z})\delta(t). \label{eqA:phi_impluse_perturb_Langevin}
\end{equation}
At the level of the Fokker--Planck generator, the perturbation can be written as
\begin{equation}
    \hat{\mathcal L}_{\bm{x}}
    \mapsto
    \hat{\mathcal L}_{\bm{x}}
    -
    \epsilon\, \delta(t)
    \nabla_{\bm{x}}^{\mathsf T}
    \delta(\bm{x}-\bm{z})\,
    \hat{\bm K}_{\phi_k,\bm{x}}\,, 
    \label{eqA:perturbed_FP_Generator}
\end{equation}
where the operator $\hat{\bm K}_{\phi_k,\bm{x}}$ depends on the perturbed quantity. For diagonal $\textsf M$ and $\textsf T$, it is given by
\begin{align}    
    &\hat{\bm K}_{F_k,\bm{x}}
    = \mu_k(\bm{x})\,\bm{e}_k, \label{eq:K_F} \\
    &\hat{\bm K}_{\ln M_k,\bm{x}} =\mu_k(\bm{x})\,\bm{e}_k\bm{e}_k^{\mathsf T}  \bigl[
        \bm{F}(\bm{x})-\textsf T(\bm{x})\nabla_{\bm{x}}
    \bigr],  \label{eq:K_mu}\\
    &\hat{\bm K}_{T_k,\bm{x}} = - \mu_k(\bm{x})\,\bm{e}_k\bm{e}_k^{\mathsf T}\nabla_{\bm{x}}, \label{eq:K_T}
\end{align}
where $\bm{e}_k$ denotes the $k$th unit vector.

Assuming the system is initially in the steady state, we expand the probability density after the perturbation as
\begin{equation}
    p(\bm{x},t)
    =
    \pi(\bm{x})
    +
    \epsilon\,q_{\phi_k}(\bm{x},t;\bm{z})
    +
    O(\epsilon^2).
\end{equation}
Substituting this expansion into the perturbed Fokker--Planck equation using~\eqref{eqA:perturbed_FP_Generator} and retaining terms up to first order in $\epsilon$, we obtain
\begin{equation}
    \partial_t q_{\phi_k}(\bm{x},t;\bm{z})
    =
    \hat{\mathcal L}_{\bm{x}}q_{\phi_k}(\bm{x},t;\bm{z})
    - \delta(t) \nabla_{\bm{x}}^{\mathsf T}
    \delta(\bm{x}-\bm{z})\,
    \bm N_{\phi_k}(\bm{z}) ,
    \label{eqA:first_ODE_q}
\end{equation}
with $ \bm N_{\phi_k}(\bm{z}) = \hat{\bm K}_{\phi_k,\bm{z}}\pi(\bm{z})$.
The solution of Eq.~\eqref{eqA:first_ODE_q} is 
\begin{align}
    q_{\phi_k}(\bm{x},t;\bm{z}) 
    &= - \int_0^t ds \, e^{\hat{\mathcal{L}}_{\bm x}(t-s)}\delta(s) \nabla_{\bm{x}}^{\mathsf T}
    \delta(\bm{x}-\bm{z})\,
    \bm N_{\phi_k}(\bm{z}) \nonumber \\
    &= - \int d{\bm y} \, e^{\hat{\mathcal{L}}_{\bm x}t} \delta(\bm x - \bm y) \nabla_{\bm{y}}^{\mathsf T}
    \delta(\bm{y}-\bm{z})\,
    \bm N_{\phi_k}(\bm{z}) \nonumber \\
    &= \int d{\bm y} \, \delta(\bm{y}-\bm{z})\,
    [\bm N_{\phi_k}(\bm{z})]^\textsf{T}  \nabla_{\bm{y}} P(\bm x,t|\bm y, 0) \nonumber \\
    &=\bigl[\bm N_{\phi_k}(\bm{z}) \bigr]^{\mathsf T}     \nabla_{\bm{z}}P(\bm{x},t|\bm{z},0).
\end{align}
Here, the third equality follows from integration by parts together with the identity $e^{\hat{\mathcal{L}}_{\bm x}t} \delta(\bm x - \bm y) = P(\bm x,t|\bm y, 0)$.

By definition, the response function of the empirical density is $R_{\phi_k(\bm z)}^{\rho(\bm x)} (t,0) = \delta \langle \rho(\bm x,t)\rangle /\delta\phi_k(\bm z, 0) = q_{\phi_k} (\bm x, t; \bm z)$. Therefore, recalling the definition of the excess propagator,
\begin{equation}
    H(\bm{x},\bm{z};\omega)
    =
    \int_0^\infty
    \bigl[
        P(\bm{x},t|\bm{z},0)-\pi(\bm{x})
    \bigr]e^{i\omega t}\,dt,
\end{equation}
the spectral response of the empirical density~\eqref{eq:spectral_response_Langevin} is 
\begin{equation}
    R^{\rho(\bm{x})}_{\phi_k(\bm{z})}(\omega)
    =
    \bigl[ \bm N_{\phi_k}(\bm{z}) \bigr]^{\mathsf T}
    \nabla_{\bm{z}}H(\bm{x},\bm{z};\omega).
    \label{eq:app_langevin_density_response}
\end{equation}

Under the steady-state initial condition, the perturbation~\eqref{eqA:phi_impluse_perturb_Langevin} modifies the current operator according to $\hat{\bm J}_{\bm x} \mapsto \hat{\bm J}_{\bm x} + \epsilon\,\delta(t)\delta(\bm x-\bm z)\, \hat{\bm K}_{\phi_k,\bm x}$. Accordingly, the probability current can be expanded as
\begin{equation}
    \bm j(\bm x, t) \approx \bigl[\hat{\bm J}_{\bm x} +\epsilon \delta(t)\delta(\bm x -\bm z) \hat{\bm K}_{\phi_k,\bm x} \bigr]\bigl[\pi(\bm x) + \epsilon q_{\phi_k}(\bm x, t;\bm z) \bigr] .
\end{equation}
Since the response function of the empirical current is $\bm R_{\phi_k(\bm z)}^{\bm \jmath(\bm x)} (t,0) = \lim_{\epsilon \rightarrow 0}  \delta \langle \bm \jmath(\bm x,t)\rangle /\epsilon$ with $\delta \langle \bm \jmath(\bm x, t) \rangle = \bm j(\bm x, t) - \bm j_{\rm ss}(\bm x)$, we obtain
\begin{equation}
    \bm R_{\phi_k(\bm z)}^{\bm \jmath(\bm x)} (t,0) = \delta(t)\delta(\bm{x}-\bm{z})\bm N_{\phi_k}(\bm{z})
    +
    \hat{\bm J}_{\bm{x}}\,q_{\phi_k}(\bm{x},t;\bm{z}) .
\end{equation}
Taking the Fourier transform in time yields
\begin{equation}
    \bm R^{\bm{\jmath}(\bm{x})}_{\phi_k(\bm{z})}(\omega)
    =
    \textsf P(\bm{x},\bm{z};\omega)\,\bm N_{\phi_k}(\bm{z}),
    \label{eq:app_langevin_current_response}
\end{equation}
where 
\begin{equation}
    \textsf P(\bm{x},\bm{z};\omega)
    \equiv 
    \textsf I\,\delta(\bm{x}-\bm{z})
    +
    \hat{\bm J}_{\bm{x}}\nabla_{\bm{z}} ^\mathsf{T} H(\bm{x},\bm{z};\omega).
    \label{eq:app_langevin_P_def}
\end{equation}

From Eqs.~\eqref{eq:K_F}, \eqref{eq:K_mu}, and \eqref{eq:K_T}, the local prefactor vectors for diagonal $\textsf M$ and $\textsf T$ are
\begin{align}
    &\bm N_{F_k}(\bm{z}) =
    \pi(\bm{z})\mu_k(\bm{z})\,\bm{e}_k, \label{eq:N_F} \\
    &\bm N_{\ln M_k}(\bm{z}) =
    j_k^{\mathrm{ss}}(\bm{z})\,\bm{e}_k, \label{eq:N_mu} \\
    &\bm N_{T_k}(\bm{z}) =
    -\mu_k(\bm{z}) \partial_{z_k}\pi(\bm{z})\,\bm{e}_k. \label{eq:N_T}
\end{align}
The second identity follows from the definition of the steady-state current,
\begin{equation}
    j_k^{\mathrm{ss}}(\bm{z})
    =
    \mu_k(\bm{z})
    \Bigl[
        F_k(\bm{z})\pi(\bm{z})
        -
        T_k(\bm{z})\partial_{z_k}\pi(\bm{z})
    \Bigr].
\end{equation}
Thus, once the perturbation-dependent vector $\bm N_{\phi_k}(\bm{z})$ is identified, both density and current responses are determined by the same excess propagator.

\subsection{Properties of the excess propagator}
\label{appsubsec:langevin_H}

In this subsection, we derive the identities for the excess propagator used in Sec.~\ref{sec:langevin_frt}. We begin by recalling its definition:
\begin{equation}
    H(\bm{x},\bm{z};\omega)
    =
    \int_0^\infty
    \bigl[
        P(\bm{x},t|\bm{z},0)-\pi(\bm{x})
    \bigr]e^{i\omega t}\,dt.
\end{equation}
The propagator satisfies both the forward and backward Fokker--Planck equations,
\begin{align}
    &\partial_t P(\bm{x},t|\bm{z},0)
    =
    \hat{\mathcal L}_{\bm{x}}P(\bm{x},t|\bm{z},0), \\
    &\partial_t P(\bm{x},t|\bm{z},0)
    =
    \hat{\mathcal L}_{\bm{z}}^\dagger P(\bm{x},t|\bm{z},0),
\end{align}
together with
\begin{equation}
    \lim_{t\to 0^+}P(\bm{x},t|\bm{z},0)=\delta(\bm{x}-\bm{z}),
    \qquad
    \lim_{t\to\infty}P(\bm{x},t|\bm{z},0)=\pi(\bm{x}),
\end{equation}
where $\hat{\mathcal L}_{\bm{z}}^\dagger$ denotes the adjoint Fokker--Planck generator acting on the initial position $\bm{z}$.

We first show that
\begin{equation}
    \int d\bm{z}\,H(\bm{x},\bm{z};\omega)\pi(\bm{z})=0.
    \label{eq:app_langevin_H_identity_1}
\end{equation}
Using the stationarity of the steady-state distribution,
\begin{equation}
    \int d\bm{z}\,P(\bm{x},t|\bm{z},0)\pi(\bm{z})=\pi(\bm{x}),
\end{equation}
we obtain
\begin{equation}
    \begin{aligned}
        &\int d\bm{z}\,H(\bm{x},\bm{z};\omega)\pi(\bm{z})
        \\ &=
        \int_0^\infty dt\,e^{i\omega t}
        \int d\bm{z}\,
        \bigl[
            P(\bm{x},t|\bm{z},0)-\pi(\bm{x})
        \bigr]\pi(\bm{z})=0. \nonumber
    \end{aligned}
\end{equation}

Next, we show that
\begin{equation}
    \bigl(i\omega+\hat{\mathcal L}_{\bm{x}}\bigr)H(\bm{x},\bm{z};\omega)
    =
    -\delta(\bm{x}-\bm{z})+\pi(\bm{x}).
    \label{eq:app_langevin_H_identity_2}
\end{equation}
Indeed, 
\begin{equation}
    \begin{aligned}
        &\bigl(i\omega+\hat{\mathcal L}_{\bm{x}}\bigr)H(\bm{x},\bm{z};\omega)
        \\ &=
        \int_0^\infty dt\,e^{i\omega t}
        \Bigl[
            i\omega\bigl\{P(\bm{x},t|\bm{z},0)-\pi(\bm{x})\bigr\}
            +
            \hat{\mathcal L}_{\bm{x}}P(\bm{x},t|\bm{z},0)
        \Bigr] \\
        &=
        \int_0^\infty dt\,
        \partial_t
        \Bigl[
            e^{i\omega t}\bigl\{P(\bm{x},t|\bm{z},0)-\pi(\bm{x})\bigr\}
        \Bigr] \\
        &=
        \Bigl.
            e^{i\omega t}[P(\bm{x},t|\bm{z},0)-\pi(\bm{x})]
        \Bigr|_{t=0}^{t=\infty} \\
        &=
        -\delta(\bm{x}-\bm{z})+\pi(\bm{x}),
    \end{aligned}
\end{equation}
where we used the forward Fokker--Planck equation and the limiting behaviors of the propagator.

Similarly,
\begin{equation}
    \bigl(i\omega+\hat{\mathcal L}_{\bm{z}}^\dagger\bigr)H(\bm{x},\bm{z};\omega)
    =
    -\delta(\bm{x}-\bm{z})+\pi(\bm{x}).
    \label{eq:app_langevin_H_identity_3}
\end{equation}
The proof is identical to that of Eq.~\eqref{eq:app_langevin_H_identity_2}, with the forward Fokker--Planck equation replaced by the backward one.

\subsection{Covariances of empirical density and current}
\label{appsubsec:langevin_covariances}

We now derive the covariance formulas for the empirical density and current. We begin with the empirical density $\rho(\bm{x},t)=\delta(\bm{x}-\bm{x}(t))$.
For $t>0$, its two-time correlation is
\begin{equation}
    \langle \rho(\bm{x},t)\rho(\bm{y},0)\rangle-\pi(\bm{x})\pi(\bm{y})
    =
    \bigl[
        P(\bm{x},t|\bm{y},0)-\pi(\bm{x})
    \bigr]\pi(\bm{y}),
\end{equation}
whereas for $t<0$,
\begin{equation}
    \langle \rho(\bm{x},t)\rho(\bm{y},0)\rangle-\pi(\bm{x})\pi(\bm{y})
    =
    \bigl[
        P(\bm{y},-t|\bm{x},0)-\pi(\bm{y})
    \bigr]\pi(\bm{x}),
\end{equation}
where the latter follows from time-translation invariance, $P(\bm{y},0|\bm{x},t) = P(\bm{y},-t|\bm{x},0)$.
Substituting these expressions into the spectral covariance definition, Eq.~\eqref{eq:covariance_definition_main}, yields 
\begin{equation}
    C_{\rho(\bm{x}),\rho(\bm{y})}(\omega)
    =
    H(\bm{x},\bm{y};\omega)\pi(\bm{y})
    +
    H(\bm{y},\bm{x};-\omega)\pi(\bm{x}).
    \label{eq:app_langevin_rho_rho_raw}
\end{equation}
As shown in Appendix~\ref{appsubsec:langevin_quadratic_identity}, this can be equivalently expressed in the quadratic form
\begin{align}
    &C_{\rho(\bm{x}),\rho(\bm{y})}(\omega) \nonumber
    \\&=
    2\int d\bm{z}\,
    \pi(\bm{z})\,
    \bigl[\nabla_{\bm{z}}H(\bm{x},\bm{z};\omega)\bigr]^{\mathsf T}
    \textsf D(\bm{z})
    \nabla_{\bm{z}}H(\bm{y},\bm{z};-\omega).
    \label{eq:app_langevin_rho_rho_quadratic}
\end{align}

We next decompose the empirical current into its drift and fluctuation contributions as
\begin{equation}
    \bm{\jmath}(\bm{x},t)
    =
    \hat{\bm J}_{\bm{x}}\rho(\bm{x},t)
    +
    \bm{\zeta}(\bm{x},t).
    \label{eq:app_langevin_current_decomp}
\end{equation}
Here, the fluctuation term $\bm{\zeta}(\bm{x},t)$ is defined as
\begin{equation}
    \bm{\zeta}(\bm{x},t)
    \equiv \,\delta(\bm{x}-\bm{x}(t)) \textsf B(\bm{x}(t)) \bullet \bm{\xi}(t),
\end{equation}
where $\bullet$ denotes the It\^o product. Although the Langevin equation is formulated in the anti-It\^o convention, the noise term is expressed here in the It\^o convention for computational convenience. Note that $\langle \bm{\zeta}(\bm{x},t) \rangle = 0$. 
The derivation of Eq.~\eqref{eq:app_langevin_current_decomp} proceeds as follows:
\begin{equation}
    \begin{aligned}
        &\boldsymbol\jmath (\bm{x},t) = \delta (\bm{x} - \bm{x}(t)) \circ \dot{\bm{x}}(t)
        \\ & = \delta (\bm{x} - \bm{x}(t)) \bullet \dot{\bm{x}}(t) + \textsf{M}(\bm{x}(t))\textsf{T}(\bm{x}(t)) \boldsymbol\nabla_{\bm{x}(t)} \delta(\bm{x}-\bm{x}(t)) 
        \\ & = \textsf{M}(\bm{x})[\bm{F}(\bm{x}) -\textsf{T}(\bm{x}) \boldsymbol\nabla_{\bm{x}}]\delta(\bm{x}-\bm{x}(t)) \\
        &\quad + \delta(\bm{x}-\bm{x}(t))\textsf{B}(\bm{x}(t))\bullet \boldsymbol\xi(t)\;,
    \end{aligned}
\end{equation}
where we used that $\boldsymbol\nabla_{\bm{x}(t)}\delta (\bm{x}-\bm{x}(t)) = -\boldsymbol\nabla_{\bm{x}} \delta (\bm{x}-\bm{x}(t))$.

In the steady state, the noise term has covariance 
\begin{equation}
    \langle \zeta_i(\bm{x},t)\zeta_j(\bm{y},t')\rangle
    =
    2\pi(\bm{x})D_{ij}(\bm{x})\delta(\bm{x}-\bm{y})\delta(t-t'),
\end{equation}
and therefore
\begin{equation}
    \textsf C_{\bm{\zeta}(\bm{x}),\bm{\zeta}(\bm{y})^{\mathsf T}}(\omega)
    =
    2\pi(\bm{x})\delta(\bm{x}-\bm{y}) \textsf D(\bm{x}).
    \label{eq:app_langevin_zeta_zeta}
\end{equation}
In addition, the density--noise covariance is given by
\begin{equation}
    \textsf C_{\rho(\bm{x}),\bm{\zeta}(\bm{y})^{\mathsf T}}(\omega)
    =
    2\pi(\bm{y})
    \nabla_{\bm{y}} ^\mathsf{T}H(\bm{x},\bm{y};\omega)
    \textsf D(\bm{y}),
    \label{eq:app_langevin_rho_zeta}
\end{equation}
which follows from the definition of $\bm{\zeta}$ and the delta-correlated noise.
Similarly,
\begin{equation}
    \textsf C_{\bm{\zeta}(\bm{x}),\rho(\bm{y})}(\omega)
    =
    2 \pi(\bm{x}) \textsf D(\bm{x})\nabla_{\bm{x}}H(\bm{y},\bm{x};-\omega).
    \label{eq:app_langevin_zeta_rho}
\end{equation}
These results are consistent with the stationarity relation $[\textsf{C}_{\bm{\zeta}(\bm{x}),\rho(\bm{y})}(\omega)]^{\mathsf T}
= \textsf{C}_{\rho(\bm{y}),\bm{\zeta}(\bm{x})^{\mathsf T}}(-\omega)$.

The derivation of Eq.~\eqref{eq:app_langevin_rho_zeta} follows the approach used in Refs.~\cite{Kwon2025_unified,Chun2026fluctuation}. For completeness, we provide the derivation here. First, for $t>t'$, note that
\begin{equation}
    \begin{aligned}
        &\langle \rho (\bm{x},t)\zeta_i (\bm{y},t')\rangle 
        \\ &= \sum_j \langle \delta \bigl(\bm{x}-\bm{x}(t) \bigr)\delta \bigl(\bm{y}-\bm{x}(t') \bigr)B_{ij} \bigl(\bm{x}(t') \bigr) \xi_j (t')\rangle
        \\ &=\sum_j \int d\boldsymbol\xi p_{\boldsymbol\xi} (\boldsymbol\xi)P(\bm{x},t|\bm{y}+d\bm{y},t'+dt') B_{ij}(\bm y)\xi_j \pi (\bm{y})\;.
    \end{aligned}
\end{equation}
In the second equality, we condition on the realization of the noise $\boldsymbol\xi(t') = \boldsymbol\xi$ at time $t'$, whose probability density is denoted by $p_{\boldsymbol\xi}(\boldsymbol\xi)$. Since $\bm{x}(t') = \bm{y}$, the conditioned trajectory satisfies $\bm{x}(t'+dt') = \bm{y}+d\bm{y}$, where
\begin{equation}
    d\bm{y} = \bm{v}(\bm{y}) dt' + \textsf{B}(\bm{y})\bullet \boldsymbol\xi dt'
    \;.
\end{equation}
Here, $\bm{v}(\bm{y})$ denotes the drift term obtained by
converting the anti-It\^o Langevin equation to the It\^o convention.
Expanding $P(\bm{x},t|\bm{y}+d\bm{y},t'+dt')$ yields
\begin{equation}
    \begin{aligned}
        &P(\bm{x},t|\bm{y}+d\bm{y},t'+dt') 
        \\ &= P(\bm{x},t|\bm{y},t') +dt'\partial_{t'}P(\bm{x},t|\bm{y},t')  +[\boldsymbol\nabla^\textsf{T}_{\bm{y}} P(\bm{x},t|\bm{y},t')]d\bm{y}.
    \end{aligned}
\end{equation}
Since $p_{\boldsymbol\xi}(\boldsymbol\xi)$ is a Gaussian with covariance $\textsf{I}/dt'$, we obtain
\begin{equation}
    \langle \rho (\bm{x},t)\zeta_i (\bm{y},t')\rangle = 2\pi (\bm{y})[\textsf{D}(\bm{y}) \boldsymbol\nabla_{\bm{y}} P(\bm{x},t|\bm{y},t')]_i
    \;.
\end{equation}
For $t<t'$, the correlation vanishes by causality.
Equation~\eqref{eq:app_langevin_rho_zeta} then follows by taking the Fourier transform.

Using the decomposition \eqref{eq:app_langevin_current_decomp}, the density--current covariance is given by
\begin{equation}
    \textsf C_{\rho(\bm{x}),\bm{\jmath}(\bm{y})^{\mathsf T}}(\omega)
    =
    [\hat{\bm{J}}_{\bm{y}} C_{\rho(\bm{x}),\rho(\bm{y})}(\omega)]^\mathsf{T}
    +
    \textsf C_{\rho(\bm{x}),\bm{\zeta}(\bm{y})^{\mathsf T}}(\omega), \label{eq:app_langevin_density_current_covariance}
\end{equation}
while the current--current covariance is given by
\begin{equation}
    \begin{aligned}
        \textsf C_{\bm{\jmath}(\bm{x}),\bm{\jmath}(\bm{y})^{\mathsf T}}(\omega)
        =
        {}&
        \hat{\bm J}_{\bm{x}} [\hat{\bm J}_{\bm{y}}
        C_{\rho(\bm{x}),\rho(\bm{y})}(\omega)]^\mathsf{T}
        +
        \hat{\bm J}_{\bm{x}}
        \textsf C_{\rho(\bm{x}),\bm{\zeta}(\bm{y})^{\mathsf T}}(\omega)
        \\
        &+
        [\hat{\bm J}_{\bm{y}} \textsf C_{\rho(\bm{y}),\bm{\zeta}(\bm{x})^{\textsf T}}(-\omega)]^\mathsf{T}
        +
        \textsf C_{\bm{\zeta}(\bm{x}),\bm{\zeta}(\bm{y})^{\mathsf T}}(\omega).
    \end{aligned} \label{eq:app_langevin_current_covariance}
\end{equation}
Substituting Eqs.~\eqref{eq:app_langevin_rho_rho_raw} and \eqref{eq:app_langevin_rho_zeta} into Eq.~\eqref{eq:app_langevin_density_current_covariance}, together with the quadratic representation~\eqref{eq:app_langevin_rho_rho_quadratic}, yields
\begin{align}
    &\textsf C_{\rho(\bm{x}),\bm{\jmath}(\bm{y})^{\mathsf T}}(\omega) \nonumber \\ 
    &=
    2\int d\bm{z}\,
    \pi(\bm{z})\,
    \nabla_{\bm{z}}^{\mathsf T}H(\bm{x},\bm{z};\omega)
    \textsf D(\bm{z})\,
    \bigl[ \textsf{P}(\bm{y},\bm{z};\omega) \bigr]^\dagger,
    \label{eq:app_langevin_rho_j_compact}
\end{align}
Similarly, substituting Eqs.~\eqref{eq:app_langevin_rho_rho_quadratic},
\eqref{eq:app_langevin_zeta_zeta}, and
\eqref{eq:app_langevin_rho_zeta} into Eq.~\eqref{eq:app_langevin_current_covariance} yields
\begin{equation}
    \textsf C_{\bm{\jmath}(\bm{x}),\bm{\jmath}(\bm{y})^{\mathsf T}}(\omega)
    =
    2\int d\bm{z}\,
    \pi(\bm{z})\,
    \textsf P(\bm{x},\bm{z};\omega)
    \textsf D(\bm{z})\,
    \bigl[ \textsf P(\bm{y},\bm{z};\omega) \bigr]^\dagger.
    \label{eq:app_langevin_j_j_compact}
\end{equation}

\subsection{Derivation of the quadratic representation of the density--density covariance, Eq.~\eqref{eq:app_langevin_rho_rho_quadratic}}
\label{appsubsec:langevin_quadratic_identity}

In this subsection, we derive the quadratic representation~\eqref{eq:app_langevin_rho_rho_quadratic} for the density--density covariance. To this end, we first rewrite $H(\bm{x},\bm{y};\omega)\pi(\bm{y})$ in terms of a delta function so that the resolvent identity can be applied:
\begin{equation}\label{eq:Hpi_delta_expr}
    H(\bm{x},\bm{y};\omega)\pi(\bm{y})
    =
    \int d\bm{z}\,
    \pi(\bm{z})H(\bm{x},\bm{z};\omega)\delta(\bm{y}-\bm{z}).
\end{equation}
Using the identity~\eqref{eq:app_langevin_H_identity_3}, we rewrite the delta function as
\begin{equation}\label{eq:delta_alt_expr}
    \delta(\bm{y}-\bm{z})
    =
    \pi(\bm{y})
    +
    i\omega H(\bm{y},\bm{z};-\omega)
    -
    \hat{\mathcal L}_{\bm{z}}^\dagger H(\bm{y},\bm{z};-\omega).
\end{equation}
Substituting Eq.~\eqref{eq:delta_alt_expr} into Eq.~\eqref{eq:Hpi_delta_expr} yields
\begin{equation}
    \begin{aligned}
        H(\bm{x},\bm{y};\omega)\pi(\bm{y})
        =
        {}&
        i\omega
        \int d\bm{z}\,
        \pi(\bm{z})H(\bm{x},\bm{z};\omega)H(\bm{y},\bm{z};-\omega)
        \\
        &-
        \int d\bm{z}\,
        \pi(\bm{z})H(\bm{x},\bm{z};\omega)
        \hat{\mathcal L}_{\bm{z}}^\dagger H(\bm{y},\bm{z};-\omega),
    \end{aligned}
\end{equation}
where the term proportional to $\pi(\bm{y})$ vanishes due to the identity~\eqref{eq:app_langevin_H_identity_1}.
Using $\hat{\mathcal L}_{\bm{z}} = -\nabla_{\bm{z}}^{\mathsf T}\hat{\bm J}_{\bm{z}}$
and integration by parts yields
\begin{equation}
    \begin{aligned}
        &H(\bm{x},\bm{y};\omega)\pi(\bm{y})
        \\ &=
        i\omega
        \int d\bm{z}\,
        \pi(\bm{z})H(\bm{x},\bm{z};\omega)H(\bm{y},\bm{z};-\omega)
        \\
        &\quad+
        \int d\bm{z}\, \pi(\bm{z})
        \nabla_{\bm{z}}^{\mathsf T} H(\bm{x},\bm{z};\omega)
        \mathsf{D}(\bm{z})
        \nabla_{\bm{z}}H(\bm{y},\bm{z};-\omega)\,
        \\
        & \quad-
        \int d\bm{z}\,
        H(\bm{x},\bm{z};\omega)\,
        \bigl[ \bm{j}_{\mathrm{ss}}(\bm{z}) \bigr]^{\mathsf T}
        \nabla_{\bm{z}}H(\bm{y},\bm{z};-\omega).
    \end{aligned}
    \label{eq:app_langevin_first_half}
\end{equation}
Similarly, $H(\bm{y},\bm{x};-\omega)\pi(\bm{x})$ is obtained by exchanging $\bm{x}\leftrightarrow\bm{y}$ and $\omega\leftrightarrow-\omega$.
Adding the two expressions, the first terms proportional to $i\omega$ cancel. The current-dependent terms (the third terms) combine into
\begin{equation}
    -\int d\bm{z}\,
    \bigl[ \bm{j}_{\mathrm{ss}}(\bm{z}) \bigr]^{\mathsf T}
    \nabla_{\bm{z}}
    \Bigl[
        H(\bm{x},\bm{z};\omega)H(\bm{y},\bm{z};-\omega)
    \Bigr],
\end{equation}
which vanishes since $\nabla_{\bm{z}}^{\mathsf T}\bm{j}_{\mathrm{ss}}(\bm{z})=0$
and the boundary term is assumed to vanish. The remaining diffusion terms are identical and therefore combine to give twice the second term in Eq.~\eqref{eq:app_langevin_first_half}, which proves Eq.~\eqref{eq:app_langevin_rho_rho_quadratic}.

\subsection{\texorpdfstring{Explicit forms of the perturbation weight matrices $\textsf A^{\bm \phi}(\bm{z})$}{Explicit forms of the perturbation weight matrices A} }
\label{appsubsec:langevin_Aphi}

In this subsection, we present the explicit forms of the perturbation weight matrices $\textsf A^{\bm \phi}(\bm{z})$ that appear in the frequency-resolved FRR. We first recall their definition:
\begin{equation}
    \textsf A^{\bm \phi}(\bm{z})
    =
    \frac{1}{2\pi(\bm{z})}
    \bigl[\textsf N^{\bm \phi}(\bm{z})\bigr]^{\mathsf T}
    \textsf D(\bm{z})^{-1}
    \textsf N^{\bm \phi}(\bm{z}).
\end{equation}
For perturbations $\bm \phi \in \{\bm{F}, \ln \textsf M, \textsf T\}$, 
using \eqref{eq:N_F}--\eqref{eq:N_T},  we obtain the following explicit forms of $\textsf A^{\bm \phi}(\bm{z})$:
\begin{align}
    &[\textsf A^{\bm F}(\bm{z})]_{kl}
    =
    \delta_{kl}\,
    \frac{\pi(\bm{z})\mu_k(\bm{z})}{2T_k(\bm{z})},
    \label{eq:app_langevin_A_F} \\
    &[\textsf A^{\ln \textsf M}(\bm{z})]_{kl}
    =
    \delta_{kl}\,
    \frac{\bigl[j_k^{\mathrm{ss}}(\bm{z})\bigr]^2}{2\pi(\bm{z})D_k(\bm{z})},
    \label{eq:app_langevin_A_lnM} \\
    &[\textsf A^{\textsf T}(\bm{z})]_{kl} =
    \delta_{kl}\,
    \frac{\mu_k(\bm{z})\bigl[\partial_{z_k}\pi(\bm{z})\bigr]^2}{2\pi(\bm{z})T_k(\bm{z})}.
    \label{eq:app_langevin_A_T}
\end{align}

\section{Technical details for Markov jump systems}
\label{app:jump}

\subsection{Perturbed master equation and local response formulas}
\label{appsubsec:jump_response}

In this subsection, we derive the local response formulas presented in Sec.~\ref{sec:jump_frt}. The derivation closely parallels that in Appendix~\ref{appsubsec:langevin_response}. The unperturbed dynamics is governed by
\begin{equation}
    \partial_t \bm{p}(t)=\textsf W \bm{p}(t),
\end{equation}
with stationary distribution $\bm{\pi}$ satisfying $\textsf W \bm{\pi}=0$. We consider a local perturbation $\phi_{kl}\mapsto \phi_{kl}+\epsilon\delta(t)$ applied at time $t=0$ to the edge $k\leftrightarrow l$ (with $k>l$).
The perturbed dynamics is then governed by
\begin{equation}
    \partial_t \bm{p}(t)
    =
    \bigl[
        \textsf W+\epsilon\, \delta(t)\textsf K_{\phi_{kl}}
    \bigr]\bm{p}(t),
\end{equation}
where $\phi\in\{F,B\}$ specifies the type of perturbation.

An entropic perturbation, $F_{kl}\mapsto F_{kl}+\epsilon\delta(t)$, modifies the antisymmetric part of the transition rates according to
\begin{equation}
    \delta W_{kl}
    =
    \frac{\epsilon}{2}W_{kl}\delta(t),
    \qquad
    \delta W_{lk}
    =
    -\frac{\epsilon}{2}W_{lk}\delta(t), \label{eqA:F_perturb_tran_rate}
\end{equation}
which gives
\begin{equation}
    \textsf K_{F_{kl}}
    =
    \frac{W_{kl}}{2}\bigl(\textsf E^{kl}- \textsf E^{ll}\bigr)
    -
    \frac{W_{lk}}{2}\bigl(\textsf E^{lk}- \textsf E^{kk}\bigr).
\end{equation}
Here, $\textsf E^{nm}$ denotes the matrix whose only nonzero entry is $1$ in row $n$ and column $m$.

Likewise, a kinetic perturbation, $B_{kl}\mapsto B_{kl}+\epsilon\delta(t)$, modifies the symmetric part of the transition rates according to
\begin{equation}
    \delta W_{kl}
    =
    \epsilon W_{kl}\delta(t),
    \qquad
    \delta W_{lk}
    =
    \epsilon W_{lk}\delta(t),
    \label{eqA:B_perturb_tran_rate}
\end{equation}
which gives 
\begin{equation}
    \textsf K_{B_{kl}}
    =
    W_{kl}\bigl(\textsf E^{kl}- \textsf E^{ll}\bigr)
    +
    W_{lk}\bigl(\textsf E^{lk}- \textsf E^{kk}\bigr).
\end{equation}

Assuming the system is initially in the steady state, we expand the probability vector as
\begin{equation}
    \bm{p}(t)
    =
    \bm{\pi}
    +
    \epsilon\,\bm{q}_{\phi_{kl}}(t)
    +
    O(\epsilon^2). \label{eqA:perturbed_p_vector}
\end{equation}
Substituting this expansion into the perturbed master equation and retaining terms up to linear order in $\epsilon$, we obtain
\begin{equation}
    \frac{d}{dt}\bm{q}_{\phi_{kl}}(t)
    =
    \textsf W \bm{q}_{\phi_{kl}}(t)
    +
    \delta(t) \textsf K_{\phi_{kl}}\bm{\pi} ,
    \qquad
    \bm{q}_{\phi_{kl}}(0^-)=\bm{0}.
\end{equation}
We define the local prefactor vector as
\begin{equation}
    \bm{N}_{\phi_{kl}}
    \equiv
    \textsf{K}_{\phi_{kl}}\bm{\pi} ,
    \label{eq:jump_Nphi_definition}
\end{equation}
which is the discrete analogue of Eq.~\eqref{eq:langevin_Nphi_definition} for Langevin systems. For $\phi \in \{ F, B\}$, the explicit expressions are 
\begin{align}
    &\bm N_{F_{kl}} =
    \textsf K_{F_{kl}}\bm{\pi} =
    \frac{a_{kl}}{2}\,(\bm e_k-\bm e_l), \\
    &\bm N_{B_{kl}} =
    \textsf K_{B_{kl}}\bm{\pi} =
    j_{kl}\,(\bm e_k-\bm e_l).
\end{align}
The linearized equation then takes the compact form
\begin{equation}
    \frac{d}{dt}\bm{q}_{\phi_{kl}}(t) =
    \textsf W \bm{q}_{\phi_{kl}}(t) +
    \delta(t) \bm N_{\phi_{kl}}.
\end{equation}
Its solution is
\begin{equation}
    \bm{q}_{\phi_{kl}}(t)
    =
    e^{\textsf W t}\bm N_{\phi_{kl}}
    \qquad (t>0).
\end{equation}

Since $\langle \eta_n(t)\rangle = p_n(t)$, the response of the empirical state indicator coincides with that of $p_n(t)$ and, therefore, is given by the $n$th component of $\bm{q}_{\phi_{kl}}(t)$:
\begin{equation}
    R_{\phi_{kl}}^{\eta_n} (t,0) = \bigl( \bm{q}_{\phi_{kl}}(t) \bigr)_n = \sum_{m} \bigl[ e^{\textsf{W}t} \bigr]_{nm} \bigl(\bm N_{\phi_{kl}} \bigr)_m
\end{equation}
Recalling the definition
\begin{equation}\label{eq:discrete_excess_propagator}
    H_{nm}(\omega)
    =
    \int_0^\infty
    \bigl(
        [e^{\mathsf{W} t}]_{nm}-\pi_n
    \bigr)e^{i\omega t}\,dt,
\end{equation}
and using the relation $\sum_m(\bm{N}_{\phi_{kl}})_m=0$ which follows from probability conservation, we obtain
\begin{equation}
    R^{\eta_n}_{\phi_{kl}}(\omega)
    =
    \sum_m H_{nm}(\omega)(\bm{N}_{\phi_{kl}})_m.
\end{equation}
The explicit expressions for $\phi \in \{ F, B\}$ are
\begin{align}
    &R^{\eta_n}_{F_{kl}}(\omega)
    =
    \frac{a_{kl}}{2}
    \bigl[
        H_{nk}(\omega)-H_{nl}(\omega)
    \bigr],
    \label{eq:app_jump_eta_response_F} \\
    &R^{\eta_n}_{B_{kl}}(\omega)
    =
    j_{kl}
    \bigl[
        H_{nk}(\omega)-H_{nl}(\omega)
    \bigr].
    \label{eq:app_jump_eta_response_B}
\end{align}
The two response functions differ only in their perturbation-dependent prefactors and therefore satisfy
\begin{equation}
    \frac{2}{a_{kl}}R^{\eta_n}_{F_{kl}}(\omega)
    =
    \frac{1}{j_{kl}}R^{\eta_n}_{B_{kl}}(\omega)
    =
    H_{nk}(\omega)-H_{nl}(\omega).
    \label{eq:app_jump_eta_response_prop}
\end{equation}

We next derive the response of the empirical current,
\begin{equation}
    \jmath_{nm}(t)=\dot N_{nm}(t)-\dot N_{mn}(t).
\end{equation}
For an observed edge different from the perturbed edge, the expectation value of the empirical current in the perturbed dynamics is
\begin{equation}
    \langle \jmath_{nm}(t)\rangle
    =
    W_{nm}p_m(t)-W_{mn}p_n(t),
\end{equation}
where $\bm{p}(t)$ is given in Eq.~\eqref{eqA:perturbed_p_vector}. Hence, the corresponding response is entirely determined by the response of $\bm{q}_{\phi_{kl}}(t)$.
If the observed edge coincides with the perturbed one, the transition rates themselves are also perturbed according to Eqs.~\eqref{eqA:F_perturb_tran_rate} and \eqref{eqA:B_perturb_tran_rate}. The expectation value of the empirical current in the perturbed dynamics is therefore
\begin{equation}
    \langle \jmath_{nm}(t)\rangle
    =
    (W_{nm}+\delta W_{nm})p_m(t)
    -
    (W_{mn}+\delta W_{mn})p_n(t).
\end{equation}
Therefore, combining the contributions from the occupation dynamics $\bm q_{\phi_{kl}}(t)$ and the direct perturbation of the transition rates $\delta W_{kl}$ and $\delta W_{lk}$, we obtain
\begin{equation}\label{eq:emp_curr_resp}
    R^{\jmath_{nm}}_{\phi_{kl}}(\omega)
    =
    \Gamma^\phi_{nm;kl}
    +
    W_{nm}R^{\eta_m}_{\phi_{kl}}(\omega)
    -
    W_{mn}R^{\eta_n}_{\phi_{kl}}(\omega),
\end{equation}
where
\begin{align}
    &\Gamma^F_{nm;kl} =
    \frac{a_{kl}}{2}
    \bigl(
        \delta_{nk}\delta_{ml}
        -
        \delta_{nl}\delta_{mk}
    \bigr), \\
    &\Gamma^B_{nm;kl} =
    j_{kl}
    \bigl(
        \delta_{nk}\delta_{ml}
        -
        \delta_{nl}\delta_{mk}
    \bigr).
\end{align}
Substituting Eqs.~\eqref{eq:app_jump_eta_response_F} and \eqref{eq:app_jump_eta_response_B} into Eq.~\eqref{eq:emp_curr_resp}, we obtain
\begin{widetext}
\begin{align}
        &R^{\jmath_{nm}}_{F_{kl}}(\omega)
        =
        \frac{a_{kl}}{2}
        \Bigl[
            \delta_{nk}\delta_{ml}
            -
            \delta_{nl}\delta_{mk}
            +
            W_{nm}\bigl\{H_{mk}(\omega)-H_{ml}(\omega)\bigr\}
            -
            W_{mn}\bigl\{H_{nk}(\omega)-H_{nl}(\omega)\bigr\}
        \Bigr],
    \label{eq:app_jump_j_response_F} \\
        &R^{\jmath_{nm}}_{B_{kl}}(\omega)
        =
        j_{kl}
        \Bigl[
            \delta_{nk}\delta_{ml}
            -
            \delta_{nl}\delta_{mk}
            +
            W_{nm}\bigl\{H_{mk}(\omega)-H_{ml}(\omega)\bigr\}
            -
            W_{mn}\bigl\{H_{nk}(\omega)-H_{nl}(\omega)\bigr\}
        \Bigr].
    \label{eq:app_jump_j_response_B}
\end{align}
\end{widetext}
Hence, the two current response functions differ only in their perturbation-dependent prefactors:
\begin{equation}
    \frac{2}{a_{kl}}R^{\jmath_{nm}}_{F_{kl}}(\omega)
    =
    \frac{1}{j_{kl}}R^{\jmath_{nm}}_{B_{kl}}(\omega).
\end{equation}

\subsection{Properties of the discrete excess propagator}
\label{appsubsec:jump_H}

We now derive several identities, Eqs.~\eqref{eq:jump_H_identity_1}--\eqref{eq:jump_H_identity_3}, satisfied by the discrete excess propagator defined in Eq.~\eqref{eq:discrete_excess_propagator}. We first prove
\begin{equation}
    \sum_m H_{nm}(\omega)\pi_m=0.
    \label{eq:app_jump_H_identity_1}
\end{equation}
Multiplying Eq.~\eqref{eq:discrete_excess_propagator} by $\pi_m$ and summing over $m$, we obtain
\begin{equation}
    \sum_m H_{nm}(\omega)\pi_m
    =
    \int_0^\infty dt\,e^{i\omega t}
    \sum_m
    \bigl(
        [e^{\textsf W t}]_{nm}-\pi_n
    \bigr)\pi_m
    =0.
\end{equation}
The last equality follows from the stationarity of the steady-state distribution,
\begin{equation}
    \sum_m[e^{\textsf W t}]_{nm}\pi_m=\pi_n.
\end{equation}

Second, we prove
\begin{equation}
    \sum_m
    \bigl(
        i\omega\delta_{nm}+[\textsf W]_{nm}
    \bigr)H_{mk}(\omega)
    =
    -\delta_{nk}+\pi_n.
    \label{eq:app_jump_H_identity_2}
\end{equation}
Using
\begin{equation}
    \frac{d}{dt}[e^{\textsf W t}]_{nk}
    =
    \sum_m [\textsf W]_{nm}[e^{\textsf W t}]_{mk},
\end{equation}
we obtain
\begin{equation}
    \begin{aligned}
        &\sum_m
        \bigl(
            i\omega\delta_{nm}+ [\textsf W]_{nm}
        \bigr)H_{mk}(\omega)
        \\ &=\int_0^\infty dt\,e^{i\omega t}
        \sum_m
        \Bigl[
            i\omega\delta_{nm}\bigl([e^{\textsf W t}]_{mk}-\pi_m\bigr)
            +
            [\textsf W]_{nm}[e^{\textsf W t}]_{mk}
        \Bigr] \\
        &=
        \int_0^\infty dt\,
        \partial_t
        \Bigl(
            e^{i\omega t}\bigl([e^{\textsf W t}]_{nk}-\pi_n\bigr)
        \Bigr) \\
        &=
        -\delta_{nk}+\pi_n.
    \end{aligned}
\end{equation}
In the last step, we used
\begin{equation}
    \lim_{t\to 0^+}[e^{\textsf W t}]_{nk}=\delta_{nk},
    \qquad
    \lim_{t\to\infty}[e^{\textsf W t}]_{nk}=\pi_n.
\end{equation}

Finally, using
\begin{equation}
    \frac{d}{dt}[e^{\textsf W t}]_{nk}
    =
    \sum_m [e^{\textsf W t}]_{nm} [\textsf W]_{mk},
\end{equation}
we similarly obtain
\begin{equation}
    \sum_m
    H_{nm}(\omega)
    \bigl(
        i\omega\delta_{mk}+ [\textsf W]_{mk}
    \bigr)
    =
    -\delta_{nk}+\pi_n.
    \label{eq:app_jump_H_identity_3}
\end{equation}

\subsection{Covariances of empirical state indicators and edge currents}
\label{appsubsec:jump_covariances}

We now derive the covariance formulas for the empirical state indicators and empirical edge currents. We begin with the empirical state indicator,
\begin{equation}
    \eta_n(t)=\delta_{X(t),n}.
\end{equation}
For $t>0$,
\begin{equation}
    \langle \eta_n(t)\eta_m(0)\rangle-\pi_n\pi_m
    =
    \bigl(
        [e^{\textsf W t}]_{nm}-\pi_n
    \bigr)\pi_m,
\end{equation}
whereas for $t<0$, time-translation invariance gives
\begin{equation}
    \langle \eta_n(t)\eta_m(0)\rangle-\pi_n\pi_m
    =
    \bigl(
        [e^{\textsf W(-t)}]_{mn}-\pi_m
    \bigr)\pi_n.
\end{equation}
Substituting these expressions into the definition of the covariance spectrum~\eqref{eq:covariance_definition_main},  we obtain
\begin{equation}
    C_{\eta_n,\eta_m}(\omega)
    =
    H_{nm}(\omega)\pi_m
    +
    H_{mn}(-\omega)\pi_n.
    \label{eq:app_jump_eta_eta_raw}
\end{equation}
As shown in Appendix~\ref{appsubsec:jump_quadratic_identity}, this can be equivalently written in the quadratic form
\begin{equation}
    \begin{aligned}
        & C_{\eta_n,\eta_m}(\omega)
        \\ &=
        \sum_{k>l}
        a_{kl}
        \bigl[
            H_{nk}(\omega)-H_{nl}(\omega)
        \bigr]
        \bigl[
            H_{mk}(-\omega)-H_{ml}(-\omega)
        \bigr].
    \end{aligned}
    \label{eq:app_jump_eta_eta_quadratic}
\end{equation}

Next, we decompose the empirical current~\eqref{eq:jump_empirical_current} into drift and fluctuating parts:
\begin{equation}
    \jmath_{nm}(t)
    =
    W_{nm}\eta_m(t)-W_{mn}\eta_n(t)+\zeta_{nm}(t),
    \label{eq:app_jump_current_decomp}
\end{equation}
where the fluctuation term is defined by
\begin{equation}
    \zeta_{nm}(t)
    \equiv \dot{\tilde N}_{nm}(t) - \dot{\tilde N}_{mn}(t) 
    \;,
\end{equation}
with $\dot{\tilde N}_{nm}(t) \equiv \dot{N}_{nm}(t) - W_{nm} \eta_m(t)$. We note that $\langle \dot{\tilde N}_{nm}(t) \rangle =0$, and thus $\langle \zeta_{nm}(t) \rangle =0$. Using the identity $\langle \dot{\tilde N}_{nm}(t) \dot{\tilde N}_{n^\prime m^\prime}(t^\prime) \rangle = W_{nm}\pi_m \delta_{nn^\prime} \delta_{mm^\prime} \delta(t-t^\prime)$~\cite{Kwon2024unified,Kwon2025_unified}, a straightforward calculation yields 
\begin{equation}
    \langle \zeta_{nm}(t)\zeta_{kl}(0)\rangle
    =
    a_{nm}
    \bigl(
        \delta_{nk}\delta_{ml}
        -
        \delta_{nl}\delta_{mk}
    \bigr)\delta(t),
\end{equation} 
and therefore
\begin{equation}
    C_{\zeta_{nm},\zeta_{kl}}(\omega)
    =
    a_{nm}
    \bigl(
        \delta_{nk}\delta_{ml}
        -
        \delta_{nl}\delta_{mk}
    \bigr).
    \label{eq:app_jump_zeta_zeta}
\end{equation}

We next compute the covariance between the state indicator and the fluctuating part of the empirical current. For $t>0$, using $\langle \eta_n(t) \dot N_{kl}(0) \rangle = \bigl[ e^{\textsf W t} \bigr]_{nk} W_{kl}\pi_l$, we obtain 
\begin{equation}
    \langle \eta_n(t)\zeta_{kl}(0)\rangle
    = 
    a_{kl}
    \bigl(
        [e^{\textsf W t}]_{nk}-[e^{\textsf W t}]_{nl}
    \bigr).
\end{equation}
For $t<0$, $\langle \eta_n(t)\zeta_{kl}(0)\rangle = 0$. Taking the Fourier transform, we find
\begin{equation}
    C_{\eta_n,\zeta_{kl}}(\omega)
    =
    a_{kl}
    \bigl[
        H_{nk}(\omega)-H_{nl}(\omega)
    \bigr].
    \label{eq:app_jump_eta_zeta}
\end{equation}
Likewise, exchanging the order of the variables and applying $t\to -t$ ($\omega\to -\omega$ in the Fourier domain), we obtain
\begin{equation}
    C_{\zeta_{nm},\eta_k}(\omega)
    =
    a_{nm}
    \bigl[
        H_{kn}(-\omega)-H_{km}(-\omega)
    \bigr].
    \label{eq:app_jump_zeta_eta}
\end{equation}

From Eq.~\eqref{eq:app_jump_current_decomp}, the state--current covariance reads
\begin{equation}
    C_{\eta_n,\jmath_{kl}}(\omega)
    =
    W_{kl}C_{\eta_n,\eta_l}(\omega)
    -
    W_{lk}C_{\eta_n,\eta_k}(\omega)
    +
    C_{\eta_n,\zeta_{kl}}(\omega).
    \label{eq:app_jump_eta_j_raw1}
\end{equation}
To express the result in an edge-resolved form, we rewrite 
Eqs.~\eqref{eq:app_jump_eta_zeta},
\eqref{eq:app_jump_zeta_zeta}, and
\eqref{eq:app_jump_zeta_eta} as
\begin{align}
    &C_{\eta_n, \zeta_{kl}} (\omega) \nonumber \\ 
    &= \sum_{k'>l'} a_{k'l'}[\delta_{kk'}\delta_{ll'} - \delta_{kl'}\delta_{lk'}][H_{nk'}(\omega) - H_{nl'}(\omega)]
    \;, \label{eq:app_jump_eta_zeta_re} \\
    &C_{\zeta_{nm}, \zeta_{kl}} (\omega) \nonumber \\
    &= \sum_{k'>l'} a_{k'l'}[\delta_{nk'}\delta_{ml'} - \delta_{nl'}\delta_{mk'}] [\delta_{kk'}\delta_{ll'} - \delta_{kl'}\delta_{lk'}], \label{eq:app_jump_zeta_zeta_re} \\
    &C_{\zeta_{nm}, \eta_k} (\omega) \nonumber \\ &= \sum_{k'>l'} a_{k'l'}[\delta_{nk'}\delta_{ml'} - \delta_{nl'}\delta_{mk'}][H_{kk'}(-\omega) - H_{kl'}(-\omega)]
    \;. \label{eq:app_jump_zeta_eta_re} 
\end{align}
Substituting Eqs.~\eqref{eq:app_jump_eta_eta_quadratic} and \eqref{eq:app_jump_eta_zeta_re} into Eq.~\eqref{eq:app_jump_eta_j_raw1}, we obtain
\begin{widetext}
\begin{equation}
    \begin{aligned}
        C_{\eta_n,\jmath_{kl}}(\omega) &= \sum_{k'>l'} a_{k'l'}[H_{nk'}(\omega)-H_{n l'} (\omega)] \\ 
        &\quad \quad \quad \times         \bigl[ \delta_{kk'}\delta_{ll'} - \delta_{kl'}\delta_{lk'}+W_{kl} \bigl\{ H_{lk'}(-\omega) - H_{ll'}(-\omega) \bigr\} - W_{lk} \bigl\{ H_{kk'}(-\omega)-H_{kl'}(-\omega)\bigr\} \bigr]
    \end{aligned} \label{eq:app_jump_eta_j_quadratic}
\end{equation}
Together with Eqs.~\eqref{eq:jump_eta_response_F}--\eqref{eq:jump_current_response_proportionality}, Eq.~\eqref{eq:app_jump_eta_j_quadratic} can be written in the FRR form~\eqref{eq:jump_local_ffrr_eta_j}. 

Similarly, the current--current covariance reads
\begin{equation}
    \begin{aligned}
        C_{\jmath_{nm},\jmath_{kl}}(\omega)
        =
        {}&
        W_{nm}W_{kl}C_{\eta_m,\eta_l}(\omega)
        -
        W_{nm}W_{lk}C_{\eta_m,\eta_k}(\omega)
        -
        W_{mn}W_{kl}C_{\eta_n,\eta_l}(\omega)
        +
        W_{mn}W_{lk}C_{\eta_n,\eta_k}(\omega)
        \\
        &+
        W_{nm}C_{\eta_m,\zeta_{kl}}(\omega)
        -
        W_{mn}C_{\eta_n,\zeta_{kl}}(\omega)
        +
        W_{kl}C_{\zeta_{nm},\eta_l}(\omega)
        -
        W_{lk}C_{\zeta_{nm},\eta_k}(\omega)
        +
        C_{\zeta_{nm},\zeta_{kl}}(\omega).
    \end{aligned}
    \label{eq:app_jump_j_j_raw}
\end{equation}
Substituting Eqs.~\eqref{eq:app_jump_eta_eta_quadratic}, \eqref{eq:app_jump_eta_zeta_re}, \eqref{eq:app_jump_zeta_eta_re}, and \eqref{eq:app_jump_zeta_zeta_re} into Eq.~\eqref{eq:app_jump_j_j_raw}, we obtain
\begin{equation}
    \begin{aligned}
        C_{\jmath_{nm},\jmath_{kl}}(\omega) &= \sum_{k'>l'} a_{k'l'} \{\delta_{nk'}\delta_{ml'} - \delta_{nl'}\delta_{mk'}+W_{nm}[H_{mk'}(\omega) - H_{ml'}(\omega)] - W_{mn}[H_{nk'}(\omega)-H_{nl'}(\omega)]\}
        \\ & \quad \quad \quad \times \{\delta_{kk'}\delta_{ll'} - \delta_{kl'}\delta_{lk'}+W_{kl}[H_{lk'}(-\omega) - H_{ll'}(-\omega)] - W_{lk}[H_{kk'}(-\omega)-H_{kl'}(-\omega)]\}
        \;.
    \end{aligned} \label{eq:app_jump_j_j_quadratic}
\end{equation}
Together with Eqs.~\eqref{eq:jump_eta_response_F}--\eqref{eq:jump_current_response_proportionality}, Eq.~\eqref{eq:app_jump_j_j_quadratic} can be written in the FRR form~\eqref{eq:jump_local_ffrr_j_j}. 

\end{widetext}

\subsection{Derivation of the quadratic representation of the state--state covariance, Eq.~\eqref{eq:app_jump_eta_eta_quadratic} }
\label{appsubsec:jump_quadratic_identity}

In this subsection, we prove the quadratic representation \eqref{eq:app_jump_eta_eta_quadratic}. To this end, we first rewrite $H_{nm}(\omega)\pi_m$ in terms of a Kronecker-delta function as
\begin{equation}\label{eq:Hpi_delta_expr_jump}
    H_{nm}(\omega)\pi_m 
    = \sum_k H_{nk}(\omega)\pi_k\delta_{km}.
\end{equation}
Using the identity \eqref{eq:app_jump_H_identity_3} at frequency $-\omega$, we rewrite the Kronecker-delta as
\begin{equation}\label{eq:kronecker_alt_expr}
    \delta_{km}
    =
    -
    \sum_l
    \bigl(
        -i\omega\delta_{lk}+ [\mathsf{W}]_{lk} 
    \bigr)H_{ml}(-\omega)
    +
    \pi_m.
\end{equation}
Substituting \eqref{eq:kronecker_alt_expr} into \eqref{eq:Hpi_delta_expr_jump} and using \eqref{eq:app_jump_H_identity_1}, we have
\begin{equation}
    \begin{aligned}
        H_{nm}(\omega)\pi_m
        &=
        i\omega\sum_k H_{nk}(\omega)H_{mk}(-\omega)\pi_k
        \\ & \quad-
        \sum_{k,l}H_{nk}(\omega) [\mathsf{W}]_{lk}
        \pi_k H_{ml}(-\omega).
    \end{aligned}
\end{equation}
Separating the $l=k$ term and using
\begin{equation}
    [\mathsf{W}]_{kk} 
    =-\sum_{l(\neq k)}W_{lk},
\end{equation}
we find
\begin{equation}
    \begin{aligned}
        &H_{nm}(\omega)\pi_m
        =
        i\omega\sum_k H_{nk}(\omega)H_{mk}(-\omega)\pi_k
        \\ &+
        \sum_{k\neq l}
        H_{nk}(\omega)W_{lk}\pi_k
        \bigl[
            H_{mk}(-\omega)-H_{ml}(-\omega)
        \bigr].
    \end{aligned}
    \label{eq:app_jump_first_half}
\end{equation}

Similarly, $H_{mn}(-\omega)\pi_n$ is obtained by exchanging $n\leftrightarrow m$ and $\omega\leftrightarrow-\omega$:
\begin{equation}
    \begin{aligned}
        &H_{mn}(-\omega)\pi_n
        =
        -i\omega\sum_k H_{nk}(\omega)H_{mk}(-\omega)\pi_k
        \\ &+
        \sum_{k\neq l}
        H_{mk}(-\omega)W_{lk}\pi_k
        \bigl[
            H_{nk}(\omega)-H_{nl}(\omega)
        \bigr].
    \end{aligned}
    \label{eq:app_jump_first_half2}
\end{equation}
The state--state covariance is given by the sum of the two expressions \eqref{eq:app_jump_first_half} and \eqref{eq:app_jump_first_half2}, where the terms proportional to $i\omega$ cancel.
\begin{widetext}
The resulting expression can be rearranged as
\begin{equation}
    \begin{aligned}\label{eq:state-state_cov_intermediate}
        C_{\eta_n,\eta_m}(\omega)
        =
        {}&
        \sum_{k\neq l}
        W_{lk}\pi_k
        \Bigl[
            2H_{nk}(\omega)H_{mk}(-\omega)
            -
            H_{nk}(\omega)H_{ml}(-\omega)
            -
            H_{nl}(\omega)H_{mk}(-\omega)
        \Bigr].
    \end{aligned}
\end{equation}
\end{widetext}

The stationary condition
\begin{equation}
    \sum_{l(\neq k)}W_{lk}\pi_k
    =
    \sum_{l(\neq k)}W_{kl}\pi_l
\end{equation}
implies
\begin{equation}
    \sum_{k\neq l}
    W_{lk}\pi_k H_{nk}(\omega)H_{mk}(-\omega)
    =
    \sum_{k\neq l}
    W_{lk}\pi_k H_{nl}(\omega)H_{ml}(-\omega),
\end{equation}
under relabelling $k\leftrightarrow l$.
Accordingly, \eqref{eq:state-state_cov_intermediate} reduces to
\begin{equation}
    \begin{aligned}
        &C_{\eta_n,\eta_m}(\omega)
        \\ &=
        \sum_{k\neq l}
        W_{lk}\pi_k
        \bigl[
            H_{nk}(\omega)-H_{nl}(\omega)
        \bigr]
        \bigl[
            H_{mk}(-\omega)-H_{ml}(-\omega)
        \bigr].
    \end{aligned}
\end{equation}
Separating the sum into contributions with $k>l$ and $k<l$, and exchanging dummy indices in the latter, yields \eqref{eq:app_jump_eta_eta_quadratic}.

\section{Derivations of the consequences of the frequency-resolved FRR}
\label{app:consequences}

\subsection{Derivation of the frequency-resolved response uncertainty relation}
\label{appsubsec:frur_derivation}

In this subsection, we derive the frequency-resolved RUR stated in Sec.~\ref{subsec:frur}. The derivation follows directly from applying the Cauchy--Schwarz inequality to the quadratic response representation of the covariance matrix.

We begin with the overdamped Langevin case. For an arbitrary vector \(\bm{u}\in\mathbb{C}^{N_{\rm O}}\), sandwiching Eq.~\eqref{eq:langevin_general_ffrr}  between \(\bm{u}^\dagger\) and
\(\bm{u}\) gives 
\begin{equation}
    \begin{aligned}
        &\bm{u}^\dagger
        \textsf C_{\bm{\theta},\bm{\theta}^{\mathsf T}}(\omega)
        \bm{u}
        \\ &=
        \sum_{k,l=1}^{N}
        \int d\bm{x}\,
        \bm{u}^\dagger \bm R^{\bm{\theta}}_{\phi_k(\bm{x})}(\omega)
        \,
        [\textsf A^{\bm{\phi}}(\bm{x})^{-1}]_{kl}\,
        \bigl[
            \bm R^{\bm{\theta}}_{\phi_l(\bm{x})}(\omega)\bigr]^\dagger\bm{u}.
    \end{aligned}
\end{equation}
We define two functions
\begin{align}
    &f_k(\bm{x},\omega)
    \equiv
    \sum_{l=1}^{N}
    [\textsf A^{\bm{\phi}}(\bm{x})^{1/2}]_{kl}\,\psi_l(\bm{x},\omega), \\
    &g_k(\bm{x},\omega;\bm{u})
    \equiv
    \sum_{l=1}^{N}
    [\textsf A^{\bm{\phi}}(\bm{x})^{-1/2}]_{kl}\,
    \bigl[\bm R^{\bm{\theta}}_{\phi_l(\bm{x})}(\omega)\bigr]^\dagger \bm{u}.
\end{align}
These functions relate the local response to the response under the global perturbation $\bm \phi \mapsto \bm \phi+ \epsilon \bm{\psi}$ through
\begin{equation}\label{eq:fg_inner_product_langevin}
    \sum_{k=1}^{N}
    \int d\bm{x}\,
    g_k(\bm{x},\omega;\bm{u})^*f_k(\bm{x},\omega)
    =
    \bm{u}^\dagger
    \bm R^{\bm{\theta}}_{\bm \phi,\bm{\psi}}(\omega).
\end{equation}
Moreover, the covariance is written as
\begin{equation}
    \sum_{k=1}^{N}
    \int d\bm{x}\,
    |g_k(\bm{x},\omega;\bm{u})|^2
    =
    \bm{u}^\dagger
    \textsf C_{\bm{\theta},\bm{\theta}^{\mathsf T}}(\omega)
    \bm{u}.
\end{equation}
Applying the Cauchy--Schwarz inequality to Eq.~\eqref{eq:fg_inner_product_langevin} yields
\begin{equation}
    \begin{aligned}\label{eq:CS_ineq_langevin}
        &\bigl|
            \bm{u}^\dagger
            \bm R^{\bm{\theta}}_{\bm \phi,\bm{\psi}}(\omega)
        \bigr|^2
        \\ &\le
        \bigl[
            \bm{u}^\dagger
            \textsf C_{\bm{\theta},\bm{\theta}^{\mathsf T}}(\omega)
            \bm{u}
        \bigr]
        \sum_{k,l=1}^{N}
        \int d\bm{x}\,
        \psi_k(\bm{x},-\omega)\,
        [\textsf A^{\bm{\phi}}(\bm{x})]_{kl}\,
        \psi_l(\bm{x},\omega),
    \end{aligned}
\end{equation}
where we used
\begin{align}
    &\sum_{k=1}^{N}
    \int d\bm{x}\,
    |f_k(\bm{x},\omega)|^2 \nonumber \\
    &=
    \sum_{k,l=1}^{N}
    \int d\bm{x}\,
    \psi_k(\bm{x},-\omega)\,
    [\textsf A^{\bm{\phi}}(\bm{x})]_{kl}\,
    \psi_l(\bm{x},\omega).
\end{align}
Assuming that \(\textsf C_{\bm{\theta},\bm{\theta}^{\mathsf T}}(\omega)\) is invertible at the frequency $\omega$, we choose
\begin{equation}
    \bm{u}
    =
    \textsf C_{\bm{\theta},\bm{\theta}^{\mathsf T}}(\omega)^{-1} \bm R^{\bm{\theta}}_{\bm \phi,\bm{\psi}}(\omega).
\end{equation}
Substituting this choice into Eq.~\eqref{eq:CS_ineq_langevin}
yields
\begin{equation}
    \begin{aligned}
    &\bigl[
        \bm R^{\bm{\theta}}_{\bm \phi,\bm{\psi}}(\omega)
    \bigr]^\dagger
    \textsf C_{\bm{\theta},\bm{\theta}^{\mathsf T}}(\omega)^{-1}
    \bm R^{\bm{\theta}}_{\bm \phi,\bm{\psi}}(\omega)
    \\ &\le
    \sum_{k,l=1}^{N}
    \int d\bm{x}\,
    \psi_k(\bm{x},-\omega)\,
    [\textsf A^{\bm{\phi}}(\bm{x})]_{kl}\,
    \psi_l(\bm{x},\omega),
    \end{aligned}
\end{equation}
which establishes the frequency-resolved RUR~\eqref{eq:langevin_frur} for overdamped Langevin systems.

We next consider Markov jump systems. For an arbitrary vector $\bm{u} \in \mathbb{C}^{N_{\rm O}}$, sandwiching Eq.~\eqref{eq:jump_general_ffrr} between \(\bm{u}^\dagger\) and \(\bm{u}\) gives
\begin{equation}
    \bm{u}^\dagger
    \textsf C_{\bm{\theta},\bm{\theta}^{\mathsf T}}(\omega)
    \bm{u}
    =
    \sum_{n>m}
    \frac{1}{A^\phi_{nm}}
        \bm{u}^\dagger \bm R^{\bm{\theta}}_{\phi_{nm}}(\omega)
    \bigl[
        \bm R^{\bm{\theta}}_{\phi_{nm}}(\omega)\bigr]^\dagger \bm{u}.
\end{equation}
We define two functions
\begin{equation}
    \begin{aligned}
        f_{nm}(\omega)
        &=
        \sqrt{A^\phi_{nm}}\,\psi_{nm}(\omega),
        \\
        g_{nm}(\omega;\bm{u})
        &=
        \frac{1}{\sqrt{A^\phi_{nm}}}\,
        [\bm R^{\bm{\theta}}_{\phi_{nm}}(\omega)]^\dagger \bm{u} .        
    \end{aligned}
\end{equation}
These functions relate the local response to the response under the global perturbation $\bm \phi \mapsto \bm \phi+ \epsilon \bm{\psi}$ through
\begin{equation}\label{eq:fg_inner_product_jump}
    \sum_{n>m} g_{nm}(\omega;\bm{u})^*f_{nm}(\omega)
    =
    \bm{u}^\dagger
    \bm R^{\bm{\theta}}_{\bm \phi,\bm{\psi}}(\omega).
\end{equation}
Moreover, the covariance is written as
\begin{equation}
    \sum_{n>m}|g_{nm}(\omega;\bm{u})|^2
    =
    \bm{u}^\dagger
    \textsf C_{\bm{\theta},\bm{\theta}^{\mathsf T}}(\omega)
    \bm{u}.
\end{equation}
Applying the Cauchy--Schwarz inequality to Eq.~\eqref{eq:fg_inner_product_jump} yields
\begin{equation}\label{eq:CS_ineq_jump}
    \bigl|
        \bm{u}^\dagger
        \bm R^{\bm{\theta}}_{\bm \phi,\bm{\psi}}(\omega)
    \bigr|^2
    \le
    \bigl[
        \bm{u}^\dagger
        \textsf C_{\bm{\theta},\bm{\theta}^{\mathsf T}}(\omega)
        \bm{u}
    \bigr]
    \sum_{n>m}A^\phi_{nm}|\psi_{nm}(\omega)|^2,
\end{equation}
where we used
\begin{equation}
    \sum_{n>m}|f_{nm}(\omega)|^2
    =
    \sum_{n>m}A^\phi_{nm}|\psi_{nm}(\omega)|^2.
\end{equation}
Assuming that
\(\textsf C_{\bm{\theta},\bm{\theta}^{\mathsf T}}(\omega)\)
is invertible at the frequency $\omega$, we choose
\begin{equation}
    \bm{u}
    =
    \textsf C_{\bm{\theta},\bm{\theta}^{\mathsf T}}(\omega)^{-1}
    \bm R^{\bm{\theta}}_{\bm \phi,\bm{\psi}}(\omega).
\end{equation}
Substituting this choice into Eq.~\eqref{eq:CS_ineq_jump}
yields
\begin{equation}
        \bigl[
            \bm R^{\bm{\theta}}_{\bm \phi,\bm{\psi}}(\omega)
        \bigr]^\dagger
        \textsf C_{\bm{\theta},\bm{\theta}^{\mathsf T}}(\omega)^{-1}
        \bm R^{\bm{\theta}}_{\bm \phi,\bm{\psi}}(\omega)
        \le
        \sum_{n>m}A^\phi_{nm}|\psi_{nm}(\omega)|^2,
\end{equation}
which establishes the frequency-resolved RUR~\eqref{eq:jump_frur} for Markov jump systems.

\subsection{Derivation of the frequency-resolved thermodynamic uncertainty relations}
\label{appsubsec:ftur_derivation}

In this subsection, we derive the frequency-resolved TURs by establishing the homogeneous-response identities \eqref{eq:langevin_homogeneous_response} and \eqref{eq:jump_homogeneous_response}. 
We begin with the overdamped Langevin case. Consider the homogeneous perturbation $\psi_k(\bm x,\omega)=1$ (for all $k$, $\bm{x}$, and $\omega$). For a current-like observable
\begin{equation}
    \bm\theta(t)=\int d\bm x\,\textsf L(\bm x)\bm\jmath(\bm x,t),
\end{equation}
the corresponding global response is
\begin{equation}
    \bm R^{\bm\theta}_{\ln \textsf M,\bm 1}(\omega)
    =
    \sum_{k=1}^{N}
    \int d\bm z\,
    \bm R^{\bm\theta}_{\ln \mu_k(\bm z)}(\omega).
\end{equation}
Using Eqs.~\eqref{eq:app_langevin_current_response}, \eqref{eq:app_langevin_P_def}, and \eqref{eq:N_mu}, we obtain
\begin{equation}
    \begin{aligned}
        & \bm R^{\bm\theta}_{\ln \textsf{M},\bm 1}(\omega)
        \\ &=
        \int d\bm x\,\textsf{L}(\bm x)
        \sum_{k=1}^{N}\int d\bm z\,
        \textsf{P}(\bm x,\bm z;\omega)\,
        j_k^{\mathrm{ss}}(\bm z)\,\bm e_k 
        \\&=
        \int d\bm x\,\textsf{L}(\bm x)
        \left[
            \bm j_{\mathrm{ss}}(\bm x)
            +
            \hat{\bm J}_{\bm x}
            \int d\bm z\,
            \nabla_{\bm z}^{\mathsf{T}}H(\bm x,\bm z;\omega)\,
            \bm j_{\mathrm{ss}}(\bm z)
        \right].
    \end{aligned}
\end{equation}
The second term vanishes after integration by parts, using the stationarity condition
\begin{equation}
    \nabla_{\bm z}^{\mathsf T}\bm j_{\mathrm{ss}}(\bm z)=0
\end{equation}
and the absence of boundary contributions. Hence, for this choice of perturbation, the response equals the mean value of the current-like observable:
\begin{equation}
    \bm R^{\bm\theta}_{\ln \textsf{M},\bm 1}(\omega)
    =
    \int d\bm x\,\textsf{L}(\bm x)\bm j_{\mathrm{ss}}(\bm x)
    =
    \langle \bm\theta\rangle_{\mathrm{ss}}.
\end{equation}
Substituting this into the frequency-domain response TUR \eqref{eq:langevin_rtur_simple} yields
\begin{equation}
    \langle \bm\theta\rangle_{\mathrm{ss}}^{\mathsf T}
    \textsf{C}_{\bm\theta,\bm\theta^{\mathsf T}}(\omega)^{-1}
    \langle \bm\theta\rangle_{\mathrm{ss}}
    \le
    \frac{\sigma}{2},
\end{equation}
which is the frequency-domain TUR for overdamped Langevin systems.

For the Markov jump case, we consider the homogeneous kinetic perturbation $\psi_{nm}(\omega)=1$ for all $n>m$ and $\omega$. For a single current-like observable $\theta(t) = \sum_{n>m} L_{nm} \jmath_{nm} (t)$, using Eq.~\eqref{eq:app_jump_j_response_B}, we have
\begin{widetext}
\begin{equation}
    \begin{aligned}
        R^{\theta}_{\bm{B},\bm 1} (\omega) &= \sum_{n>m} \sum_{k>l} L_{nm} R^{\jmath_{nm}}_{B_{kl}} (\omega)
        \\ & = \sum_{n>m} \sum_{k>l} L_{nm}
        j_{kl}
        \Bigl[
            \delta_{nk}\delta_{ml}
            -
            \delta_{nl}\delta_{mk}
            +
            W_{nm}\bigl(H_{mk}(\omega)-H_{ml}(\omega)\bigr)
            -
            W_{mn}\bigl(H_{nk}(\omega)-H_{nl}(\omega)\bigr)
        \Bigr].
    \end{aligned}
\end{equation}
\end{widetext}
The terms containing the Kronecker deltas give $\langle \theta\rangle_{\mathrm{ss}}$. 
The remaining terms are linear combinations of expressions of the form $\sum_{k>l} j_{kl}(A_k - A_l)$, with $A_k$ being either $\sum_{n>m} L_{nm} W_{nm} H_{mk}$ or $\sum_{n>m} L_{nm}  W_{mn} H_{nk}$.
They vanish since, for any function $A_k$ of the state, 
\begin{equation}
    \begin{aligned}
\sum_{k>l} j_{kl}(A_k-A_l)
=
\sum_k A_k \sum_{l(\neq k)} j_{kl}
=
0,
    \end{aligned}
\end{equation}
where we used the stationarity condition $\sum_{l(\neq k)} j_{kl}=0$.
Hence, for this choice of perturbation, the response equals the mean value of the current-like observable: $R^\theta_{\bm{B},\bm 1} (\omega) = \langle \theta \rangle_\text{ss}$.
This relation extends componentwise to a vector of current-like observables 
\begin{equation}
    \bm R^{\bm\theta}_{\bm B,\bm 1}(\omega)
    =
    \langle \bm\theta\rangle_{\mathrm{ss}}.
\end{equation}
Substituting this into \eqref{eq:jump_pseudoep_bound_simple} yields
\begin{equation}
    \langle \bm\theta\rangle_{\mathrm{ss}}^{\mathsf T}
    \textsf{C}_{\bm\theta,\bm\theta^{\mathsf T}}(\omega)^{-1}
    \langle \bm\theta\rangle_{\mathrm{ss}}
    \le
    \frac{\sigma^{\mathrm{ps}}}{2},
\end{equation}
which is the frequency-resolved TUR for Markov jump systems.

\subsection{Proof of the equilibrium reciprocity identities}
\label{appsubsec:fdt_reciprocity}

In this subsection, we derive the reciprocity identities used in Sec.~\ref{subsec:fdt} from equilibrium detailed balance. We begin with the overdamped Langevin case.
At equilibrium, we assume a spatially homogeneous temperature, $T_k(\bm{x})=T$, and a conservative force, $F_k(\bm{x})=-\partial_{x_k}U(\bm{x})$, so that the steady-state distribution is given by $\pi(\bm{x})\propto e^{-U(\bm{x})/T}$.
The propagator then satisfies the detailed-balance relation
\begin{equation}
    P(\bm{x},t|\bm{z},0)\pi(\bm{z})
    =
    P(\bm{z},t|\bm{x},0)\pi(\bm{x}).
\end{equation}
Subtracting $\pi(\bm{x})\pi(\bm{z})$ from both sides and integrating this relation over \(t\in[0,\infty)\) with the factor \(e^{i\omega t}\), we obtain
\begin{equation}
    H(\bm{x},\bm{z};\omega)\pi(\bm{z})
    =
    H(\bm{z},\bm{x};\omega)\pi(\bm{x}).
    \label{eq:app_langevin_H_detailed_balance}
\end{equation}

We consider the local current response to a force perturbation. 
Using Eqs.~\eqref{eq:app_langevin_current_response}--\eqref{eq:N_F} gives the \(ij\) component of the response matrix as
\begin{equation}\label{eq:langevin_current_response_F}
    \begin{aligned}
        &\bigl[
            \textsf{R}^{\bm{\jmath}(\bm{x})}_{\bm{F}(\bm{z})}(\omega)
        \bigr]_{ij}
        = \mu_i(\bm{x})\pi(\bm{x})\delta_{ij}\delta(\bm{x}-\bm{z})
        \\
        &+
        \mu_i(\bm{x})\mu_j(\bm{z}) \pi(\bm{z})
        \bigl[
            F_i(\bm{x})-T\partial_{x_i}
        \bigr]
        \partial_{z_j}H(\bm{x},\bm{z};\omega)\, .
    \end{aligned}
\end{equation}
Using Eq.~\eqref{eq:app_langevin_H_detailed_balance} together with
\begin{equation}
    \bigl[
        F_i(\bm{x})-T\partial_{x_i}
    \bigr]\pi(\bm{x})
    =0,
\end{equation}
the second term can be rewritten as
\begin{equation}
    -\mu_i(\bm{x}) \mu_j(\bm{z})T\pi(\bm{x}) \pi(\bm{z})\,\partial_{x_i}\partial_{z_j} \bigl[ H(\bm{z},\bm{x};\omega)/\pi(\bm z) \bigr].
\end{equation}
The same result is obtained by exchanging \((i,\bm{x})\leftrightarrow(j,\bm{z})\). Therefore, Eq.~\eqref{eq:langevin_current_response_F} satisfies the reciprocity identity:
\begin{equation}
    \bigl[
        \textsf{R}^{\bm{\jmath}(\bm{x})}_{\bm{F}(\bm{z})}(\omega)
    \bigr]_{ij}
    =
    \bigl[
        \textsf{R}^{\bm{\jmath}(\bm{z})}_{\bm{F}(\bm{x})}(\omega)
    \bigr]_{ji}.
\end{equation}
In matrix form, this becomes
\begin{equation}
    \bigl[
        \textsf{R}^{\bm{\jmath}(\bm{x})}_{\bm{F}(\bm{z})}(\omega)
    \bigr]_{\mathrm{eq}}
    =
    \bigl[
        \textsf{R}^{\bm{\jmath}(\bm{z})}_{\bm{F}(\bm{x})}(\omega)
    \bigr]_{\mathrm{eq}}^{\mathsf{T}}.
\end{equation}

We now turn to Markov jump systems and derive the reciprocity relation
\begin{equation}\label{eq:jump_reciprocity}
    \bigl[
        R^{\jmath_{nm}}_{F_{kl}}(\omega)
    \bigr]_{\mathrm{eq}}
    =
    \bigl[
        R^{\jmath_{kl}}_{F_{nm}}(\omega)
    \bigr]_{\mathrm{eq}}.
\end{equation}
It is convenient to work in the time domain. The time-domain counterpart of Eq.~\eqref{eq:app_jump_j_response_F} is
\begin{equation}\label{eq:jump_j_response_F_in_time}
    \begin{aligned}
       & R^{\jmath_{nm}}_{F_{kl}}(t,0)
        =
        \frac{a_{kl}}{2}
        \Bigl[
        \bigl(
            \delta_{nk}\delta_{ml}
            -
            \delta_{nl}\delta_{mk}
        \bigr)\delta(t)
        \\
        &+
        W_{nm}\bigl(P_{mk}(t)-P_{ml}(t)\bigr)
        -
        W_{mn}\bigl(P_{nk}(t)-P_{nl}(t)\bigr)
        \Bigr],
    \end{aligned}
\end{equation}
where $P_{ab}(t)\equiv [e^{\textsf{W} t}]_{ab}$.
At equilibrium, the detailed-balance condition implies
\begin{equation}\label{eq:jump_DB}
    W_{ab}\pi_b=W_{ba}\pi_a,
    \qquad
    P_{ab}(t) \pi_b=P_{ba}(t) \pi_a,
\end{equation}
which further implies
\begin{equation}\label{eq:traffic_eq}
    a_{ab}=2W_{ab}\pi_b=2W_{ba}\pi_a.
\end{equation}

To prove the reciprocity relation~\eqref{eq:jump_reciprocity}, we examine how the terms in the response function~\eqref{eq:jump_j_response_F_in_time} transform under the exchange $(nm)\leftrightarrow(kl)$ at equilibrium.
The singular term proportional to \(\delta(t)\) is manifestly invariant under this exchange.
For the regular part, the individual terms are not separately invariant but instead transform into one another.
As an example, consider the term $(a_{kl}/2)W_{nm}P_{mk}(t)$.
Using the detailed-balance relations \eqref{eq:jump_DB} and \eqref{eq:traffic_eq} repeatedly, we obtain
\begin{equation}\label{eq:reciprocity_regular_term1}
    \begin{aligned}
        \frac{a_{kl}}{2}W_{nm}P_{mk}(t)
        &= W_{kl}\pi_l\,W_{nm}P_{mk}(t)
        \\& =W_{mn}\pi_n W_{lk}P_{km}(t)
        \\& =\frac{a_{nm}}{2}W_{lk}P_{km}(t).
    \end{aligned}
\end{equation}
This reflects the balance between the trajectory $(l \to k \to \cdots \to m \to n)$ and its time reversal $(n \to m \to \cdots \to k \to l)$.
The remaining three regular terms are transformed in the same way:
\begin{equation}
    \frac{a_{kl}}{2}W_{nm}P_{ml}(t)
    =
    \frac{a_{nm}}{2}W_{kl}P_{lm}(t),
\end{equation}
\begin{equation}
    \frac{a_{kl}}{2}W_{mn}P_{nk}(t)
    =
    \frac{a_{nm}}{2}W_{lk}P_{kn}(t),
\end{equation}
and
\begin{equation}\label{eq:reciprocity_regular_term4}
    \frac{a_{kl}}{2}W_{mn}P_{nl}(t)
    =
    \frac{a_{nm}}{2}W_{kl}P_{ln}(t).
\end{equation}
Taken together, Eqs.~\eqref{eq:reciprocity_regular_term1}--\eqref{eq:reciprocity_regular_term4} show that the regular part of Eq.~\eqref{eq:jump_j_response_F_in_time} is invariant under the exchange \((nm)\leftrightarrow(kl)\).
Consequently, the full time-domain response satisfies
\begin{equation}
    R^{\jmath_{nm}}_{F_{kl}}(t,0)
    =
    R^{\jmath_{kl}}_{F_{nm}}(t,0)
    \qquad
    (t\ge 0).
\end{equation}
Taking the Fourier transform with respect to \(t\) then yields the frequency-resolved reciprocity relation~\eqref{eq:jump_reciprocity}.

\subsection{Alternative covariance--response identities}
\label{appsubsec:alternative_cov_response}

In this subsection, we derive the alternative covariance--response identities, Eqs.~\eqref{eq:langevin_covariance_alternative} and \eqref{eq:jump_covariance_alternative}, used in Secs.~\ref{subsec:fdt} and \ref{subsec:harada_sasa}.

We begin with the overdamped Langevin case.  Using Eqs.~\eqref{eq:app_langevin_current_response} and \eqref{eq:N_F}, we obtain the local force-response relation
\begin{equation}
    \textsf{R}^{\bm{\jmath}(\bm{x})}_{\bm F(\bm y)}(\omega)
    =
    \textsf P(\bm{x},\bm y;\omega) \textsf M(\bm y)\pi(\bm y).
\end{equation}
Substituting this relation together with
\(\textsf D(\bm y)=\textsf M(\bm y)\textsf T(\bm y)\) into Eq.~\eqref{eq:app_langevin_j_j_compact} yields
\begin{widetext}
\begin{equation}
    \begin{aligned}
        \textsf{C}_{\bm{\jmath}(\bm{x}),\bm{\jmath}(\bm{y})^{\mathsf T}}(\omega)
        & =
        \int d\bm{z}~ \textsf{R}_{\bm{F}(\bm{z})}^{\bm{\jmath}(\bm{x})}(\omega) \textsf{T} (\bm{z}) \bigl[ \textsf{P}(\bm{y},\bm{z};\omega) \bigr]^\dagger 
        + \int d\bm{z} ~ \textsf{P}(\bm{x},\bm{z};\omega) \bigl[\textsf R_{\bm{F}(\bm{z})}^{\bm{\jmath}(\bm{y})}(\omega) \textsf{T}(\bm{z}) \bigr]^\dagger
        \\& = \textsf{R}^{\bm{\jmath}(\bm{x})}_{\bm F(\bm y)}(\omega)\,\textsf{T}(\bm y)
        +
        \textsf{T}(\bm x)\bigl[
            \textsf{R}^{\bm{\jmath}(\bm y)}_{\bm F(\bm x)}(\omega)
        \bigr]^{\dagger}
        \\& ~~~~ + \int d\bm{z}~ \textsf{R}_{\bm{F}(\bm{z})}^{\bm{\jmath}(\bm{x})}(\omega) \textsf{T}(\bm{z}) \bigl[ \hat{\bm{J}}_{\bm{y}}\bm{\nabla}_{\bm{z}}^{\mathsf{T}} H(\bm{y},\bm{z};\omega)
        \bigr]^\dagger
        + \int d\bm{z} ~ \hat{\bm{J}_{\bm{x}}} \bigl[\bm{\nabla}_{\bm{z}}^{\mathsf{T}} H(\bm{x},\bm{z};\omega) \bigr] \textsf T(\bm{z})  \bigl[ \textsf R_{\bm{F}(\bm{z})}^{\bm{\jmath}(\bm{y})}(\omega) \bigr]^\dagger
    \end{aligned}
    \label{eq:app_langevin_current_covariance_expanded}
\end{equation}
Using Eqs.~\eqref{eq:app_langevin_rho_rho_raw}
and \eqref{eq:app_langevin_rho_rho_quadratic},
the integral terms reduce to
\begin{equation}\label{eq:alt_expr}
    \begin{aligned}
        & \hat{\bm{J}}_{\bm{x}} 
            \Bigl( 
                \pi(\bm{y}) \bm{\nabla}_{\bm{y}}^{\mathsf{T}} H(\bm{x},\bm{y};\omega) \textsf{D}(\bm{y})
                + \hat{\bm J}_{\bm y}^{\mathsf{T}} \bigl[ \pi(\bm{y})H(\bm{x},\bm{y};\omega) \bigr]
            \Bigr)  
        + \Bigl[
            \hat{\bm{J}}_{\bm{y}}
                \Bigl( 
                \pi(\bm{x}) \bm{\nabla}_{\bm{x}}^{\mathsf{T}} H(\bm{y},\bm{x};\omega)^* \textsf{D}(\bm{x})
                + \hat{\bm J}_{\bm x}^{\mathsf{T}} \bigl[ \pi(\bm{x})H(\bm{y},\bm{x};\omega)^*
                \bigr]
            \Bigr)
            \Bigr]^{\mathsf{T}}
        \\& = \bigl[
            \hat{\bm J}_{\bm x}H(\bm{x},\bm y;\omega)
        \bigr]
        \bm j_{\mathrm{ss}}(\bm y)^{\mathsf T}
        +
        \bm j_{\mathrm{ss}}(\bm x)
        \bigl[
            \hat{\bm J}_{\bm y}H(\bm y,\bm x;\omega)
        \bigr]^{\dagger},
    \end{aligned}
\end{equation}
where the operator $\hat{\bm{J}}_{\bullet}$ ($\bullet \in \{ \bm{x},\bm{y} \})$ in the first line acts on the entire bracketed expression. Substituting Eq.~\eqref{eq:alt_expr} into
Eq.~\eqref{eq:app_langevin_current_covariance_expanded}, we obtain the alternative covariance--response identity
\begin{equation}
    \begin{aligned}
        &\textsf{C}_{\bm{\jmath}(\bm{x}),\bm{\jmath}(\bm{y})^{\mathsf T}}(\omega)
        = \textsf{R}^{\bm{\jmath}(\bm{x})}_{\bm F(\bm y)}(\omega)\,\textsf{T}(\bm y)
        +
        \textsf{T}(\bm x)\bigl[
            \textsf{R}^{\bm{\jmath}(\bm y)}_{\bm F(\bm x)}(\omega)
        \bigr]^{\dagger}
        + \bigl[
            \hat{\bm J}_{\bm x}H(\bm{x},\bm y;\omega)
        \bigr]
        \bm j_{\mathrm{ss}}(\bm y)^{\mathsf T}
        +
        \bm j_{\mathrm{ss}}(\bm x)
        \bigl[
            \hat{\bm J}_{\bm y}H(\bm y,\bm x;\omega)
        \bigr]^{\dagger}.
    \end{aligned}
    \label{eq:app_langevin_alt_identity}
\end{equation}

We next turn to the Markov jump system. 
Substituting Eqs.~\eqref{eq:app_jump_eta_eta_raw}, \eqref{eq:app_jump_zeta_zeta}, \eqref{eq:app_jump_eta_zeta}, and \eqref{eq:app_jump_zeta_eta} into Eq.~\eqref{eq:app_jump_j_j_raw}, the current--current covariance $C_{\jmath_{nm},\jmath_{n'm'}}$ can be decomposed into three contributions: (i) the $\omega$-independent term $C_{\zeta_{nm},\zeta_{n'm'}} = a_{nm} (\delta_{nn'} \delta_{mm'} - \delta_{nm'} \delta_{mn'})$, (ii) terms involving $H(\omega)$, and (iii) terms involving $H(-\omega)$. The terms involving $H(\omega)$ are collected as
\begin{equation}
    \begin{aligned}
        W_{n'm'} \pi_{m'}  \bigl[
              W_{nm} H_{mn'}(\omega) - W_{mn} H_{nn'}(\omega) 
        \bigr]
        - W_{m'n'} \pi_{n'} \bigl[ W_{nm} H_{mm'}(\omega) - W_{mn} H_{nm'}(\omega) 
        \bigr]
    \end{aligned}
\end{equation}
Using $W_{n'm'}\pi_{m'}=(a_{n'm'}+j_{n'm'})/2$ and $W_{m'n'}\pi_{n'}=(a_{n'm'}-j_{n'm'})/2$, this can be rewritten as
\begin{equation}\label{eq:contribution_plus_omega}
    \begin{aligned}
        & \frac{a_{n'm'}}{2}  
            \Bigl(
                W_{nm} \bigl[H_{mn'}(\omega) - H_{mm'}(\omega) \bigr]
                -  W_{mn} \bigl[ H_{nn'}(\omega) - H_{nm'}(\omega) \bigr]
            \Bigr)
        \\ & + \frac{j_{n'm'}}{2} 
            \Bigl(
                W_{nm} \bigl[ H_{mn'}(\omega) + H_{mm'}(\omega) \bigr] 
                - W_{mn} \bigl[ H_{nn'}(\omega) + H_{nm'}(\omega) \bigr]
            \Bigr)
    \end{aligned}
\end{equation}
Together with half of the $\omega$-independent term $C_{\zeta_{nm},\zeta_{n'm'}}$, the first line of Eq.~\eqref{eq:contribution_plus_omega} reproduces $R^{\jmath_{nm}}_{F_{n'm'}}(\omega)$ in Eq.~\eqref{eq:app_jump_j_response_F}.
For convenience, we define
\begin{equation}\label{eq:def_Z}
    Z_{nm,n'm'}(\omega)
    \equiv
    \frac{1}{2}W_{nm}
    \bigl[
        H_{mn'}(\omega)+H_{mm'}(\omega)
    \bigr] j_{n'm'}
\end{equation}
so that the second line of Eq.~\eqref{eq:contribution_plus_omega} can be compactly written as $Z_{nm,n'm'}(\omega) - Z_{mn,n'm'}(\omega)$.

The terms involving \(H(-\omega)\) are obtained from the same expression by exchanging $(n,m,\omega)\leftrightarrow (n',m',-\omega)$. Combining all contributions, we obtain
\begin{equation}
    \begin{aligned}
        C_{\jmath_{nm},\jmath_{n'm'}}(\omega)
        & =
        R^{\jmath_{nm}}_{F_{n'm'}}(\omega)
        +
        R^{\jmath_{n'm'}}_{F_{nm}}(-\omega)
        +
        Z_{nm,n'm'}(\omega)
        -
        Z_{mn,n'm'}(\omega) 
        +
        Z_{n'm',nm}(-\omega)
        -
        Z_{m'n',nm}(-\omega)
    \end{aligned}
    \label{eq:app_jump_alt_identity}.
\end{equation}
We denote the contribution proportional to the steady-state current by
\begin{equation}\label{eq:def_Delta}
\begin{aligned}
    \Delta_{nm,n'm'}(\omega)
    & \equiv
    Z_{nm,n'm'}(\omega)
    -
    Z_{mn,n'm'}(\omega)
    +
    Z_{n'm',nm}(-\omega)
    -
    Z_{m'n',nm}(-\omega),
\end{aligned}
\end{equation}
so that
\begin{equation}
    C_{\jmath_{nm},\jmath_{n'm'}}(\omega)
    =
    R^{\jmath_{nm}}_{F_{n'm'}}(\omega)
    +
    R^{\jmath_{n'm'}}_{F_{nm}}(-\omega)
    +
    \Delta_{nm,n'm'}(\omega).
\end{equation}

\subsection{Proof of the generalized Harada--Sasa relation} \label{appsubsec:general_harada_sasa}

In this subsection, we provide a detailed derivation of Eq.~\eqref{eq:langevin_harada_sasa_general}. We introduce
the current-like observable $\bm w(t)\equiv \textsf M(\bm x(t))^{-1}\circ\dot{\bm x}(t)$.
We then begin with Eq.~\eqref{eq:langevin_covariance_alternative} at constant temperature $T$, weight it by $\textsf{M}^{-1}(\bm{z})$, and integrate over  $\bm{x}$ and $\bm{z}$. This yields
\begin{align}
    \textsf{C}_{\dot{\bm{x}},\bm{w}^\textsf{T}} (\omega) &= T \int d\bm{x} d\bm{z} \textsf{R} ^{\boldsymbol\jmath (\bm{x})} _{\bm{F}(\bm{z})} (\omega) \textsf{M}^{-1} (\bm{z}) +  T \left[\int d\bm{x} d\bm{z} \textsf{M}^{-1} (\bm{z}) \textsf{R} ^{\boldsymbol\jmath (\bm{z})} _{\bm{F}(\bm{x})} (\omega)  \right]^\dagger \\ \nonumber 
    &\quad +\int d\bm{x}d\bm{z} \left[ \hat{\bm{J}}_{\bm{x}}  H(\bm{x},\bm{z};\omega) \right] \bm{j}_\mathrm{ss} (\bm{z})^\textsf{T} \textsf{M}^{-1}(\bm{z}) + \int d\bm{x}d\bm{z} \bm{j}_\mathrm{ss} (\bm{x}) \left[ \textsf{M}^{-1}(\bm{z})\hat{\bm{J}}_{\bm{z}} H(\bm{z},\bm{x};\omega) \right]^\dagger  
    \;.
\end{align}
To proceed, we take the trace, integrate over $\omega$, and extract the real part. We define
\begin{align}
    I_1 &\equiv \mathrm{Re} \text{Tr} \left[  T \int \frac{d\omega}{2\pi} \int d\bm{x} d\bm{z} \textsf{R} ^{\boldsymbol\jmath (\bm{x})} _{\bm{F}(\bm{z})} (\omega) \textsf{M}^{-1} (\bm{z})\right] \\ \nonumber 
    I_2 &\equiv \mathrm{Re} \text{Tr} \left[ T \int \frac{d\omega}{2\pi} \int d\bm{x} d\bm{z} \textsf{M}^{-1} (\bm{z}) \textsf{R} ^{\boldsymbol\jmath (\bm{z})} _{\bm{F}(\bm{x})} (\omega)  \right] \\ \nonumber 
    I_3 & \equiv \mathrm{Re} \text{Tr} \left[\int \frac{d\omega}{2\pi} \int d\bm{x}d\bm{z} \left[ \hat{\bm{J}}_{\bm{x}}  H(\bm{x},\bm{z};\omega) \right] \bm{j}_\mathrm{ss} (\bm{z})^\textsf{T} \textsf{M}^{-1}(\bm{z}) \right] \\ \nonumber 
    I_4 &\equiv \mathrm{Re} \text{Tr} \left[ \int \frac{d\omega}{2\pi }\int d\bm{x}d\bm{z} \bm{j}_\mathrm{ss} (\bm{x}) \left[ \textsf{M}^{-1}(\bm{z})\hat{\bm{J}}_{\bm{z}} H(\bm{z},\bm{x};\omega) \right]^\dagger \right]
    \;.
\end{align}
Next, we evaluate the difference between $I_1$  and $I_2$. Using the explicit expression $\textsf{R}_{\bm{F}(\bm{z})}^{\bm{\jmath}(\bm{x})} = \mathsf{P}(\bm{x},\bm{z};\omega) \mathsf{M}(\bm{z}) \pi(\bm{z})= [\textsf{I}\,\delta(\bm{x}-\bm{z}) + \hat{\bm{J}}_{\bm{x}}\nabla_{\bm{z}} ^\mathsf{T} H(\bm{x},\bm{z};\omega)] \textsf{M}(\bm{z}) \pi (\bm{z})$, we obtain
\begin{align}
    I_1 - I_2 &= T \mathrm{Re} \mathrm{Tr} \left[ \int \frac{d\omega}{2\pi} \int d\bm{x}d\bm{z} \left[\hat{\bm{J}}_{\bm{x}} \nabla^\textsf{T} _{\bm{z}} H(\bm{x},\bm{z};\omega)\right] \pi (\bm{z}) - \int \frac{d\omega}{2\pi} \int d\bm{x}d\bm{z} \textsf{M} ^{-1} (\bm{x}) \left[\hat{\bm{J}}_{\bm{x}} \nabla^\textsf{T} _{\bm{z}} H(\bm{x},\bm{z};\omega)  \right] \textsf{M}(\bm{z})\pi (\bm{z})\right] \nonumber
    \\ \nonumber
    & = \frac{T}{2} \mathrm{Re} \mathrm{Tr} \left[ \int d\bm{x} d\bm{z} \left[ \hat{\bm{J}}_{\bm{x} }\nabla^\textsf{T} _{\bm{z}} \delta (\bm{x}-\bm{z}) \right] \pi (\bm{z}) - \int d\bm{x} d\bm{z} \textsf{M}^{-1} (\bm{x}) \left[ \hat{\bm{J}}_{\bm{x}} \nabla ^\textsf{T}_{\bm{z}} \delta(\bm{x}-\bm{z}) \right] \textsf{M}(\bm{z})\pi (\bm{z}) \right] 
    \\ \nonumber 
    & = \frac{T}{2} \mathrm{Re}\left[ - \sum_{k,l=1}^N \int d\bm{x} M_{kl} (F_l - T\partial_l)  \partial_k \pi (\bm{x}) +  \sum_{k,l=1}^N \int d\bm{x} (F_k - T\partial_k) \partial_l (M_{lk} \pi )  \right] 
    \\ \nonumber 
    & = \frac{T}{2} \left[ \int d\bm{x} [\nabla_{\bm{x}} \cdot \textsf{M}(\bm{x})]^\textsf{T} \textsf{M}^{-1} (\bm{x})\bm{j}_\mathrm{ss} (\bm{x}) \right]
    \;.
\end{align}
We note that 
\begin{align}
    I_2 = T  \sum_{k=1} ^N\int\frac{d\omega}{2\pi} \mathrm{Re}  R^{w_k} _{F_k ,1}(\omega)
    \;.
\end{align}
Combining this with the above result, we obtain
\begin{align}
    I_1 = T  \sum_{k=1} ^N\int\frac{d\omega}{2\pi} \mathrm{Re}  R^{w_k} _{F_k ,1}(\omega) + \frac{T}{2} \int d\bm{x} [\nabla_{\bm{x}} \cdot \textsf{M}(\bm{x})]^\textsf{T} \textsf{M}^{-1} (\bm{x}) \bm{j}_\mathrm{ss} (\bm{x}) 
    \;.
\end{align}

Next, we evaluate \(I_3\) and \(I_4\). We begin with \(I_4\), whose evaluation is straightforward:
\begin{align}
    I_4 &= \frac{1}{2}  \int d\bm{x} d\bm{z} \bm{j}_\mathrm{ss} (\bm{x}) ^\textsf{T}  \textsf{M}^{-1} (\bm{z}) \hat{\bm{J}}_{\bm{z}} [\delta (\bm{z}-\bm{x}) - \pi (\bm{z})] 
    \\ \nonumber
    & = \frac{1}{2} \int d\bm{x} \bm{j} _\mathrm{ss} (\bm{x}) ^\textsf{T} \bm{F} (\bm{x}) - \frac{1}{2} \int d\bm{x} \bm{j}_\mathrm{ss} (\bm{x})^\textsf{T} \int d\bm{z} \textsf{M}^{-1}(\bm{z}) \bm{j}_\mathrm{ss} (\bm{z}) 
    \\ \nonumber 
    &= \frac{1}{2} \langle \dot{Q} \rangle_\mathrm{ss} - \frac{1}{2} \langle \dot{\bm{x}} \rangle _\mathrm{ss}^\textsf{T} \langle \bm{w} \rangle _\mathrm{ss}
    \;.
\end{align}
The evaluation of \(I_3\) proceeds analogously:
\begin{align}
    I_3 &= \frac{1}{2} \int d\bm{x} d\bm{z} \bm{j}_\mathrm{ss} (\bm{z})^\textsf{T} \textsf{M}^{-1}(\bm{z}) \textsf{M}(\bm{x}) [\bm{F}(\bm{x}) - T\nabla_{\bm{x}}][\delta (\bm{x}-\bm{z}) - \pi (\bm{x})] \\ \nonumber
    & = \frac{1}{2}  \int d\bm{x} \bm{j}_\mathrm{ss} (\bm{x}) ^\textsf{T} \bm{F} (\bm{x}) + \frac{T}{2} \int d\bm{x}  [\nabla \cdot \textsf{M}(\bm{x})]^\textsf{T} \textsf{M}^{-1} (\bm{x}) \bm{j}_\mathrm{ss} (\bm{x}) - \frac{1}{2} \int d\bm{z} \bm{j}_\mathrm{ss} (\bm{z})^\textsf{T} \textsf{M}^{-1}(\bm{z}) \int d\bm{x}  \bm{j}_\mathrm{ss} (\bm{x}) 
    \\ \nonumber 
    & = \frac{1}{2} \langle \dot{Q} \rangle_\mathrm{ss} - \frac{1}{2} \langle \dot{\bm{x}} \rangle _\mathrm{ss}^\textsf{T} \langle \bm{w} \rangle _\mathrm{ss} +  \frac{T}{2} \int d\bm{x}  [\nabla \cdot \textsf{M}(\bm{x})]^\textsf{T} \textsf{M}^{-1} (\bm{x}) \bm{j}_\mathrm{ss} (\bm{x})
    \;.
\end{align}
Combining the above results, we obtain
\begin{align}
    \sum_{k=1}^N \int \mathrm{Re}\left[ C_{\dot{x}_k, w_k} (\omega ) - 2T R^{w_k} _{F_k,1} (\omega)  \right] \frac{d\omega}{2\pi} = \langle \dot{Q} \rangle_\mathrm{ss} - \langle \dot{\bm{x}} \rangle _\mathrm{ss}^\textsf{T} \langle \bm{w} \rangle _\mathrm{ss} + T \int d\bm{x}  [\nabla \cdot \textsf{M}(\bm{x})]^\textsf{T} \textsf{M}^{-1} (\bm{x}) \bm{j}_\mathrm{ss} (\bm{x})
\end{align}
For a divergence-free mobility matrix, ${\bm\nabla} \cdot \mathsf{M} = \bm{0}$, the last term vanishes, and the above identity reduces to Eq.~\eqref{eq:langevin_harada_sasa_general}.

\end{widetext}
\end{appendix}

\bibliography{bibliography}

\end{document}